\documentclass[journal]{IEEEtran}
\usepackage{amsmath,amssymb,amsfonts}
\usepackage{algorithmic}
\usepackage{graphicx}
\usepackage{textcomp}
\usepackage{siunitx}
\usepackage{booktabs}
\usepackage[caption=false]{subfig}
\usepackage{textcomp}
\usepackage{stfloats}
\usepackage{url}
\usepackage{verbatim}
\usepackage[export]{adjustbox}
\usepackage{enumitem}
\usepackage{xcolor}
\usepackage[edges]{forest}
\usepackage{pgfplots}
\pgfplotsset{compat=1.18}

\usepackage[english,vlined,linesnumbered,algoruled]{algorithm2e}
\DontPrintSemicolon

\usepackage{silence}
\usepackage[url=true,doi=false,backend=biber,bibstyle=ieee,citestyle=numeric-comp,isbn=false,sorting=none,defernumbers=true]{biblatex}

\definecolor{TagColor0}{HTML}{6847b2}
\definecolor{TagColor1}{HTML}{2de530}
\definecolor{TagColor2}{HTML}{b20031}
\definecolor{TagColor3}{HTML}{5ba9e5}
\definecolor{TagColor4}{HTML}{9eb223}
\definecolor{TagColor5}{HTML}{c200e5}
\definecolor{TagColor6}{HTML}{47b283}
\definecolor{TagColor7}{HTML}{e55e2d}
\definecolor{TagColor8}{HTML}{0004b2}
\definecolor{TagColor9}{HTML}{87e55b}
\definecolor{TagColor10}{HTML}{b2237a}
\definecolor{TagColor11}{HTML}{00cee5}
\definecolor{TagColor12}{HTML}{b29d47}
\definecolor{TagColor13}{HTML}{8c2de5}
\definecolor{TaxonomyHeader}{HTML}{fff2cc}
\definecolor{TaxonomyEntry}{HTML}{fffbeb}

\newcommand\submittedtext{%
  \footnotesize This work has been submitted to the IEEE for possible publication. Copyright may be transferred without notice, after which this version may no longer be accessible.}

\newcommand\submittednotice{%
    \begin{tikzpicture}[remember picture,overlay]
    \node[anchor=south,yshift=10pt] at (current page.south) {\fbox{\parbox{\dimexpr0.65\textwidth-\fboxsep-\fboxrule\relax}{\submittedtext}}};
    \end{tikzpicture}%
}

\begin{document}

\title{Decentralized Latency Models and Information Flow for TSN Shapers -- A Tutorial}
\author{
    {Alexej Grigorjew}\IEEEauthorrefmark{1},
    {Stefan Geissler}\IEEEauthorrefmark{1},
    {Jürgen Schmitt}\IEEEauthorrefmark{2},
    and {Tobias Hossfeld}\IEEEauthorrefmark{1}, \IEEEmembership{Senior Member,~IEEE}%
    \thanks{This work was partly funded by the Bavarian Ministry of Economic Affairs, Regional Development and Energy under grant number DIK0250/02 (project KOSINU5), and by the Deutsche Forschungsgesellschaft DFG (German Reasearch Agency) under grant number 316878574 (project SDN-App).}%
    \thanks{\IEEEauthorrefmark{1}{Chair of Communication Networks, University of Wuerzburg, Germany}}%
    \thanks{\IEEEauthorrefmark{2}{Siemens AG, Germany}}%
}%

\markboth
{A. Grigorjew \MakeLowercase{\textit{et al.}}: Decentralized Latency Models and Information Flow for TSN Shapers}
{A. Grigorjew \MakeLowercase{\textit{et al.}}: Decentralized Latency Models and Information Flow for TSN Shapers}


\maketitle

\begin{abstract}
Recent advancements in the fields of industrial automation and in-vehicle communication have also increased the requirements on their underlying networks. Traditional field bus technologies are being replaced by regular Ethernet devices with features from the IEEE Time-Sensitive Networking working group. Their standards are designed to achieve zero congestion loss and deterministic latency bounds, but the specific guarantees can differ depending on the shapers used and how the bounds are calculated. After several contributions to the standardization efforts, this work presents a number of decentralized latency models that can be applied locally on the switches, without requiring global knowledge from the rest of the network. It includes a formal definition of delay segments that allows to incorporate interoperability between different shapers and latency models in the future. It presents pseudo-code suggestions for the latency bounds computation of Strict Priority, Credit-Based Shaping, Asynchronous Traffic Shaping, and Cyclic Queuing and Forwarding, all of which can be applied locally on the devices. Finally, five models were implemented and evaluated with respect to their efficiency in terms of latency, jitter, and number of accepted streams. The results illustrate how different shapers are suited for different networks and traffic distributions.
\end{abstract}


\submittednotice


\section{Introduction}
\label{sec:introduction}

Real-time networking has recently become a major focus in scientific literature and standardization efforts. Real-time systems provide guaranteed transmissions with deterministic upper bounds on latency and jitter, which has been a critical requirement in industrial automation and in-vehicle communication for decades. Traditional field bus solutions providing deterministic guarantees are migrating to Ethernet-based networks to satisfy increasing bandwidth demands. However, these second-generation industrial Ethernet systems (e.g., EtherCAT, PROFINET) suffer from closed, proprietary, and homogeneous designs, making them incompatible and requiring gateways for inter-domain communication. Furthermore, unlike main Ethernet standards that scale with technology, proprietary systems often remain locked at their original transfer speeds.

\begin{figure*}
  \centering
  \subfloat[Fully decentralized resource reservation\label{fig:10-intro-examples-decentralized}]{%
    \includegraphics[width=0.3\textwidth,page=1]{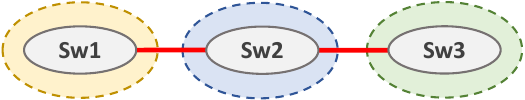}%
  }\hfill%
  \subfloat[Resource reservation between CNC domains\label{fig:10-intro-examples-cnc-domains}]{%
    \includegraphics[width=0.3\textwidth,page=2]{figs/10-decentral-examples-crop.pdf}%
  }\hfill%
  \subfloat[Reshaping when merging redundant FRER streams\label{fig:10-intro-examples-frer}]{%
    \includegraphics[width=0.35\textwidth,page=3]{figs/10-decentral-examples-crop.pdf}%
  }
  \caption{Example use cases for decentralized and interoperable latency models}
  \label{fig:10-intro-examples}
\end{figure*}

To bridge these gaps, the third generation is being developed by the IEEE Time-Sensitive Networking (TSN) task group within the Ethernet ecosystem. Instead of proprietary solutions, TSN introduces openly standardized tools for regular Ethernet switches to enable \emph{converged networking}, allowing Best-Effort and real-time traffic to coexist on shared infrastructure. These mechanisms encompass time synchronization, various shapers, transparent redundancy, and diverse configuration protocols. However, these are individual tools; their correct combination, implementation, and management are left to vendors and operators. While standard profiles suggest feature subsets for specific use cases (e.g., industrial automation \cite{iecieee60802}), they lack full-fledged communication system implementations comparable to legacy technologies.

Often, these individual tools are orchestrated by a Central Network Controller (CNC) \cite{8021Qcc} for resource management and switch configuration. A core CNC task is resource reservation and admission control based on accurate worst-case latency bounds for reserved streams. While many publications address worst-case latency computation for various shapers \cite{Maile_2020_NetworkCalculus}\cite{jiang2024network} targeting CNC use cases, their models typically require global topology and stream information. Latency models for decentralized systems are rare, and papers addressing protocol adaptation \cite{Maile_2023_OnThe} to acquire this information are scarcer. Nevertheless, decentralized latency models offer significant benefits. Crucially, a solid decentralized mathematical foundation can foster true interoperability, i.e., guaranteeing latency across disparate technologies, which is essential for fully converged networking.

Figure~\ref{fig:10-intro-examples} illustrates three scenarios highlighting the benefits of decentralized models. First, Fig.\,\ref{fig:10-intro-examples-decentralized} depicts a fully decentralized setup where each switch manages its own allocations (colored areas) via decentralized control logic. This avoids the CNC single point of failure and accommodates heterogeneous switch architectures, where a single CNC might have trouble if the internal implementations are different from what it expects. Conversely, Fig.\,\ref{fig:10-intro-examples-cnc-domains} uses two CNCs for separate TSN domains. Each domain is configured homogeneously by its CNC but may use different mechanisms. Here, the inter-domain link can leverage decentralized latency models for end-to-end guarantees. Finally, Fig.\,\ref{fig:10-intro-examples-frer} shows a FRER (Frame Replication and Elimination for Reliability) \cite{8021CB} example. The green switch on the left replicates traffic for redundancy, and the red switch on the right eliminates duplicates. However, the remaining traffic can still exceed original burstiness \cite[Sec.\,C.9]{8021CB}, doubling downstream resource demands. Asynchronous Traffic Shaping (ATS) on the red node can mitigate this, but it is rarely implemented due to complex queuing requirements. Transitioning from other shapers to ATS momentarily solves this issue cost-effectively.

Concurrently, the TSN group is developing the Resource Allocation Protocol (RAP) \cite{8021Qdd}, a publish-subscribe reservation protocol for both centralized and decentralized systems. RAP announces streams, allows end devices to attach, and performs local latency calculations and configurations.
To contribute to these efforts, this tutorial addresses the following \textbf{research questions}:
(RQ1, Sec.\,\ref{sec:latency-bounds}) How can worst-case delay bounds be derived locally and decentralized, without global topology and stream knowledge?
(RQ2, Sec.\,\ref{sec:information-flow}) What information is required for local delay bound computation, and how is it acquired during reservation?
(RQ3, Sec.\,\ref{sec:delay-segments}\&\ref{sec:discussion}) How can decentralized latency models facilitate shaper transitions in heterogeneous environments?
(RQ4, Sec.\,\ref{sec:evaluation}) How can different latency models with varying assumptions be compared fairly without bias?
(RQ5, Sec.\,\ref{sec:evaluation}) Which shaper is recommended for specific application scenarios?


The remainder of this work is structured as follows. Section~\ref{sec:related} reviews related work and identifies the research gap. Section~\ref{sec:notation} introduces the notation, symbols, and terms. Section~\ref{sec:delay-segments} defines \emph{delay segments}, which redefine the borders of responsibilities for switch, while Section~\ref{sec:information-flow} details the decentralized information flow. Section~\ref{sec:latency-bounds} presents formal models and pseudo-code implementations for latency computation. Section~\ref{sec:evaluation} evaluates their efficiency in an empirical reservation study. Finally, Section~\ref{sec:discussion} provides a discussion and future outlook, and Section~\ref{sec:conclusion} concludes the paper.


\section{Overview on Related Work}
\label{sec:related}

\begin{table}
    \centering
    \caption{Number of tagged publications by venue.}
    \label{tab:20-venues}
    \begin{tabular}{c>{\hangindent=0.7em}p{24em}}
        \toprule
        \textbf{\# Pubs} & \textbf{Venue} \\
        \midrule
        44 & IEEE Access \\
        30 & IEEE Transactions on Industrial Informatics \\
        10 & IEEE Transactions on Network and Service Management \\
        5 & IEEE International Conference on Emerging Technologies and Factory Automation (ETFA) \\
        5 & IEEE/ACM Transactions on Networking \\
        4 & IEEE International Conference on Communications \\
        4 & IEEE Transactions on Vehicular Technology \\
        3 & IEEE Communications Surveys \& Tutorials \\
        3 & IEEE Global Communications Conference \\
        3 & IEEE International Workshop on Factory Communication Systems (WFCS) \\
        40 & Other ($<3$ papers each) \\
        \bottomrule
    \end{tabular}
\end{table}

\begin{figure}
    \centering
    \resizebox{\columnwidth}{!}{%
    \begin{tikzpicture}
        \begin{axis}[
            ybar,
            bar width=15pt,
            width=12cm,
            height=4cm,
            ylabel={Number of tagged\\publications}, 
            ylabel style={align=center}, 
            ymin=0,
            ymax=35,
            symbolic x coords={2009, 2015, 2016, 2017, 2018, 2019, 2020, 2021, 2022, 2023, 2024, 2025, 2026},
            xtick=data,
            x tick label style={rotate=45, anchor=east},
            nodes near coords,
            nodes near coords align={vertical},
            ymajorgrids=true,
            major grid style={dashed, gray!30},
            axis lines*=left,
            enlarge x limits=0.1,
        ]
        \addplot[
            fill=blue!40, 
            draw=blue!80!black,
            thick
        ] coordinates {
            (2009, 1)
            (2015, 1)
            (2016, 1)
            (2017, 2)
            (2018, 7)
            (2019, 8)
            (2020, 18)
            (2021, 26)
            (2022, 32)
            (2023, 24)
            (2024, 23)
            (2025, 7)
            (2026, 1)
        };
        \end{axis}
    \end{tikzpicture}
    }
    \caption{Number of tagged publications by year.}
    \label{fig:20-publications-by-year}
\end{figure}
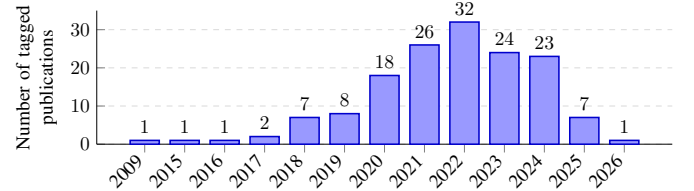

Recent literature tackles deterministic Ethernet latency from multiple directions. For systematic overviews, readers may refer to existing surveys. Specifically, \cite{Nasrallah_2019_UltralowLatency} covers TSN standards and research up to 2018, while more recent surveys focus on distinct use cases like industrial automation~\cite{Zhang_2024_TimesensitiveNetworking}, vehicular networks~\cite{peng2023surveyinvehicle}, or TSN-5G integration~\cite{adil2025integrated}. Other surveys target specific methodologies, such as Time-Aware Shaping scheduling~\cite{xue2025survey} or Network Calculus-based worst-case latency models~\cite{Maile_2020_NetworkCalculus}.

Crucially, none of these surveys address decentralized applicability or the signaling details required to gather information. To map recent research trends, this section structurally overviews publications selected via relevant search terms, which were manually categorized. The initial list was built by querying the OpenAlex\footnote{\url{https://openalex.org/} is a bibliographic catalogue of scientific papers} database using search terms including \emph{Time-Sensitive Networking}, \emph{TSN}, \emph{shaper}, \emph{shaping}, \emph{admission}, \emph{scheduling}, \emph{latency}, \emph{cyclic}, \emph{worst-case delay}, \emph{Network Calculus}, \emph{trajectory approach}, \emph{queuing theory}, \emph{delay bound}, and \emph{decentralized}. Queries used various keyword combinations, each retrieving the top 60 most-cited results from the past decade. Merging results and removing duplicates yielded 311~publications. Each was manually reviewed and assigned the following tags: (i)~\textbf{`survey/tutorial'} for general overviews; (ii)~\textbf{`standards/technical'} for introducing standards or proposing related algorithmic advancements like new shapers; (iii)~\textbf{`5G/6G/wireless'} for combining TSN with wireless technologies. The tags (iv)~\textbf{`central'} and (v)~\textbf{`decentral'} specified the architecture. Specifically, {`central'} was applied if a central authority (e.g., SDN controller, TSN CNC) or global topology knowledge was required. Conversely, {`decentral'} was assigned only if the authors explicitly noted applicability to decentralized networks. We did not infer potential applicability if unstated. Papers matching neither were classified as {`not mentioned'}. Furthermore, (vi)~\textbf{`scheduling'} denoted Gate Control List (GCL) configurations; (vii)~\textbf{`routing'} indicated path-finding contributions; and (viii)~\textbf{`worst-case delay analysis'} marked papers analyzing deterministic upper delay bounds, including but not limited to (ix)~\textbf{`Network Calculus'}. Additionally, (x)~\textbf{`reservation'} signified explicit stream admission control rather than just pre-determined delay minimization, while (xi)~\textbf{`measurement'} required actual data packet transmissions in physical testbeds or simulations, excluding purely algorithmic evaluations. Lastly, (xii)~\textbf{`shaper comparison'} indicated performance evaluations of multiple shaping strategies under identical scenarios; (xiii)~\textbf{`signaling'} marked papers describing the utilization or adaptation of reservation protocols; and (xiv)~\textbf{`acc. max latency'} specified explicit use of the \emph{accumulated maximum latency} field in SRP or RAP for calculations. For context, this tutorial would be tagged with {`survey/tutorial'}, {`decentral'}, {`worst-case delay analysis'}, {`Network Calculus'}, {`reservation'}, {`shaper comparison'}, {`signaling'}, and {`acc. max latency'}.

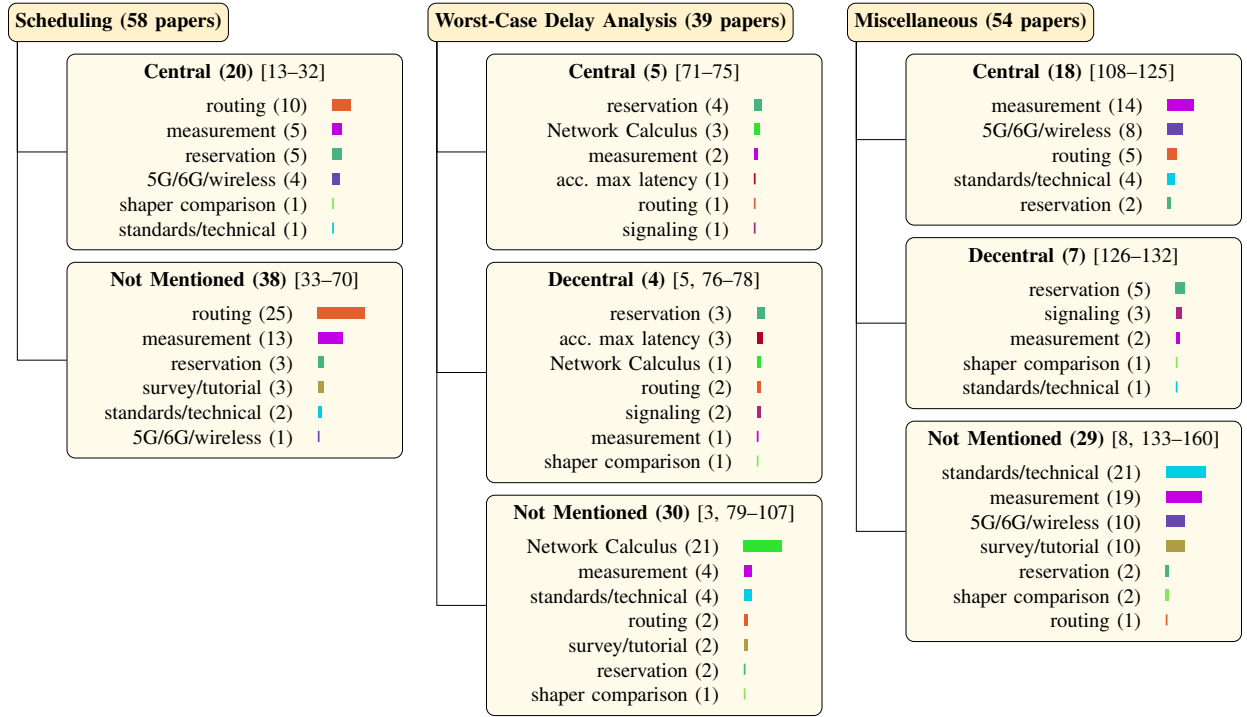
\begin{figure*}
\centering
\resizebox{0.9\textwidth}{!}{%
    \begin{tabular}[t]{@{}ccc@{}}
        \begin{forest}
          baseline,
          for tree={
            folder,
            grow'=0,
            draw,
            rounded corners,
            align=center,
            l sep=1cm,
            s sep=0.2cm,
            if n children=0{fill=TaxonomyEntry}{fill=TaxonomyHeader}
          }
          [{\textbf{Scheduling (58 papers)}}
            [{\textbf{Central (20)} \cite{Nasrallah_2019_PerformanceComparison,Yang_2023_BurstawareTimetriggered,Pang_2021_FlowScheduling,Lu_2023_AnIntelligent,Wang_2023_ReinforcementLearningbased,Sudhakaran_2022_WirelessTime,Guo_2023_MssExploiting,Huang_2023_FlexibleCyclic,Zhang_2021_QosawareMapping,Kong_2021_RuntimeRecovery,Zhang_2024_SeamlessScheduling,Alvarez_2022_ImplementingA,Xue_2024_SchedulingTimecritical,Zhang_2023_DynamicRapid,Bringhenti_2022_ATwofold,Zhang_2024_ScalableScheduling,Zhang_2025_PrioritydominatedTraffic,Lin_2025_FlextasFlexible,Feng_2025_AnEfficient,Zhang_2021_WirelesswiredIntegrated}\\[1ex]\parbox{5.5cm}{\centering\begin{tabular}{rl}
        routing (10) & \textcolor{TagColor7}{\rule[0.2ex]{0.32cm}{1.2ex}} \\
        measurement (5) & \textcolor{TagColor5}{\rule[0.2ex]{0.16cm}{1.2ex}} \\
        reservation (5) & \textcolor{TagColor6}{\rule[0.2ex]{0.16cm}{1.2ex}} \\
        5G/6G/wireless (4) & \textcolor{TagColor0}{\rule[0.2ex]{0.13cm}{1.2ex}} \\
        shaper comparison (1) & \textcolor{TagColor9}{\rule[0.2ex]{0.03cm}{1.2ex}} \\
        standards/technical (1) & \textcolor{TagColor11}{\rule[0.2ex]{0.03cm}{1.2ex}} \\
        \end{tabular}}}]
            [{\textbf{Not Mentioned (38)} \cite{Atallah_2020_RoutingAnd,Gavrilut_2018_AvbawareRouting,St_ber_2023_ASurvey,Zhang_2022_EfficientFlow,Patti_2023_DeadlineawareOnline,Jin_2020_RealtimeScheduling,Wang_2023_JointRouting,Yu_2020_AdaptiveGroup,Yu_2020_OnlineScheduling,Seijo_2021_WsharpImplementation,Zhang_2020_PacketsizeAware,Sun_2024_GraphAttention,Barzegaran_2021_CommunicationScheduling,Alexandris_2022_ConfigurationAnd,Bhattacharjee_2022_LatencyawareFunction,Xu_2024_EfficientTasknetwork,Zhang_2022_ScalableNowait,Fiori_2024_ProposalAnd,Sun_2024_IrfsA,Akram_2024_EnhancingReliability,Akram_2023_JointScheduling,Nakayama_2021_RealtimeReconfiguration,St_ber_2024_PerformanceComparison,Tan_2022_LargescaleDeterministic,Li_2024_DeterminacyorientedTask,Hong_2024_ResourceawareOnline,Zhou_2024_LooselyCoupled,Tian_2025_GtsnetA,Arestova_2023_OptimizationOf,Xu_2025_ScalableScheduling,Cao_2021_EnhancedSystem,Chahed_2023_TsnNetwork,Gavrilut_2018_SchedulingIn,Jin_2021_JointAlgorithm,Reusch_2020_WindowbasedSchedule,Kim_2021_DevelopmentOf,Mai_2019_AHybrid,Bujosa_2022_HermesHeuristic}\\[1ex]\parbox{5.5cm}{\centering\begin{tabular}{rl}
        routing (25) & \textcolor{TagColor7}{\rule[0.2ex]{0.81cm}{1.2ex}} \\
        measurement (13) & \textcolor{TagColor5}{\rule[0.2ex]{0.42cm}{1.2ex}} \\
        reservation (3) & \textcolor{TagColor6}{\rule[0.2ex]{0.10cm}{1.2ex}} \\
        survey/tutorial (3) & \textcolor{TagColor12}{\rule[0.2ex]{0.10cm}{1.2ex}} \\
        standards/technical (2) & \textcolor{TagColor11}{\rule[0.2ex]{0.06cm}{1.2ex}} \\
        5G/6G/wireless (1) & \textcolor{TagColor0}{\rule[0.2ex]{0.03cm}{1.2ex}} \\
        \end{tabular}}}]
          ]
        \end{forest}
        & 
        \begin{forest}
          baseline,
          for tree={
            folder,
            grow'=0,
            draw,
            rounded corners,
            align=center,
            l sep=1cm,
            s sep=0.2cm,
            if n children=0{fill=TaxonomyEntry}{fill=TaxonomyHeader}
          }
          [{\textbf{Worst-Case Delay Analysis (39 papers)}}
            [{\textbf{Central (5)} \cite{Zhao_2024_MinimumBandwidth,Wang_2022_ADelayoptimized,Prados_Garzon_2020_LearnetReinforcement,Maile_2022_DelayguaranteeingAdmission,salomon2026negotiating}\\[1ex]\parbox{5.5cm}{\centering\begin{tabular}{rl}
        reservation (4) & \textcolor{TagColor6}{\rule[0.2ex]{0.13cm}{1.2ex}} \\
        Network Calculus (3) & \textcolor{TagColor1}{\rule[0.2ex]{0.10cm}{1.2ex}} \\
        measurement (2) & \textcolor{TagColor5}{\rule[0.2ex]{0.06cm}{1.2ex}} \\
        acc. max latency (1) & \textcolor{TagColor2}{\rule[0.2ex]{0.03cm}{1.2ex}} \\
        routing (1) & \textcolor{TagColor7}{\rule[0.2ex]{0.03cm}{1.2ex}} \\
        signaling (1) & \textcolor{TagColor10}{\rule[0.2ex]{0.03cm}{1.2ex}} \\
        \end{tabular}}}]
            [{\textbf{Decentral (4)} \cite{Miserez_2024_ExploitingQueue,Desai_2022_MalocBuilding,Maile_2023_OnThe,Grigorjew_2020_BoundedLatency}\\[1ex]\parbox{5.5cm}{\centering\begin{tabular}{rl}
        reservation (3) & \textcolor{TagColor6}{\rule[0.2ex]{0.13cm}{1.2ex}} \\
        acc. max latency (3) & \textcolor{TagColor2}{\rule[0.2ex]{0.10cm}{1.2ex}} \\
        Network Calculus (1) & \textcolor{TagColor1}{\rule[0.2ex]{0.06cm}{1.2ex}} \\
        routing (2) & \textcolor{TagColor7}{\rule[0.2ex]{0.06cm}{1.2ex}} \\
        signaling (2) & \textcolor{TagColor10}{\rule[0.2ex]{0.06cm}{1.2ex}} \\
        measurement (1) & \textcolor{TagColor5}{\rule[0.2ex]{0.03cm}{1.2ex}} \\
        shaper comparison (1) & \textcolor{TagColor9}{\rule[0.2ex]{0.03cm}{1.2ex}} \\
        \end{tabular}}}]
            [{\textbf{Not Mentioned (30)} \cite{Li_2022_BandwidthAllocation,Zhao_2018_WorstcaseLatency,Zhao_2022_QuantitativePerformance,Thomas_2022_WorstcaseDelay,Joung_2021_ZeroJitter,Zhang_2019_AFeasibility,Finzi_2020_WorstcaseTiming,Shalghum_2022_WorstcaseLatency,Fukui_2022_DelaybasedShaper,Mohammadpour_2022_AnalysisOf,Hasan_2021_ImproveService,Specht_2016_UrgencybasedScheduler,Joung_2019_RegulatingScheduler,Maxim_2017_DelayAnalysis,He_2017_ImpactAnalysis,Zhao_2018_TimingAnalysis,Kalor_2018_NetworkSlicing,Zhao_2021_LatencyAnalysis,Mohammadpour_2018_LatencyAnd,Zhao_2021_ImprovingLatency,Maile_2020_NetworkCalculus,Zhang_2019_AnalysisOf,Thiele_2015_FormalTiming,Zhou_2020_ASurvey,Gao_2023_OptimalControl,Li_2024_FormalModeling,Yuan_2022_AdaptivePriority,Zhou_2024_HybridScheduling,Shalghum_2021_CriticalOffset,Dai_2020_FixedpriorityScheduling}\\[1ex]\parbox{5.5cm}{\centering\begin{tabular}{rl}
        Network Calculus (21) & \textcolor{TagColor1}{\rule[0.2ex]{0.65cm}{1.2ex}} \\
        measurement (4) & \textcolor{TagColor5}{\rule[0.2ex]{0.13cm}{1.2ex}} \\
        standards/technical (4) & \textcolor{TagColor11}{\rule[0.2ex]{0.13cm}{1.2ex}} \\
        routing (2) & \textcolor{TagColor7}{\rule[0.2ex]{0.06cm}{1.2ex}} \\
        survey/tutorial (2) & \textcolor{TagColor12}{\rule[0.2ex]{0.06cm}{1.2ex}} \\
        reservation (2) & \textcolor{TagColor6}{\rule[0.2ex]{0.03cm}{1.2ex}} \\
        shaper comparison (1) & \textcolor{TagColor9}{\rule[0.2ex]{0.03cm}{1.2ex}} \\
        \end{tabular}}}]
          ]
        \end{forest}
        & 
        \begin{forest}
          baseline,
          for tree={
            folder,
            grow'=0,
            draw,
            rounded corners,
            align=center,
            l sep=1cm,
            s sep=0.2cm,
            if n children=0{fill=TaxonomyEntry}{fill=TaxonomyHeader}
          }
          [{\textbf{Miscellaneous (54 papers)}}
            [{\textbf{Central (18)} \cite{Val_2022_Ieee8021as,Seijo_2022_TacklingThe,Bhattacharjee_2021_NetworkSlicing,Balasubramanian_2023_FedtsnJoint,Miranda_2023_EnablingTimesensitive,Nikhileswar_2022_TimesensitiveNetworking,Luque_Schempp_2023_AutomadaptZero,Avila_Campos_2024_UnlockingMobility,Bulbul_2022_ReinforcementLearning,Miranda_2022_TimesensitiveNetworking,Hagargund_2023_DtpfAlgorithm,Hu_2021_MultidomainTime,Miranda_2022_EvaluatingTimesensitive,Hackel_2023_SecureTimesensitive,Zhang_2022_SvtiSfcdriven,Prados_Garzon_2021_OptimizationOf,Li_2020_PracticalImplementation,Diederich_2025_LcdnProviding}\\[1ex]\parbox{5.5cm}{\centering\begin{tabular}{rl}
        measurement (14) & \textcolor{TagColor5}{\rule[0.2ex]{0.45cm}{1.2ex}} \\
        5G/6G/wireless (8) & \textcolor{TagColor0}{\rule[0.2ex]{0.26cm}{1.2ex}} \\
        routing (5) & \textcolor{TagColor7}{\rule[0.2ex]{0.16cm}{1.2ex}} \\
        standards/technical (4) & \textcolor{TagColor11}{\rule[0.2ex]{0.13cm}{1.2ex}} \\
        reservation (2) & \textcolor{TagColor6}{\rule[0.2ex]{0.06cm}{1.2ex}} \\
        \end{tabular}}}]
            [{\textbf{Decentral (7)} \cite{Bujosa_2021_CsrpAn,grigorjew2021mlassisted_MlassistedLatency,Grigorjew_2021_DynamicRealtime,Grigorjew_2021_DistributedImplementation,Alvarez_2020_ComparingAdmission,Raunecker_2024_CentralizedVs,jasperneite2018investigation_InvestigationOn}\\[1ex]\parbox{5.5cm}{\centering\begin{tabular}{rl}
        reservation (5) & \textcolor{TagColor6}{\rule[0.2ex]{0.16cm}{1.2ex}} \\
        signaling (3) & \textcolor{TagColor10}{\rule[0.2ex]{0.10cm}{1.2ex}} \\
        measurement (2) & \textcolor{TagColor5}{\rule[0.2ex]{0.06cm}{1.2ex}} \\
        shaper comparison (1) & \textcolor{TagColor9}{\rule[0.2ex]{0.03cm}{1.2ex}} \\
        standards/technical (1) & \textcolor{TagColor11}{\rule[0.2ex]{0.03cm}{1.2ex}} \\
        \end{tabular}}}]
            [{\textbf{Not Mentioned (29)} \cite{Nasrallah_2019_UltralowLatency,Seol_2021_TimelySurvey,Romanov_2021_APrecise,Sharma_2023_TowardDeterministic,Sasiain_2025_TowardThe,Wang_2022_IndustedgeA,Agarwal_2019_MultilevelTimesensitive,Ulbricht_2024_TsnflextestFlexible,Mutalemwa_2020_AClassification,Liu_2024_ImprovingTsn,Bosk_2022_MethodologyAnd,Sanchez_Garrido_2020_DigitalElectrical,Walrand_2023_AConcise,Garbugli_2022_AFramework,Gomez_2023_StrategiesFor,Fernandez_2019_AnalysisAnd,Guan_2021_SoftwarizedIndustrial,Fu_2023_ASurvey,Maile_2022_EnsuringReliable,Haxhibeqiri_2022_SafetyrelatedApplications,Zhu_2020_MeasuresTo,Ihle_2024_P4psfpP4based,Zhou_2020_MappingTsn,Ben_Ahmed_2024_HardwareDesign,Weichlein_2023_DataFlow,Zhou_2021_SimulatingTsn,Behnke_2024_RealtimePerformance,Luo_2022_DesignMethodology,Bilstrup_2009_OnThe}\\[1ex]\parbox{5.5cm}{\centering\begin{tabular}{rl}
        standards/technical (21) & \textcolor{TagColor11}{\rule[0.2ex]{0.68cm}{1.2ex}} \\
        measurement (19) & \textcolor{TagColor5}{\rule[0.2ex]{0.61cm}{1.2ex}} \\
        5G/6G/wireless (10) & \textcolor{TagColor0}{\rule[0.2ex]{0.32cm}{1.2ex}} \\
        survey/tutorial (10) & \textcolor{TagColor12}{\rule[0.2ex]{0.32cm}{1.2ex}} \\
        reservation (2) & \textcolor{TagColor6}{\rule[0.2ex]{0.06cm}{1.2ex}} \\
        shaper comparison (2) & \textcolor{TagColor9}{\rule[0.2ex]{0.06cm}{1.2ex}} \\
        routing (1) & \textcolor{TagColor7}{\rule[0.2ex]{0.03cm}{1.2ex}} \\
        \end{tabular}}}]
          ]
        \end{forest}
    \end{tabular}%
}
    \caption{Distribution of topics in related publications; (parantheses) indicate number of publications; [brackets] contain reference IDs}
    \label{fig:20-related-work-taxonomy}
\end{figure*}

From the initial 311~publications, 160~unrelated papers were excluded. Table~\ref{tab:20-venues} outlines the remaining publications by venue, which is incidentally dominated by IEEE journals, while Figure~\ref{fig:20-publications-by-year} illustrates their chronological distribution. Notably, over 50\% of these papers were published between 2022--2026. Figure~\ref{fig:20-related-work-taxonomy} details the final tag distribution, where papers are categorized under {`scheduling'}, {`worst-case delay analysis'}, or {`miscellaneous'}. These categories are further subdivided into {`central'}, {`decentral'}, and {`not mentioned'}. Each subgroup lists its respective references alongside the distribution of the 10~remaining tags.

The `scheduling' category contains the majority of publications and lacks a `decentral' subgroup. Its most common supplementary tags are `routing' and `measurement', reflecting the prevalent research focus: resolving joint scheduling and routing problems via GCL computation and validating the configurations empirically. Although 38 scheduling papers omit mentioning a central authority, most of them likely depend on it due to the nature of the scheduling problem: temporal isolation of streams on their end-to-end path. The next largest group comprises 54 `miscellaneous' papers. As they are neither configuring GCL schedules nor computing worst-case delays, they primarily address measurements and technical advancements like new shaping mechanisms and protocols. Within the `decentral' subgroup, seven publications focus heavily on resource reservation and signaling protocols \cite{Bujosa_2021_CsrpAn,Grigorjew_2021_DistributedImplementation}, compare centralized and decentralized admission control \cite{Alvarez_2020_ComparingAdmission,Raunecker_2024_CentralizedVs,jasperneite2018investigation_InvestigationOn}, evaluate optimal resource configuration \cite{grigorjew2021mlassisted_MlassistedLatency}, or adapt GCL configurations dynamically \cite{Grigorjew_2021_DynamicRealtime}.
The `worst-case delay analysis' group is heavily dominated by `Network Calculus' (NC) publications, which provide a convenient framework for deterministic latency calculations. While most delay analysis papers omit explicit mention of architectural applicability, their arrival curves typically require preceding shaper and path details, and their service curves rely on per-stream concatenation and leftover-service properties \cite{Maile_2020_NetworkCalculus}. Consequently, these works implicitly assume a central controller unless local information acquisition is explicitly addressed. Furthermore, most shapers, excluding Asynchronous Traffic Shaping (ATS), suffer from hop-by-hop burstiness cascading (cf.\,Sec.\,\ref{ssec:sp-simple}). Accounting for this effect in worst-case bounds is possible, but it requires global recalculations whenever new streams enter the network. To prevent signaling storms in decentralized architectures, traffic-independent bounds for burstiness cascading are required. Following our previous work~\cite{Grigorjew_2020_BoundedLatency}, this tutorial utilizes the reservation protocol's `accumulated maximum latency' field to achieve this -- a technique shared by only three other listed publications \cite{Miserez_2024_ExploitingQueue,Maile_2023_OnThe,salomon2026negotiating}. Only four publications feature a explicitly distinguished `decentral' latency analysis \cite{Miserez_2024_ExploitingQueue,Desai_2022_MalocBuilding,Maile_2023_OnThe,Grigorjew_2020_BoundedLatency}; two of these focus primarily on path finding in delay-constrained networks, while the others restrict their scope to CBS and SP shaping, respectively. Generally, most delay analyses investigate only a single shaper, though an extensive NC-based performance comparison of SP, ATS, TAS, and CBS is available in \cite{Zhao_2022_QuantitativePerformance}, albeit without a decentralized focus.

\subsection*{Lessons learned}
Most existing scientific publications focus on GCL scheduling optimization or Network Calculus-based latency analysis. However, current latency analysis literature leaves a significant gap between formal models and practical deployment. Implementing Network Calculus models can be highly challenging when the required information sources remain unclarified. This tutorial complements prior work by strengthening the decentralized dimension of worst-case latency analysis. It explicitly details the mathematical models, demonstrates local information acquisition, provides reference implementations for each model, and compares their admission control efficiency. No other known publication spans these categories.



\section{Notion and Notation}
\label{sec:notation}

\newcommand{\tabsection}[2]{
  \multicolumn{3}{c}{
    \rule[0.5ex]{#1}{0.4pt}
    \textbf{#2}
    \rule[0.5ex]{#1}{0.4pt}
  }
}

\begin{table}
    \centering
    \caption{Overview of used symbols and abbreviations}
    \vspace{-0.5em}
    \label{tab:20-symbols}
    \begin{tabular}{>{\hangindent=0.7em}p{2.5em}
                    >{\hangindent=0.7em}p{3.5em}
                    >{\hangindent=0.7em}p{19.5em}} \toprule
        \textbf{Symbol} & \textbf{Ref.} & \textbf{Description} \\ \midrule
        
        TSN & Sec.\,\ref{sec:introduction} & Time-Sensitive Networking, an IEEE standardization working group \\
        RAP & Sec.\,\ref{sec:introduction} & Resource Allocation Protocol, IEEE 802.1DD \\
        SP & Sec.\,\ref{ssec:sp-simple} & Strict Priority (no shaping) \\
        ATS & Sec.\,\ref{ssec:ats} & Asynchronous Traffic Shaping \\
        CBS/A & Sec.\,\ref{ssec:cbs} & Credit-Based Shaping (Algorithm) \\
        CQF & Sec.\,\ref{ssec:cqf} & Cyclic Queuing and Forwarding \\
        TAS & Sec.\,\ref{ssec:cqf} & Time-Aware Shaping \\
        NC & Sec.\,\ref{ssec:sp-nc} & Network Calculus, delay calculation framework \\

        \tabsection{2.2cm}{Priorities and delay segments} \\

        $\mathcal{P}$ & Sec.\,\ref{sec:notation} & Priority $0..7$ of a stream (4,5,6,7 for real-time) \\
        $\mathcal{C}$ & Sec.\,\ref{sec:notation} & Traffic class of a stream in the data plane of a specific device, based on $\mathcal{P}$ (conversion table) \\
        $\mathcal{C}_i^\textit{up}$ & Sec.\,\ref{ssec:per-stream-info} & Traffic class of stream~$i$ in the previous \emph{upstream} device \\
        $\hat{\mathcal{C}}$ & Sec.\,\ref{ssec:sp-simple} & Highest traffic class of observed device \\
        \textit{prv} & Fig.\,\ref{fig:30-delay-segments} & Previous delay segment \\
        \textit{seg} & Fig.\,\ref{fig:30-delay-segments} & Currently observed delay segment \\
        \textit{nxt} & Fig.\,\ref{fig:30-delay-segments} & Next delay segment \\
        $\delta_\mathcal{P}^\textit{seg}$ & Sec.\,\ref{ssec:a-priori-info} & Pre-configured delay threshold for priority~$\mathcal{P}$ on delay segment~\textit{seg} \\[0.2em]
        $d_\mathcal{P}^\textit{seg}$ & Sec.\,\ref{ssec:a-priori-info} & Calculated delay bound for successfully reserved streams for priority~$\mathcal{P}$ on delay segment~\textit{seg} \\

        \tabsection{2.25cm}{Traffic Specification (TSpec)} \\

        $\mathcal{L}_i^-$ & Sec.\,\ref{ssec:per-stream-info} & Accumulated minimum latency of stream~$i$ \\
        $\mathcal{L}_i^+$ & Sec.\,\ref{ssec:per-stream-info} & Accumulated maximum latency of stream~$i$ \\
        $\ell_i^\textit{min}$ & Sec.\,\ref{ssec:per-stream-info} & Minimum frame length of stream~$i$ \\
        $\ell_i^\textit{max}$ & Sec.\,\ref{ssec:per-stream-info} & Maximum frame length of stream~$i$ \\
        $\ell^\textit{max}$ & Sec.\,\ref{ssec:sp-simple} & Maximum frame length of all streams on all priorities (typically MTU, e.g., \SI{1542}{bytes}) \\
        $R^\textit{seg}$ & Sec.\,\ref{ssec:per-stream-info} & Bandwidth (service rate) of delay segment~\textit{seg} \\
        $b_i$ & Sec.\,\ref{ssec:per-stream-info} & Committed burstiness of stream~$i$ (at the talker) \\
        $r_i$ & Sec.\,\ref{ssec:per-stream-info} & Committed data rate of stream~$i$ (at the talker) \\

        \tabsection{2.65cm}{Latency Computation} \\

        $w_i(\Delta t)$ & Eq.\,\ref{eq:token-bucket} & Traffic volume (workload) generated by stream~$i$ during time duration $\Delta t$ \\
        $b_i^\textit{eff}$ & Eq.\,\ref{eq:effective-burstiness} & Maximum \emph{effective} burstiness of stream~$i$ at observed device \\
        $\mathcal{H}_{\mathcal{C}}$ & Eq.\,\ref{eq:sp-simple-delay} & Set of all streams~$i$ with $\mathcal{C}_i > \mathcal{C}$ \\
        $b_{\mathcal{H}_\mathcal{C}}$ & Eq.\,\ref{eq:sp-service-curve-general-2} & Sum of all effective bursts~$b_i^\textit{eff}$ with $i \in \mathcal{H}_\mathcal{C}$ \\
        $r_{\mathcal{H}_\mathcal{C}}$ & Eq.\,\ref{eq:sp-service-curve-general-2} & Sum of all data rates~$r_i$ with $i \in \mathcal{H}_\mathcal{C}$ \\
        $\mathcal{S}^\textit{seg}$ & Alg.\,\ref{alg:spSimple} & Set of all successfully reserved streams at observed device \\
        $\hat{\mathcal{S}}$ & Eq.\,\ref{eq:tbhat} & Abbreviation for $\mathcal{S}_\mathcal{C}^\textit{seg}$: set of all streams that traverse link~\textit{seg} with traffic class~$\mathcal{C}$ \\
        $\hat{\mathcal{S}}_k$ & Alg.\,\ref{alg:wc-delay-sp} & Set of streams $\mathcal{S}_\mathcal{C}^{\textit{prv}_k}$ from the $k$-th ingress segment~$\textit{prv}_k$ \\
        $t_b$ & Eq.\,\ref{eq:tb} & Bending position of a single stream arrival curve~$\alpha_i$, after depleting the effective burst $b_i^\textit{eff}$ \\
        $t_{\hat{b}}$ & Eq.\,\ref{eq:tbhat} & Bending position of an aggregate arrival curve~$\alpha_{\hat{\mathcal{S}}}$ for all streams of an ingress \textit{prv} \\
        $t_{\hat{cr}}$ & Eq.\,\ref{eq:cbs-tcr} & 2nd bending position for CBS arrival curves, after depleting positive credits $\textit{maxcr}_\mathcal{C}^\textit{prv}$ \\
        $\alpha_i(\Delta t)$ & Eq.\,\ref{eq:arrival-curve-single-stream} & Arrival curve function for a single stream~$i$ \\
        $\alpha_{\hat{\mathcal{S}}}(\Delta t)$ & Eq.\,\ref{eq:arrival-curve-single-port} & Aggregate arrival curve of all streams in~$\hat{\mathcal{S}}$ \\[0.2em]
        $\beta_\mathcal{C}^\textit{seg}(\Delta t)$ & Eq.\,\ref{eq:spnc-beta} & Service curve for delay segment~\textit{seg} and class~$\mathcal{C}$ \\[0.2em]
        $\tau_\mathcal{C}^\textit{seg}$ & Eq.\,\ref{eq:spnc-beta} & Initial delay of rate-latency service curve $\beta_\mathcal{C}^\textit{seg}$ \\
        $\textit{isl}_\mathcal{C}^\textit{seg}$ & Eq.\,\ref{eq:spnc-alphaS2} & Idle slope of CBS delay segment~\textit{seg} and class~$\mathcal{C}$ \\
        $maxcr_\mathcal{C}^\textit{seg}$ & Eq.\,\ref{eq:maxcr} & Upper bound for accumulated, positive CBS credits of class~$\mathcal{C}$ on delay segment~\textit{seg} \\

        \tabsection{3.33cm}{Evaluation} \\

        $\mathcal{S}^\mathcal{P}_a$ & Sec.\,\ref{ssec:kpis} & Set of all \emph{accepted} streams in priority~$\mathcal{P}$ \\[0.2em]
        $b_a^\mathcal{P}$ & Sec.\,\ref{ssec:kpis} & Sum of all accepted streams' bursts in priority~$\mathcal{P}$ \\[0.2em]
        $r_a^\mathcal{P}$ & Sec.\,\ref{ssec:kpis} & Sum of all accepted streams' data rates in prio.~$\mathcal{P}$ \\
        $\mathcal{L}^\Delta_i$ & Sec.\,\ref{ssec:kpis} & Max. latency jitter at receiver, $\mathcal{L}^\Delta_i = \mathcal{L}_i^+ - \mathcal{L}_i^-$ \\
        $t_\textit{comp}$ & Sec.\,\ref{ssec:kpis} & Time needed to compute all latency bounds of a scenario \\

        \bottomrule
    \end{tabular}
\end{table}

This section serves as a reference for commonly used terms and symbols throughout this paper. The tutorial is designed to be self-contained, hence this section does not require active reading, but it can be consulted if subsequent equations need clarification.

First, several terms are used synonymously to improve readability. \emph{Frame} and \emph{packet} both refer to a continuous layer-2 Ethernet protocol data unit. Similarly, \emph{delay} and \emph{latency} are used interchangeably. The term \emph{latency jitter} denotes the \emph{maximum possible jitter}, i.e., the total range between maximum and minimum end-to-end latency, rather than standard deviations or differences between consecutive frames. Here, \emph{shaper} refers to a Transmission Selection Algorithm (TSA) from IEEE 802.1Q \cite{8021Q}, specifically Strict Priority (SP), Asynchronous Traffic Shaping (ATS), Cyclic Queuing and Forwarding (CQF), or Credit-Based Shaping (CBS). The latter is also known as the Credit-Based Shaping Algorithm (CBSA). The acronym \emph{NC} stands for \emph{Network Calculus}. Most latency models in this tutorial rely on a pre-configured \emph{latency threshold}, also called a \emph{latency configuration} or \emph{latency budget} in literature. It represents the pre-configured upper bound~$\delta_{\mathcal{P}}^\textit{seg}$ for worst-case latency on an observed link~\textit{seg} and priority~$\mathcal{P}$.

Important distinctions exist between seemingly similar terms. Unlike the threshold $\delta_{\mathcal{P}}^\textit{seg}$, the computed delay $d_{\mathcal{P}}^\textit{seg}$ is the active worst-case estimate for already reserved traffic. Admission control enforces $d_{\mathcal{P}}^\textit{seg} \leq \delta_{\mathcal{P}}^\textit{seg}$; any reservation exceeding this is rejected. Furthermore, \emph{priority}~$\mathcal{P}$ and \emph{traffic class}~$\mathcal{C}$ are related, but not the same thing. Each stream belongs to a specific priority~$\mathcal{P}$ included in its reservation announcement, and each latency threshold $\delta_{\mathcal{P}}^\textit{seg}$ is configured for a specific priority~$\mathcal{P}$. Both terms operate in the control plane before or during reservation. While IEEE 802.1Q supports up to eight priorities \cite{8021Q}, switches may not implement all eight in the data plane. A local mapping table maps priority $\mathcal{P}$ to a local traffic class $\mathcal{C}$ on each device. During queuing and forwarding, packets within the same class $\mathcal{C}_i$ are treated equally important, even if they map to different priorities $\mathcal{P}_x, \mathcal{P}_y$ in the control plane. Consequently, single-device latency analysis focuses on the local traffic class $\mathcal{C}$. Without loss of generality, this paper assumes switches support all eight traffic classes, with the highest four (7, 6, 5, 4) dedicated to real-time traffic.

Table~\ref{tab:20-symbols} provides a comprehensive reference list of symbols used in the equations throughout this paper.


\section{Delay Model and Delay Segments}
\label{sec:delay-segments}

\begin{figure}
	\centering
	\includegraphics[scale=0.5,page=1]{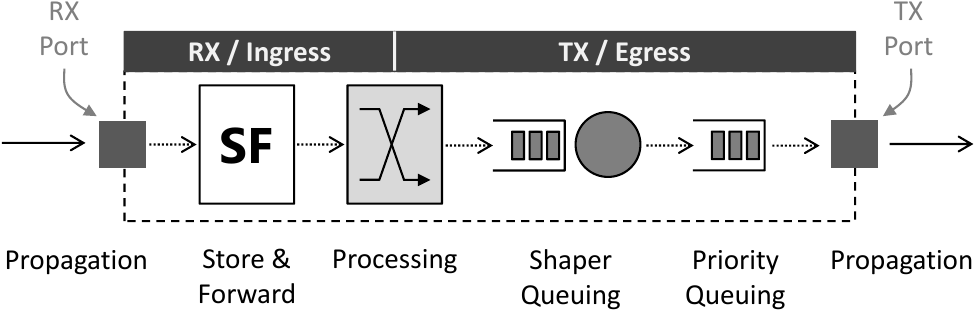}
	\caption{Switch delay model and different delay types}
	\label{fig:30-delay-types}
\end{figure}

To derive worst-case latency upper bounds in switched Ethernet networks, a formal service model covering all relevant latency sources is necessary. For decentralized execution, devices must independently calculate the worst-case delay of adjacent links, which requires well-defined boundaries of responsibility. Accordingly, Section~\ref{ssec:types-of-delay} details the latency model and its delay types, and Section~\ref{ssec:delay-segments} defines the per-hop delay segment boundaries.

\subsection{Types of Delay}
\label{ssec:types-of-delay}

Switch latency is traditionally divided into four components~\cite[Chapter\,1.4.1]{KuroseRoss16}: transmission (store and forward), propagation, queuing, and processing delay. Figure~\ref{fig:30-delay-types} maps these components to the forwarding process from the perspective of a frame.

\emph{Transmission delay} (store and forward) is the time required to push a frame's bits onto the wire, or read them at the receiver. This paper accounts for it solely at the receiver. It equals the frame size divided by link bandwidth. \emph{Propagation delay} is the time needed to traverse the physical distance between nodes, determined by link length and the medium's propagation speed. \emph{Queuing delay} represents the time a frame waits in an egress queue before transmission, which depends on the service discipline and competing traffic in other queues. Finally, \emph{processing delay} covers internal switch tasks like filtering, metering, table lookups, egress port selection, and MAC/PHY transitions. This component varies heavily based on switch architecture, configuration, and traffic composition.

In Figure~\ref{fig:30-delay-types}, queuing delay is split into two phases. \emph{Shaper queuing} encompasses artificial delays introduced by IEEE 802.1Q Transmission Selection Algorithms (TSAs)~\cite{8021Q}, including Credit-Based, Asynchronous Traffic, and Time-Aware Shaping. Once a frame becomes eligible for transmission under the shaper and any timed-gating mechanisms, it enters \emph{priority queuing}. Here, the queue of the highest traffic class that is currently not empty is serviced at line rate. Isolating these two phases is necessary for the definition of delay segments for decentralized scenarios.

Note that this model serves as a behavioral abstraction for latency analysis, without implying certain physical switch implementations. The two logical queues can be implemented in the same physical buffer, the model only separates the queuing process by its cause.

\begin{figure*}
	\centering
	\includegraphics[scale=0.5,page=2]{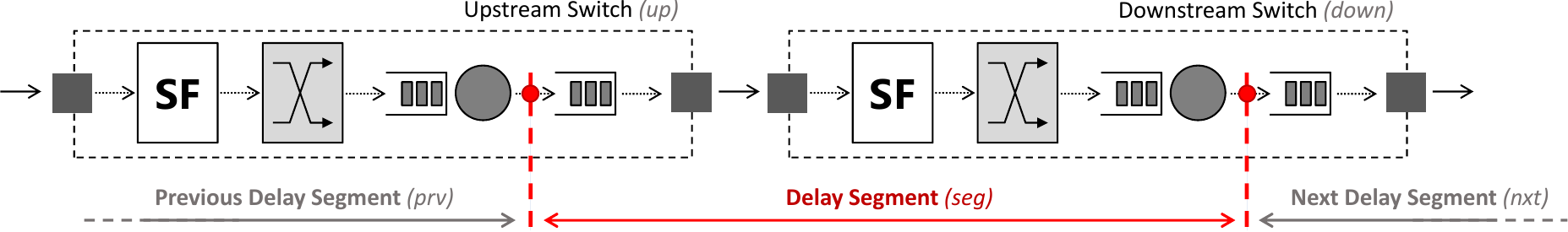}
	\caption{Delay segments: borders of responsibility for the decentralized per-hop latency computation (components are introduced in Fig.\,\ref{fig:30-delay-types})}
	\label{fig:30-delay-segments}
\end{figure*}

\subsection{Delay Segments}
\label{ssec:delay-segments}

Decentralized latency models require clear boundaries of responsibility along a path. The following approach has been proposed as a contribution \cite{grigorjew2022measurementpoints,grigorjew2023latencymodel} to the IEEE P802.1DD RAP standard \cite{8021Qdd}.

Intuitively, segments would be split at physical device boundaries, e.g., from ingress to ingress. However, insights from the analysis of Asynchronous Traffic Shaping~\cite{specht2016urgency} show that placing measurement points inside devices can facilitate latency computation. Specifically, it analyzed the delay from shaper to shaper: from the moment a frame becomes eligible at an upstream shaper until it becomes eligible at the downstream shaper. This ensures well-defined traffic workloads at each segment boundary. For instance, traffic entering a segment from a token-bucket shaper strictly conforms to those shaping parameters before experiencing interference in the priority queue. The resulting segments are illustrated in Figure~\ref{fig:30-delay-segments}. During reservation, the downstream switch computes the worst-case latency bound for its incoming segment.

Beyond aligning with existing ATS analyses, this approach offers broader benefits. Shapers inherently restore formal traffic properties, such as enforcing token-bucket volume bounds (ATS) or reducing jitter to combat burstiness cascading (CQF). Although subsequent priority queuing reintroduces jitter, the downstream shaper can compensate for it. Defining segments post-shaping allows analyses to leverage these formal traffic properties directly, turning the shaper's output into a well-defined input for the next segment. This uniform definition also simplifies shaper transitions in heterogeneous networks.

\subsection{Lessons learned}
The assumption that the borders of latency analysis must match with the physical borders of the devices can be a limitation for tight hop-by-hop latency bounds.
This paper separates the queuing process into two stages: shaping and priority queuing.
The delay segment of a single hop is defined from one shaper's end to following shaper's end. This shift allows to analyze the delay of that segment in a decentralized architecture, while profiting from the formal, predictable behavior of the shaper in the analysis.


\section{Information Flow}
\label{sec:information-flow}

\newcommand{\tabsectiontwo}[2]{
  \multicolumn{4}{c}{
    \hspace{-0.7em}\rule[0.5ex]{#1}{0.4pt}
    \textbf{#2}
    \rule[0.5ex]{#1}{0.4pt}\hspace{-0.7em}
  }
}

\begin{table*}
	\centering
	\caption{Information exchanged by switches before and during stream reservation}
	\label{tab:35-info-flow}
	\begin{tabular}{>{\hangindent=0.7em}p{17em}
	                >{\hangindent=0.7em}p{21.5em}
	                >{\hangindent=0.7em}p{8em}
	                >{\hangindent=0.7em}p{11em}} \toprule
	\textbf{Shared Information} & \textbf{Purpose} & \textbf{Origin} & \textbf{Notes} \\
	\tabsectiontwo{7cm}{A Priori / Before Reservation} \\
	Latency threshold~$\delta_\mathcal{P}^\textit{seg}$ per delay segment~\textit{seg} and per priority~$\mathcal{P}$
		& Added to each reservation's AccMaxLatency, and used as an upper bound for resource consumption
		& Pre-configured
		& Not actually shared, just required locally
		\\
	Used shaper per traffic class
		& Estimation of burstiness of incoming traffic
		& Neighbor's config.
		& --
		\\
	CBS idle slope (shaper parameter)
		& Improved CBS latency bound
		& Neighbor's config.
		& Only applicable to CBS
		\\
	CQF cycle time and GCL offsets (shaper parameter)
		& Verify matching configuration
		& Neighbor's config.
		& Only applicable to CQF
		\\
	\tabsectiontwo{7.7cm}{During Reservation} \\
	Stream ID and Ethernet header fields
		& Identification, priority, matching of incoming packets
		& Per stream
		& --
		\\
	Traffic Specification (TSpec)
		& Specification of traffic amount
		& Per stream
		& --
		\\
	Accumulated latency (min $\mathcal{L}^-$, max $\mathcal{L}^+$)
		& Estimation of burstiness increase from latency jitter
		& Per stream
		& Increased at every hop
		\\
	Upstream traffic class $\mathcal{C}^\textit{up}$
		& Computation of worst-case queuing delay in the upstream switch, based on its internal priorities
		& Per stream
		& --
		\\
	Ingress port number of upstream device
		& Advanced Strict Priority latency bound (Sec.\,\ref{ssec:sp-nc})
		& Per stream
		& Only for improved SP
		\\
	Previous delay segment link speed $R^\textit{prv}$
		& Advanced Strict Priority latency bound (Sec.\,\ref{ssec:sp-nc})
		& Per link
		& Only for improved SP
		\\
	\bottomrule
	\end{tabular}
\end{table*}

When streams require deterministic service guarantees, this is typically achieved by reserving network resources for them. These resources include physical capacities (bandwidth, buffer space) and abstract metrics like worst-case latency. Every new reservation increases worst-case latency, effectively consuming the available `latency budget'. Consequently, dynamic admission control rejects reservations that would push worst-case latency beyond a pre-defined threshold, much like it manages bandwidth limits.

To calculate the exact latency consumed by each stream, the models from Section~\ref{sec:latency-bounds} are used. These models rely on stream and neighbor device information to compute upper bounds for individual delay segments. This section details this information and its propagation, presenting a simplified signaling model based on the Resource Allocation Protocol (RAP)~\cite{8021Qdd} draft, and is meant to further support its development.

\subsection{A Priori Information}
\label{ssec:a-priori-info}

\begin{figure}
	\centering
	\includegraphics[scale=0.5,page=1]{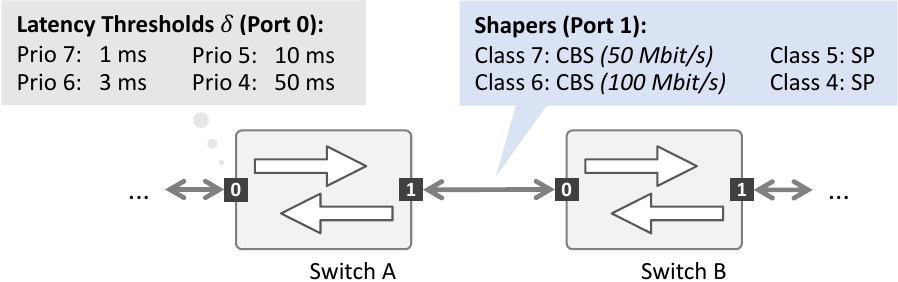}
	\caption{Exemplary a priori information flow: Switch~A shares its shaper configuration of port~1 with Switch~B, its direct neighbor}
	\label{fig:35-info-flow-example}
\end{figure}

Before signaling information about streams, certain device-specific parameters must be exchanged first. These are summarized in the first part of Table~\ref{tab:35-info-flow}, with an exchange example shown in Figure~\ref{fig:35-info-flow-example}, and detailed below.

\paragraph*{Latency Threshold}
Most latency algorithms in Section~\ref{sec:latency-bounds} rely on pre-configured latency thresholds $\delta_\mathcal{P}^\textit{seg}$ per delay segment~\textit{seg} and priority~$\mathcal{P}$. For each segment, the downstream switch handles latency computations. During reservation, it verifies that the active worst-case latency $d_\mathcal{P}^\textit{seg}$ does not exceed $\delta_\mathcal{P}^\textit{seg}$. This threshold is added to each stream's accumulated maximum latency $\mathcal{L}_i^+$, which represents a traffic-independent upper bound for the end-to-end latency of that stream which remains valid even after future reservations. This accumulated latency is later used to evaluate future stream reservations and compute latency bounds further down the stream's path (cf. Equation~\ref{eq:effective-burstiness} in Section~\ref{ssec:sp-simple}). Notably, $d_\mathcal{P}^\textit{seg}$ is calculated dynamically for admission control to measure resource consumption, whereas the configured threshold $\delta_\mathcal{P}^\textit{seg}$ serves as the absolute resource limit and is advertised to neighbors as a guarantee. Modifying $\delta_\mathcal{P}^\textit{seg}$ post-hoc could invalidate existing reservations.

In this work, thresholds are defined per priority $\mathcal{P}$ rather than per traffic class $\mathcal{C}$ because a delay segment spans two devices that may not support identical internal traffic classes. Although the RAP draft~\cite[Sec.\,6.4.2.2]{8021Qdd} introduces a shared neighbor-to-neighbor class called \emph{RAClass}, we omit it here for simplicity and use the Ethernet VLAN priority $\mathcal{P}$ instead.

\paragraph*{Upstream Shaper}
Computing tight bounds requires the downstream switch to know the shaper configurations of the upstream device. In Figure~\ref{fig:35-info-flow-example}, Switch~A informs Switch~B that port~1 uses the Credit-Based Shaper for classes 7 and 6, and Strict Priority for classes 5 and 4 (classes 3--0 handle Best-Effort traffic and are omitted here). Certain shapers require additional parameter sharing: the Credit-Based Shaper requires the idle slope for tight bounds (cf. Section~\ref{ssec:cbs} and~\cite{Maile_2023_OnThe}), while Cyclic Queuing and Forwarding requires matching cycle timings (cf. Section~\ref{ssec:cqf}).

Note that shaper configurations only need to be shared between direct neighbors to let the downstream switch estimate incoming burstiness. While the RAP draft supports this via \emph{RA class templates}~\cite[Annex\,D]{8021Qdd} in the \emph{RA Class Descriptor}~\cite[Sec.\,6.4.2.2]{8021Qdd}, it does not yet distribute specific shaper parameters.

\subsection{Per-Stream Information}
\label{ssec:per-stream-info}

\begin{figure}
	\centering
	\includegraphics[scale=0.5,page=2]{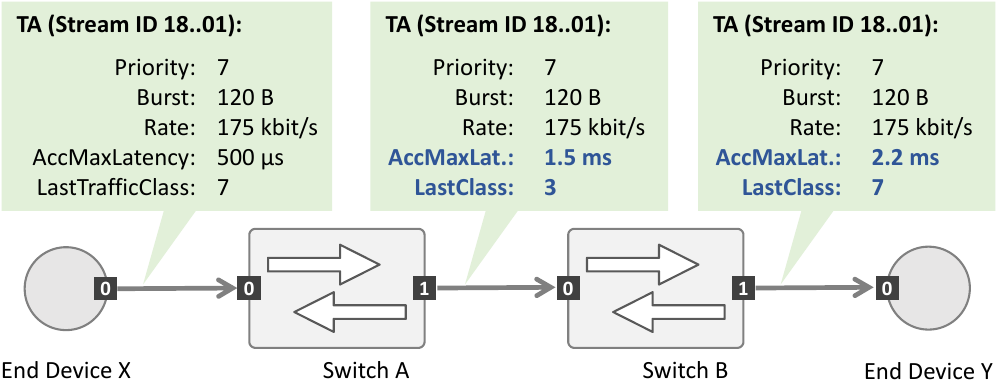}
	\caption{Exemplary information flow during stream reservation: Talker Announce (TA) from End Device~X to End Device~Y (not all info shown)}
	\label{fig:35-info-flow-example-2}
\end{figure}

Calculating worst-case queuing latency requires bounding the queue traffic volume. End devices must share stream characteristics to estimate traffic volume along the path. This information, as summarized in Table~\ref{tab:35-info-flow} and illustrated in Figure~\ref{fig:35-info-flow-example-2}, falls into four categories: stream identification, Traffic Specification, accumulated latency, and upstream traffic class.

\paragraph*{Stream Identification}
Streams must be uniquely identified during reservation and subsequent data forwarding. RAP uses a unique stream ID combining the sender's MAC address and a unique number. Because data plane Ethernet headers lack a dedicated stream ID field, active packets are matched using previously signaled header fields: destination MAC address, VLAN ID, and VLAN priority code point. Figure~\ref{fig:35-info-flow-example-2} simplifies this ID to \texttt{18..01}. The assigned priority $\mathcal{P}$ identifies the relevant delay threshold $\delta_\mathcal{P}^\textit{seg}$ during reservation and maps to a local traffic class $\mathcal{C}$ for queue selection.

\paragraph*{Traffic Specification}
To estimate queue occupancy, the Traffic Specification must bound the traffic volume over any time interval. While the legacy SRP standard~\cite{8021Qat} used frame counts within a class measurement interval, this fits poorly with modern, token-bucket-based latency algorithms. Translating between these models is possible, but it introduces an overhead and reduces network efficiency. Thus, RAP defaults to a token-bucket model, which is also adopted here. Each stream $i$ specifies a burst size $b_i$, data rate $r_i$, and maximum/minimum frame lengths ($\ell_i^\textit{max}, \ell_i^\textit{min}$). For any duration $\Delta t$, the traffic volume $w_i$ of stream $i$ is bounded by~\cite{cruz1991calculus}:
\begin{equation}
	\label{eq:token-bucket}
	w_i(\Delta t) \leq b_i + r_i \cdot \Delta t.
\end{equation}
Here, $r_i$ is the long-term rate and $b_i$ is the maximum short-term burst deviation from this rate. In Figure~\ref{fig:35-info-flow-example-2}, only $b_i$ and $r_i$ are shown. The `burst' does not necessarily need to arrive immediately in one piece. In graphical terms, it is the maximum deviation of the cumulative traffic from the linear function $y(\Delta t) = r_i \cdot \Delta t$. Note that $b_i \ge \ell_i^\textit{max}$, and all parameters ($b_i$, $r_i$, $\ell_i^\textit{max}$) must include layer-1 overhead -- typically \SI{20}{B} for preamble, CRC, and inter-frame gap.

\paragraph*{Accumulated Latency}
The above Traffic Specification only captures the behavior at the sender, but a stream's traffic characteristics can change as it encounters network delays. To track this, reservation protocols carry the \emph{accumulated maximum latency} $\mathcal{L}_i^+$ and \emph{accumulated minimum latency} $\mathcal{L}_i^-$ to compute the effective burstiness inside the network (cf. Equation~\ref{eq:token-bucket}). Each hop updates these fields before propagation:
\begin{align}
	\label{eq:acc-max-latency}
	\mathcal{L}_i^+(\textit{Downstream}) = & ~
	  \mathcal{L}_i^+(\textit{Upstream}) + \delta_{\mathcal{P}_i}^\textit{seg},
	\\
	\label{eq:acc-min-latency}
	\mathcal{L}_i^-(\textit{Downstream}) = & ~
	  \mathcal{L}_i^-(\textit{Upstream}) + \ell_i^\textit{min} / R^\textit{seg}.
\end{align}
Here, $R^\textit{seg}$ is the physical link speed of the segment, $\ell_i^\textit{min}$ is the minimum frame size of stream~$i$, and $\delta_{\mathcal{P}_i}^\textit{seg}$ is the local delay threshold for priority~$\mathcal{P}$. The difference $(\mathcal{L}_i^+ - \mathcal{L}_i^-)$ represents the maximum latency jitter accumulated along the path.

\paragraph*{Previous Delay Segment Link Speed}
For the improved Strict Priority latency bound in Section~\ref{ssec:sp-nc}, additionally, the ingress link speed~$R^\textit{prv}$ of the link~$\textit{prv}$ to the previous upstream switch is required. This value is carried in the reservation data and updated at each hop, using $R^\textit{prv} = \infty$ for the first segment. Note that the basic SP model does not require this parameter (cf. Eq.\,\ref{eq:sp-simple-delay}).

\paragraph*{Upstream Traffic Class}
Because the downstream switch calculates the upstream priority queuing delay, it must know each stream's upstream traffic class assignment $\mathcal{C}_i^\textit{up}$. Switches do not always support all eight standard classes~\cite{8021Q}, and priority-to-class mapping can vary across devices. Thus, reservations should explicitly carry the upstream traffic class. Each switch overrides this field with its local value for the next hop. In Figure~\ref{fig:35-info-flow-example-2}, Switch~A (with limited class support) maps the stream to class 3, while Switch~B maps it to class 7. Switch~B must use Switch~A's actual class mapping rather than assuming its own configuration during calculations.

\subsection{Lessons learned}
Legacy reservation protocols like SRP only convey limited information, and although \emph{accumulated maximum latency} already exists in SRP, it was not strictly enforced via admission control. There was a selection of example latency formulas in the standards~\cite[Sec.\,6.5]{8021BA}, but they lacked formal proof and were shown to be inaccurate~\cite{boiger2010class,Maile_2023_OnThe}. This shows that signaling must be based on a solid mathematical foundation.

Decentralized worst-case delay computation requires signaling of local state parameters. Switches must know their direct neighbors' shaping configurations. For individual streams, signaling must convey the original Traffic Specification and the accumulated maximum latency on its path to calculate the localized, effective traffic volume. Because the real, tight accumulated latency would fluctuate frequently when reservations change, a static, pre-configured maximum threshold is propagated instead. Under shaper-to-shaper segments, the downstream switch calculates the upstream queuing delay, which means the upstream device must share its local priority-to-class mappings. Tighter delay bounds can be achieved by also leveraging the previous device's ingress port speed, though this is not yet supported by standard signaling protocols.

\section{Latency Bounds}
\label{sec:latency-bounds}

\begin{figure*}
	\centering
	\subfloat[Order of packets in the queue\label{fig:40-local-burst-queue}]{%
      \qquad
      \raisebox{2em}{
        \includegraphics[scale=0.5,page=5]{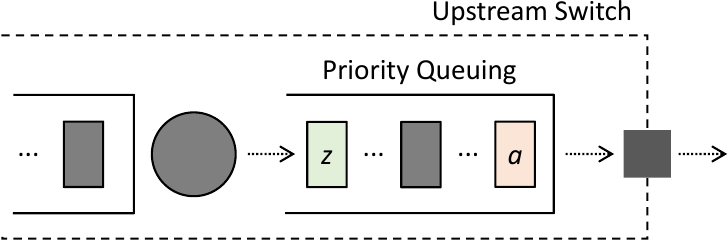}
      }
      \qquad
    }
    \qquad
    \subfloat[Timing of packets at sender and when entering the queue\label{fig:40-local-burst-timings}]{%
      \includegraphics[scale=0.5,page=6]{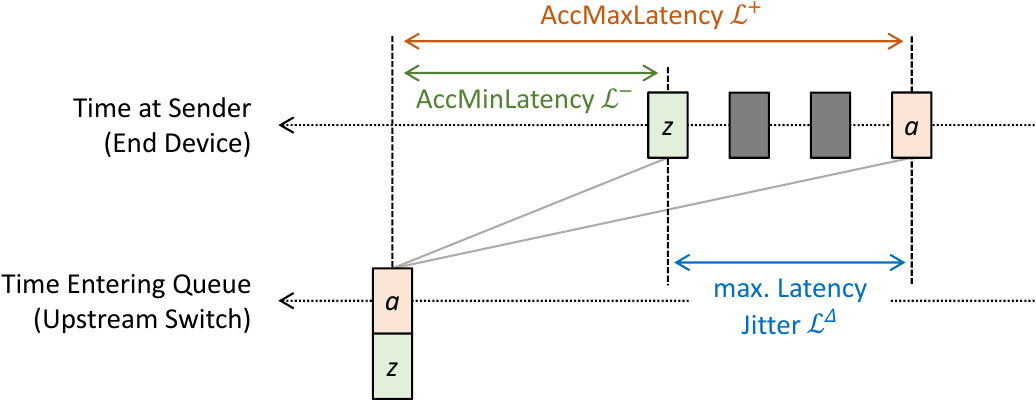}
    }
    \caption{Timings and delays of packets that can be in the priority queue at the same time}
    \label{fig:40-local-burst}
\end{figure*}

With the signaling information from Section~\ref{sec:information-flow} established, all prerequisites for decentralized worst-case latency calculations are met. This section details the mathematical models and provides pseudo-code implementations to evaluate the worst-case latency of reserved traffic. Note that these models focus on the \emph{transmission} and \emph{queuing delays} described in Section~\ref{ssec:types-of-delay}. For \emph{propagation} and \emph{processing delays}, devices are assumed to possess independent upper bounds derived from their specific hardware configurations.

First, Section~\ref{ssec:sp-simple} addresses the basic scenario of Strict Priority transmission selection without shaping, deriving a simple formula for practical implementation. Section~\ref{ssec:sp-nc} then refines this bound using the Network Calculus framework~\cite{cruz1991calculus,jiang2008stochastic,le2001network,van2016network} by accounting for physical link speed serialization. Specifically, frames arriving from the same upstream port cannot be received simultaneously. Next, Section~\ref{ssec:cqf} adapts the basic Strict Priority formula for Cyclic Queuing and Forwarding, while Section~\ref{ssec:ats} outlines the latency analysis for Asynchronous Traffic Shaping proposed by Specht and Samii~\cite{specht2016urgency}. Finally, Section~\ref{ssec:cbs} adapts the refined Strict Priority model to accommodate the unique characteristics of the Credit-Based Shaper.


\subsection{Strict Priority -- Simple Bound}
\label{ssec:sp-simple}

Under Strict Priority (SP) transmission selection, the shaper queuing delay from Figure~\ref{fig:30-delay-types} is zero, meaning downstream devices introduce no artificial delays. Consequently, the delay of that segment is primarily caused by the priority queue of the upstream switch within the same segment. To compute this delay, the downstream switch needs to know the maximum \emph{effective burstiness} of all reserved streams when they enter that priority queue.

While end devices declare their original burstiness in the Traffic Specification, the effective burstiness can increase further downstream due to accumulated latency jitter. In simple terms, frames with lower latency travel \emph{`faster'} and catch up to frames with higher latency, which travel \emph{`slower'}. This can cause packets that were originally sent separately to cluster in downstream queues. The maximum data accumulation from this effect can be bounded using the maximum latency jitter, given by the difference between the accumulated maximum and minimum latency fields.

Our prior work~\cite{Grigorjew_2020_BoundedLatency} used the same accumulated latency fields under a more constrained Traffic Specification, based on packet inter-arrival times. In contrast, the effective burstiness bound formulated in this work is more generic and directly applicable to standard token-bucket-shaped traffic.

\paragraph*{Maximum Effective Burstiness}
First, this paragraph analyzes the maximum traffic volume that can arrive `instantaneously' in the upstream switch's priority queue, i.e., at a rate exceeding the Traffic Specification rate~$r_i$. This burstiness bound assumes that end-device transmissions conform to their Traffic Specification, and that the worst-case latency of each delay segment remains within its pre-configured threshold, which is enforced via admission control during stream reservation.

For any stream~$i$ with Traffic Specification~$w_i$ (Eq.~\ref{eq:token-bucket}), the accumulated maximum latency $\mathcal{L}_i^+$, and accumulated minimum latency $\mathcal{L}_i^-$ from the previous delay segment, the maximum effective burst $b_i^\textit{eff}$ is bounded by:
\begin{equation}
  \label{eq:effective-burstiness}
  b_i^\textit{eff} \leq w_i(\mathcal{L}_i^+ - \mathcal{L}_i^-) \\
  = b_i + r_i \cdot (\mathcal{L}_i^+ - \mathcal{L}_i^-).
\end{equation}
Here, $b_i^\textit{eff}$ represents the amount of bytes that can theoretically be enqueued simultaneously in the priority queue, after shaping, as illustrated in Figure~\ref{fig:40-local-burst}. This priority queue is depicted in Figure~\ref{fig:40-local-burst-queue}, where packets are ordered from earliest arrival~$a$ to latest~$z$. Furthermore, Figure~\ref{fig:40-local-burst-timings} illustrates the packet timings from both the sender's and the priority queue's perspectives. Note that the time advances from right to left to retain the order from the queue. Any packet sent after~$z$ cannot contribute to $b_i^\textit{eff}$ because its minimum latency would fall below the \textit{AccMinLatency} $\mathcal{L}_i^-$. Similarly, packets sent before~$a$ cannot contribute to $b_i^\textit{eff}$ because their latency would exceed the \textit{AccMaxLatency} $\mathcal{L}_i^+$. Consequently, the relevant time interval $\Delta t$ for $w_i(\Delta t)$ in Equation~\ref{eq:token-bucket} is defined by the difference between these two accumulated latency fields. A comparable property can be formally derived using the Network Calculus framework, cf. \cite[Eq.\,20]{Maile_2023_OnThe}.

\paragraph*{Latency Bound for the Highest Traffic Class}
From the perspective of the highest traffic class~$\hat{\mathcal{C}}$ of the upstream switch, the following observations can be made.

\newcounter{delayobsv}
\newcommand\delayobsvpp{\refstepcounter{delayobsv}\thedelayobsv}

(\delayobsvpp) During any busy period of class~$\hat{\mathcal{C}}$, only a single lower-class transmission can occur, which must begin before the first higher-class packet arrives in the queue. Thereafter, the higher-class queue is completely emptied before any lower-class packet is transmitted again.

(\delayobsvpp) If the maximum lower-class packet length is $\ell^\textit{max}$ and the segment link speed is $R^\textit{seg}$, this lower-class transmission introduces a latency of $\ell^\textit{max} / R^\textit{seg}$.

(\delayobsvpp) During this lower-class transmission, at most $b_i^\textit{eff}$ bytes arrive from any stream $i \in \hat{\mathcal{C}}$, so their worst-case latency becomes $(\sum_{i \in \hat{\mathcal{C}}} b_i^\textit{eff} + \ell^\textit{max}) / R^\textit{seg}$ until now.

(\delayobsvpp) While servicing these initial bursts, additional higher-class packets may arrive with rate~$r_i$. These subsequent packets do not experience higher latency because admission control ensures that the link speed exceeds the aggregate rate: $\sum_{i \in \hat{\mathcal{C}}} r_i < R^\textit{seg}$. This allows the backlog to clear faster than it accumulates. Furthermore, these packets do not delay preceding packets as they are enqueued behind them.

In summary, for the highest traffic class~$\hat{\mathcal{C}}$, the worst case latency~$d^\textit{seg}_{\hat{\mathcal{C}}}$ is bounded by:
\begin{equation}
  \label{eq:spsimple-highest-c}
  d^\textit{seg}_{\hat{\mathcal{C}}} \leq \frac{\sum_{i \in \hat{\mathcal{C}}} b_i^\textit{eff} + \ell^\textit{max}}{R^\textit{seg}}.
\end{equation}

\paragraph*{Latency Bound for Lower Traffic Classes}
Consider any observed priority $\mathcal{P}$, whose delay bound is desired. Using information from prior signaling, the downstream switch maps it to the upstream traffic class $\mathcal{C} = \mathcal{C}_\mathcal{P}^\textit{up}$. For any traffic class $\mathcal{C}$ below the highest class, the analysis is extended as follows.

(\delayobsvpp) Ignoring all higher-priority traffic classes $\mathcal{H}_\mathcal{C}$ yields the same latency bound: $(\sum_{i \in \mathcal{C}} b_i^\textit{eff} + \ell^\textit{max}) / R^\textit{seg}$.

(\delayobsvpp) Packets from bursts in higher classes $\mathcal{H}_\mathcal{C}$ are prioritized over class $\mathcal{C}$, adding an extra delay of $\sum_{i \in \mathcal{H}_\mathcal{C}} b_i^\textit{eff} / R^\textit{seg}$.

(\delayobsvpp) Unlike same- or lower-priority packets, new packets from higher classes $\mathcal{H}_\mathcal{C}$ take precedence over $\mathcal{C}$ even if they arrive later. This increases the worst-case delay of class $\mathcal{C}$ based on the higher streams' data rates~$r_i$. During any interval $\Delta t$, the workload increases by $\Delta t \cdot r_i$. Due to admission control, the relevant time interval $\Delta t$ will never exceed the delay threshold $\delta_\mathcal{P}^\textit{seg}$, so the maximum interference from any higher stream~$i$ is bounded by $\delta_\mathcal{P}^\textit{seg} \cdot r_i$. Note that this bound holds when the actual worst-case delay $d^\textit{seg}_\mathcal{P}$ remains below $\delta_\mathcal{P}^\textit{seg}$, a condition verified for every reservation during admission control.

In summary, for any priority~$\mathcal{P}$ and its corresponding upstream traffic class $\mathcal{C} = \mathcal{C}_\mathcal{P}^\textit{up}$, the worst case latency~$d^\textit{seg}_{\mathcal{P}}$ is bounded by:
\begin{equation}
  \label{eq:sp-simple-delay}
  d^\textit{seg}_{\mathcal{P}} \leq \frac{\sum_{i \in \mathcal{H}_\mathcal{C}} \left( b_i^\textit{eff} + r_i \cdot \delta_\mathcal{P}^\textit{seg} \right) + \sum_{i \in \mathcal{C}} b_i^\textit{eff} + \ell^\textit{max}}{R^\textit{seg}}
\end{equation}
Here, $\mathcal{C}$ is the set of streams in the upstream traffic class of the observed
priority~$\mathcal{P}$, $\mathcal{H}_\mathcal{C}$ represents higher-priority streams, $b_i^\textit{eff}$ is the effective burstiness of stream~$i$ (Eq.\,\ref{eq:effective-burstiness}), $\delta_\mathcal{P}^\textit{seg}$ is the pre-configured latency threshold of the delay segment, $\ell^\textit{max}$ is the maximum length of any lower-priority packet, and $R^\textit{seg}$ is the egress link speed.

\paragraph*{Pseudo Code Implementation}
Implementing Equation~\ref{eq:sp-simple-delay} is relatively straightforward, provided the downstream and upstream traffic classes and their associated delay segments are not confused. For the lower-priority packet length, a full Maximum Transmission Unit (MTU) must typically be assumed to account for unknown Best-Effort traffic. Here, $\ell^\textit{max} = \SI{1542}{Bytes}$ is used, comprising a layer-2 MTU of \SI{1522}{Bytes} and \SI{20}{Bytes} of layer-1 overhead (preamble, CRC, and inter-frame gap).

\begin{algorithm}
  \caption{$\mathtt{max\_interf}$}
  \label{alg:maxInterference}
  \SetKwInOut{Input}{Input}

    \Input{Stream $i$, Priority $\mathcal{P}$, Delay Segment $\textit{seg}$}

    \uIf(\tcp*[f]{same class}){$\mathcal{C}_i^\textit{up} == \mathcal{C}_\mathcal{P}^\textit{up}$}{
      \Return $w_i(\mathcal{L}_i^+ - \mathcal{L}_i^-)$ \;
    }
    \uElseIf(\tcp*[f]{higher class}){$\mathcal{C}_i^\textit{up} > \mathcal{C}_\mathcal{P}^\textit{up}$}{
      \Return $w_i(\mathcal{L}_i^+ - \mathcal{L}_i^- + \delta_\mathcal{P}^\textit{seg})$ \;
    }
    \Else(\tcp*[f]{lower class}){
      \Return $0$ \;
    }
\end{algorithm}

First, Algorithm~\ref{alg:maxInterference} computes the individual contribution of every stream to the latency of the observed priority $\mathcal{P}$ across three cases. If stream $i$ and priority $\mathcal{P}$ share the same class, the algorithm returns the effective burstiness $b_i^\textit{eff}$ from Equation~\ref{eq:effective-burstiness}. If the stream's class is higher, it incorporates the extra interference $\delta_\mathcal{P}^\textit{seg} \cdot r_i$ by adding $\delta_\mathcal{P}^\textit{seg}$ to the interval $\Delta t$ in $w_i(\Delta t)$. If the class is lower, it returns zero, as lower-class interference is handled separately in Algorithm~\ref{alg:spSimple}. Note that all traffic classes $\mathcal{C}$ (including the target priority's class $\mathcal{C}_\mathcal{P}^\textit{up}$) refer to the upstream switch. Additionally, the accumulated latencies $\mathcal{L}_i^+$ and $\mathcal{L}_i^-$ denote the values prior to being updated via Equations~\ref{eq:acc-max-latency} and~\ref{eq:acc-min-latency}.

\begin{algorithm}
  \caption{$\mathtt{wc\_delay\_sp\_simple}$}
  \label{alg:spSimple}
  \SetKwInOut{Input}{Input}

  	\Input{Priority $\mathcal{P}$, Delay Segment $\textit{seg}$, Streams $\mathcal{S}^\textit{seg}$, Additional Stream $s$}

  	$interf \leftarrow \ell^\textit{max}$ \;
    \For{Stream $i \in \mathcal{S}^\textit{seg} \cup \{s\}$}{
    	$interf \leftarrow interf + \mathtt{max\_interf}(i, \mathcal{P})$ \;
    }
    \Return $interf / R^\textit{seg}$ \;
\end{algorithm}

Next, Algorithm~\ref{alg:spSimple} evaluates the numerator of Equation~\ref{eq:sp-simple-delay} by summing $\mathtt{max\_interf}(i, \mathcal{P})$ from Algorithm~\ref{alg:maxInterference} for all streams in the segment. Lower class interference $\ell^\textit{max}$ is only added once. This total interference is then divided by link speed $R^\textit{seg}$ to complete Equation~\ref{eq:sp-simple-delay}. This process evaluates the existing reservations $\mathcal{S}^\textit{seg}$ in addition to a candidate stream~$s$ awaiting admission control.

This simple implementation highlights a key benefit of the basic Strict Priority bound: it supports incremental computation. Because subsequent stream reservations do not alter $\ell^\textit{max}$ or $\mathtt{max\_interf}(i, \mathcal{P})$ for existing streams, implementations can cache the latency bounds of $\mathcal{S}^\textit{seg}$ and perform only a single addition for a new candidate stream~$s$. This can significantly reduce overhead on low-power switch CPUs, but this optimization is omitted in this paper for simplicity. In the following section, a tighter bound is suggested at the expense of a higher computation time.


\subsection{Strict Priority -- Tighter Bound}
\label{ssec:sp-nc}

The latency bound in Equation~\ref{eq:sp-simple-delay} simplifies the computation by assuming that all bursts from all streams arrive simultaneously. In practice, without shaping, this is only possible if every burst originates from a different ingress port. The opposite extreme is a system with one ingress port and one egress port operating at identical link speeds. Here, the maximum latency remains $\ell^\textit{max} / R^\textit{seg}$ regardless of the stream count or burstiness, because the egress queue drains at the same rate that incoming data arrives. This section presents a generalized latency bound that accounts for this data serialization at ingress ports across all intermediate configurations.

This is achieved by analyzing the maximum arriving bits and the minimum serviced bits over time as separate cumulative functions, where the worst-case latency is the maximum difference between them. This approach is commonly formalized using the Network Calculus framework. To keep this paper self-contained, the relevant Network Calculus theorems are only stated briefly. For comprehensive proofs and derivations, see \cite{cruz1991calculus,jiang2008stochastic,le2001network,van2016network}.

In short, the core idea is to model the priority queue's arrival and service processes using piecewise linear curves. The maximum horizontal distance between these curves yields the worst-case delay. The closer these curves represent the real system, the tighter the delay bound becomes. The arrival curves are constructed from the Traffic Specifications (Section~\ref{ssec:per-stream-info}) while accounting for stream distribution across physical ports. The service curves are modeled as rate-latency curves, where the initial latency captures the waiting time imposed by higher-priority traffic.

\paragraph*{Arrival Curves}

\begin{figure}
  \centering
  \subfloat[Single stream arrival curve function $\alpha_i(\Delta t)$ (Eq.\,\ref{eq:arrival-curve-single-stream})\label{fig:40-arrival-curve-single-stream}]{%
    \includegraphics[scale=0.50,page=1]{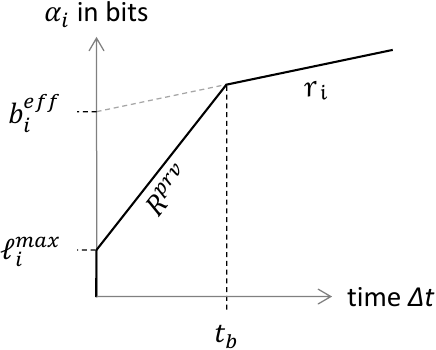}
  }
  \hfill
  \subfloat[Aggregate arrival curve function $\alpha_{\hat{\mathcal{S}}}(\Delta t)$ for a port (Eq.\,\ref{eq:arrival-curve-single-port})\label{fig:40-arrival-curve-single-port}]{%
    \includegraphics[scale=0.50,page=2]{figs/40-network-calculus-curves-crop.pdf}
  }
  \caption{Piecewise linear arrival curves for token bucket traffic models}
  \label{fig:40-arrival-curve-sp}
\end{figure}

Let $\mathcal{A}$ be the arrival process of one or more streams at a port, where $\mathcal{A}(t_1, t_2) = \mathcal{A}(t_2) - \mathcal{A}(t_1)$ denotes the data volume arriving in the interval $(t_1, t_2 \rbrack$. This process is upper-bounded by an arrival curve $\alpha(\Delta t) = \alpha(t_2 - t_1)$ if, for all $0 \leq t_1 \leq t_2$:
\begin{equation}
  \label{eq:arrival-curve-general}
  \mathcal{A}(t_1, t_2) \leq \alpha(t_2 - t_1).
\end{equation}
For a token-bucket-shaped stream $i$ with burst $b_i$ and rate $r_i$, the baseline arrival curve at the source is $\alpha_i(\Delta t) = w_i(\Delta t) = b_i + r_i \cdot \Delta t$ for $\Delta t \geq 0$. However, at any downstream switch, the effective burstiness increases to $b_i^\textit{eff}$ (Eq.~\ref{eq:effective-burstiness}). Furthermore, the instantaneous arrival of data is bounded by the maximum length of a single frame $\ell_i^\textit{max}$. The remainder of a burst will arrive at the link speed $R^\textit{prv}$ of the upstream segment \textit{prv}. Thus, the arrival curve for stream $i$ in segment $\textit{seg}$ is modeled as:
\begin{equation}
  \label{eq:arrival-curve-single-stream}
  \alpha_i(\Delta t) =
    \begin{cases}
      \ell_i^\textit{max} + R^\textit{prv} \cdot \Delta t & \text{if $\Delta t \leq t_b$,} \\
      b_i^\textit{eff} + r_i \cdot \Delta t & \text{else.}
    \end{cases}
\end{equation}
Here, $t_b$ is the time when the arrival rate transitions from line rate $R^\textit{prv}$ to the stream rate $r_i$ (Fig.~\ref{fig:40-arrival-curve-single-stream}). It is determined by the intersection of two simpler curves. Both $\alpha_{i,1}(\Delta t) = b_i^\textit{eff} + r_i \cdot \Delta t$ and $\alpha_{i,2}(\Delta t) = \ell_i^\textit{max} + R^\textit{prv} \cdot \Delta t$ are valid upper bounds individually, where $\alpha_{i,1}$ is the limit of produced data at the source, and $\alpha_{i,2}$ is the limit of transmission speed on the link. Their minimum is also an upper bound, and the intersection point $t_b$ of these functions is:
\begin{align}
  & b_i^\textit{eff} + r_i \cdot t_b = \ell_i^\textit{max} + R^\textit{prv} \cdot t_b \\
  \label{eq:tb}
  \Leftrightarrow \quad & t_b = \frac{b_i^\textit{eff} - \ell_i^\textit{max}}{R^\textit{prv} - r_i}.
\end{align}
For the first delay segment where the upstream device is the traffic source, $R^\textit{prv} = \infty$ can be assumed, simplifying the model to $\alpha_{i,1}(\Delta t) = b_i^\textit{eff} + r_i \cdot \Delta t$.

For the full arrival curve of the entire delay segment~\textit{seg}, the superposition property of Network Calculus could be used, which would basically sum up all individual stream arrival curves. However, this would overestimate the immediate arrivals as $\sum_{i \in \mathcal{C}} \ell_i^\textit{max}$, i.e., the sum of all streams' frame lengths. In reality, simultaneous frame arrivals are physically limited by the number of ingress ports. To obtain a tighter bound, one aggregate arrival curve per ingress port is constructed (Fig.~\ref{fig:40-arrival-curve-single-port}), and the full arrival curve will be the sum of those curves instead of the individual stream curves.

Let $\hat{\mathcal{S}} = \mathcal{S}_\mathcal{C}^\textit{prv}$ represent all streams of class $\mathcal{C}$ arriving from the previous segment \textit{prv}. Because \textit{prv} represents a single port, immediate data arrival is limited to a single maximum frame length, $\max_{i \in \hat{\mathcal{S}}} \ell_i^\textit{max}$. Thereafter, data can arrive at line rate $R^\textit{prv}$ for a duration of $t_{\hat{b}}$, after which the upstream queue is empty and the arrival rate drops to the aggregate source rate $\sum_{i \in \hat{\mathcal{S}}} r_i$. Similarly to before, the aggregate arrival curve $\alpha_{\hat{\mathcal{S}}}$ is the minimum of two curves. First, $\alpha_{\hat{\mathcal{S}},1}(\Delta t) = \sum_{i \in \hat{\mathcal{S}}} \left( b_i^\textit{eff} + r_i \cdot \Delta t \right)$ is the upper bound of cumulative data produced by all traffic sources. Second, $\alpha_{\hat{\mathcal{S}},2}(\Delta t) = \max_{i \in \hat{\mathcal{S}}} \ell_i^\textit{max} + R^\textit{prv} \cdot \Delta t$ is the limit enforced by the ingress link speed. These two curves intersect at:
\begin{align}
  \label{eq:tbhat}
  t_{\hat{b}} = \frac{\sum_{i \in \hat{\mathcal{S}}} b_i^\textit{eff} - \max_{i \in \hat{\mathcal{S}}} \ell_i^\textit{max}}{R^\textit{prv} - \sum_{i \in \hat{\mathcal{S}}} r_i}.
\end{align}
The resulting piecewise linear curve (Fig.~\ref{fig:40-arrival-curve-single-port}) is defined by:
\begin{equation}
  \label{eq:arrival-curve-single-port}
  \alpha_{\hat{\mathcal{S}}}(\Delta t) =
    \begin{cases}
      \max_{i \in \hat{\mathcal{S}}} \ell_i^\textit{max} + R^\textit{prv} \cdot \Delta t & \text{if $\Delta t \leq t_{\hat{b}}$,} \\
      \sum_{i \in \hat{\mathcal{S}}} b_i^\textit{eff} + \sum_{i \in \hat{\mathcal{S}}} r_i \cdot \Delta t & \text{else.}
    \end{cases}
\end{equation}
The total arrival curve $\alpha_{\mathcal{C}}^\textit{seg}$ for segment \textit{seg} and class $\mathcal{C}$ is the sum of the aggregate arrival curves across all $n$ active ingress ports $\textit{prv}_k$:
\begin{equation}
  \label{eq:sp-arrival-curve-full}
  \alpha_{\mathcal{C}}^\textit{seg}(\Delta t) =
    \sum_{1 \leq k \leq n} \alpha_{\hat{\mathcal{S}}_k}(\Delta t).
\end{equation}
Here, the sum iterates over all $n$ ingress ports $\textit{prv}_k$ (cf. Fig.~\ref{fig:30-delay-segments}) using stream sets $\hat{\mathcal{S}}_k = \mathcal{S}_\mathcal{C}^{\textit{prv}_k}$. For strict priority systems, the upstream switch's priority queue delay is analyzed. This requires all streams from $\bigcup_\textit{prv} \mathcal{S}_\mathcal{C}^\textit{prv}$ across all previous segments $\textit{prv}$. These streams do not necessarily traverse the observed segment $\textit{seg}$ downstream, they can continue on different delay segments, but they share the same upstream priority queue.

\paragraph*{Service Curves}
Now, after the arrival process is determined, a formal model for worst-case service is required, given by a \emph{minimum} cumulative service curve. As before, it is based on the signaling information from Section~\ref{sec:information-flow}. The general Network Calculus methodologies are detailed in \cite{cruz1991calculus,le2001network,van2016network}, and briefly presented in the following.

Let~$\mathcal{A}$ be the arrival process of one or more streams at a service unit,
and $\mathcal{A}(t_1, t_2) = \mathcal{A}(t_2) - \mathcal{A}(t_1)$
be the amount of data that arrives from that process
in the time interval $(t_1, t_2 \rbrack$.
Further, let $\mathcal{A}^*$ be the departure process of that service unit,
and $\mathcal{A}^*(t_1, t_2) = \mathcal{A}^*(t_2) - \mathcal{A}^*(t_1)$
be the amount of data that departs from that process
in the time interval $(t_1, t_2 \rbrack$.
The service unit has a service curve $\beta(\Delta t) = \beta(t_2 - t_1)$ if,
for all $0 \leq t_1 \leq t_2$, the following equation holds:

Let $\mathcal{A}$ and $\mathcal{A}^*$ be the cumulative arrival and departure processes at a service unit, and let $(t_1, t_2\rbrack$ be an arbitrary time interval. During that time, $\mathcal{A}(t_2) - \mathcal{A}(t_1)$ bytes arrived at the service unit, and $\mathcal{A}^*(t_2) - \mathcal{A}^*(t_1)$ bytes were serviced and departed. The unit offers a service curve $\beta(\Delta t) = \beta(t_2 - t_1)$ if, for all $0 \leq t_1 \leq t_2$:
\begin{equation}
  \label{eq:service-curve-def}
  \mathcal{A}^*(t_2) \geq \inf_{t_1 \leq t_2} \left\lbrace \mathcal{A}(t_1) + \beta(t_2 - t_1) \right\rbrace.
\end{equation}
Less formally, Equation~\ref{eq:service-curve-def} guarantees that at least $\beta(\Delta t)$ bits are serviced during any backlogged interval $\Delta t$.

A typical, linear service model is the rate-latency service curve~$\beta^\textit{seg}_\mathcal{C}$
with a constant rate~$R^\textit{seg}_\mathcal{C}$ and an initial latency~$\tau^\textit{seg}_\mathcal{C}$:
\begin{equation}
  \label{eq:spnc-beta}
  \beta^\textit{seg}_\mathcal{C}(\Delta t) = R^\textit{seg}_\mathcal{C} \cdot (\Delta t - \tau^\textit{seg}_\mathcal{C}).
\end{equation}
To capture the physical service of link $\textit{seg}$ to class $\mathcal{C}$, the two parameters $R^\textit{seg}_\mathcal{C}$ and $\tau^\textit{seg}_\mathcal{C}$ must be defined. These parameters are influenced by higher-priority traffic arrivals, which reduce lower-class service guarantees. Note that service curves typically cannot become negative. Without loss of generality, terms like $\max(0, ...)$ are omitted here for simplicity, with no impact on their applicability.

\paragraph*{Service Curve for the Highest Traffic Class}

\begin{figure}
  \centering
  \subfloat[Highest class service curve\label{fig:40-service-curve-highest-class}]{%
    \includegraphics[scale=0.50,page=3,valign=t]{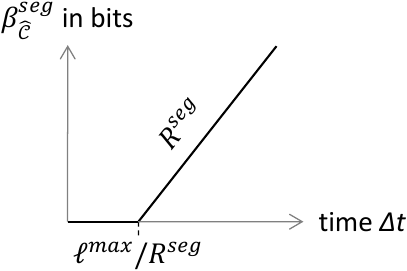}
    \vphantom{\includegraphics[scale=0.50,page=4,valign=t]{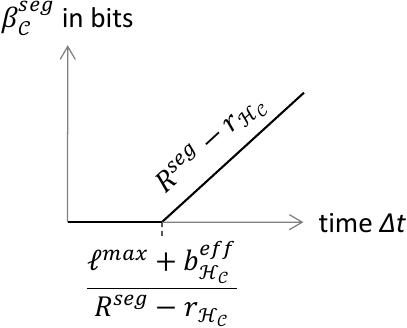}}
  }
  \hspace{2em}
  \subfloat[General service curve\label{fig:40-service-curve-general}]{%
    \includegraphics[scale=0.50,page=4,valign=t]{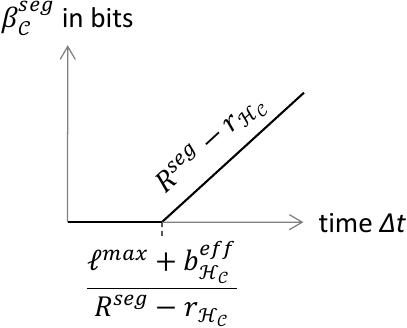}
  }
  \caption{Service curves for Strict Priority and Token Bucket traffic model}
  \label{fig:40-service-curve-sp}
\end{figure}

For the highest class $\hat{\mathcal{C}}$, the service rate $R^\textit{seg}_{\hat{\mathcal{C}}}$ is the full link speed $R^\textit{seg}$, as its queue $\mathcal{Q}_{\hat{\mathcal{C}}}$ drains without interruption once it is selected for transmission. However, if a lower-class packet \emph{just} started its own transmission before $\mathcal{Q}_{\hat{\mathcal{C}}}$ contained any packets, it can delay the start of $\hat{\mathcal{C}}$ up to the length of a maximum transmission unit $\ell^\textit{max}$. The highest-class service curve is therefore given by:
\begin{equation}
  \label{eq:sp-service-curve-highest-class}
  \beta^\textit{seg}_{\hat{\mathcal{C}}}(\Delta t) = R^\textit{seg} \cdot \left( \Delta t - \frac{\ell^\textit{max}}{R^\textit{seg}} \right) = R^\textit{seg} \cdot \Delta t - \ell^\textit{max}.
\end{equation}
This simple linear function is illustrated in Figure~\ref{fig:40-service-curve-highest-class}.

\paragraph*{Service Curve for Lower Traffic Classes}
For a general traffic class $\mathcal{C}$, let $\mathcal{H}_\mathcal{C}$ be the set of higher-class streams. Packets from class $\mathcal{C}$ must wait until any ongoing transmission finishes and all queues from higher classes $\mathcal{H}_\mathcal{C}$ are empty. Based on Equation~\ref{eq:effective-burstiness}, higher-class queues hold at most $b_{\mathcal{H}_\mathcal{C}}^\textit{eff} = \sum_{i \in \mathcal{H}_\mathcal{C}} b_i^\textit{eff}$ bytes. In the worst case, queue $\mathcal{C}$ waits for these $\ell^\textit{max} + b_{\mathcal{H}_\mathcal{C}}^\textit{eff}$ bytes to transmit at rate $R^\textit{seg}$. Meanwhile, new higher-class data arrives at rate $r_{\mathcal{H}_\mathcal{C}} = \sum_{i \in \mathcal{H}_\mathcal{C}} r_i$, which must also be served first. Thus, higher-class queues empty at an effective rate of $R^\textit{seg} - r_{\mathcal{H}_\mathcal{C}}$. Applying this rate to both the service rate $R_\mathcal{C}^\textit{seg}$ and the delay term $\tau_\mathcal{C}^\textit{seg}$ yields:
\begin{align}
  \label{eq:sp-service-curve-general}
  \beta^\textit{seg}_\mathcal{C}(\Delta t) 
  &= R_\mathcal{C}^\textit{seg} \cdot \left(\Delta t - \tau_\mathcal{C}^\textit{seg} \right) \\
  \label{eq:sp-service-curve-general-2}
  &= \left( R^\textit{seg} - r_{\mathcal{H}_\mathcal{C}} \right) \cdot \left( \Delta t - \frac{\ell^\textit{max} + b_{\mathcal{H}_\mathcal{C}}^\textit{eff}}{R^\textit{seg} - r_{\mathcal{H}_\mathcal{C}}} \right). 
\end{align}
Equation~\ref{eq:sp-service-curve-general-2} is further illustrated in Figure~\ref{fig:40-service-curve-general}.{}

\paragraph*{Latency Bound}

\begin{figure}
  \centering
  \includegraphics[scale=0.50,page=5]{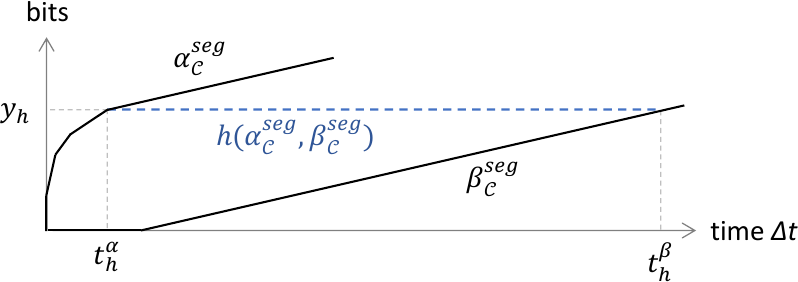}
  \caption{Max. horizontal distance between arrival and service curve}
  \label{fig:40-sp-delay-bound}
\end{figure}

The Network Calculus delay theorem states that the worst-case delay for class $\mathcal{C}$ on segment $\textit{seg}$ is bounded by the maximum horizontal deviation $h$ between $\alpha_{\mathcal{C}}^\textit{seg}$ and $\beta_{\mathcal{C}}^\textit{seg}$:
\begin{align}
  \label{eq:sp-delay-bound-general}
  d_{\mathcal{P}}^\textit{seg} & \leq h(\alpha_{\mathcal{C}}^\textit{seg}, \beta_{\mathcal{C}}^\textit{seg}) \\
  & = \sup_{s \geq 0} \left\lbrace \inf\lbrace \tau \geq 0 : \alpha_{\mathcal{C}}^\textit{seg}(s) \leq \beta_{\mathcal{C}}^\textit{seg}(s + \tau) \rbrace \right\rbrace.
\end{align}
Practically, this bound can be found by comparing slopes, cf. Figure~\ref{fig:40-sp-delay-bound}. The piecewise arrival curve~$\alpha_{\mathcal{C}}^\textit{seg}$ has a continuously decreasing slope. Consider sweeping an imaginary horizontal line from the bottom to the top of Figure~\ref{fig:40-sp-delay-bound}. The distance between $\alpha_{\mathcal{C}}^\textit{seg}$ and $\beta_{\mathcal{C}}^\textit{seg}$ increases as long as the slope of $\alpha_{\mathcal{C}}^\textit{seg}$ is higher than the service rate $R_{\mathcal{C}}^\textit{seg}$. Assuming the segment is not overprovisioned, the arrival slope eventually drops below or meets $R_{\mathcal{C}}^\textit{seg}$, after which the horizontal distance no longer increases. Thus, the maximum delay occurs where the arrival slope equals $R_{\mathcal{C}}^\textit{seg}$.
This peak is marked by $t_h^\alpha$ on the x-axis and $y_h = \alpha_{\mathcal{C}}^\textit{seg}(t_h^\alpha)$ on the y-axis. The corresponding service curve point is $t_h^\beta$, where $\beta_{\mathcal{C}}^\textit{seg}(t_h^\beta) = y_h$ holds. The maximum delay is then given by their horizontal difference $d_{\mathcal{P}}^\textit{seg} \leq t_h^\beta - t_h^\alpha$.

In this work, a closed form description of the maximum horizontal distance is omitted.
Instead, the pseudo code implementation obtains the delay bound numerically
by looping through the individual pieces of $\alpha_{\mathcal{C}}^\textit{seg}$
as explained above, until the slope is smaller or equal to $R_{\mathcal{C}}^\textit{seg}$.

\paragraph*{Pseudo Code Implementation}
First, data structures for the arrival and service curves are required. The arrival curve $\alpha_{\hat{\mathcal{S}}}(\Delta t)$ represents streams from a single ingress segment \textit{prv} and a single class~$\mathcal{C}$. Based on Equation~\ref{eq:arrival-curve-single-port} and Figure~\ref{fig:40-arrival-curve-single-port}, it is implemented as a tuple: $ac_{\hat{\mathcal{S}}} = (\textit{burst}, \textit{slope1}, \textit{slope2}, \textit{bendpos})$. Here, $\textit{burst} = \max_{i \in \hat{\mathcal{S}}} \ell_i^\textit{max}$, $\textit{slope1} = R^\textit{prv}$, and $\textit{slope2} = \sum_{i \in \hat{\mathcal{S}}} r_i$. The two slopes transition at $\textit{bendpos} = t_{\hat{b}}$, as defined in Equation~\ref{eq:tbhat}. The function value is computed as follows:
\begin{equation}
  \label{eq:arrival-curve-single-port-pseudocode}
  \alpha_{\hat{\mathcal{S}}}(\Delta t) =
    \begin{cases}
      \textit{burst} + \textit{slope1} \cdot \Delta t & \text{if $\Delta t \leq t_{\hat{b}}$,} \\
      \alpha_{\hat{\mathcal{S}}}(t_{\hat{b}}) + \textit{slope2} \cdot \left( \Delta t - t_{\hat{b}} \right) & \text{else.}
    \end{cases}
\end{equation}
Note that the \textit{burst} in Equation~\ref{eq:arrival-curve-single-port-pseudocode} is based on immediately arriving traffic from the perspective of the ingress port. It is given by the frame lengths $\ell_i^\textit{max}$, and not by the burstiness~$b_i$ of streams, which is incorporated in $t_{\hat{b}}$ (Eq.\,\ref{eq:tbhat}). The slope of~$\alpha_{\hat{\mathcal{S}}}$ at any point~$\Delta t$ is a simple case statement:
\begin{equation}
  \mathtt{slope}_{\hat{\mathcal{S}}}(\Delta t) =
  \begin{cases}
    \textit{slope1} & \text{if $\Delta t < t_{\hat{b}}$,} \\
    \textit{slope2} & \text{else.}
  \end{cases}
\end{equation}

Similarly, the service curve $\beta_\mathcal{C}^\textit{seg}(\Delta t)$ from Equation~\ref{eq:sp-service-curve-general-2} is represented by $sc_\mathcal{C}^\textit{seg} = (\textit{slope}, \textit{offset})$, where $\textit{slope} = R_\mathcal{C}^\textit{seg}$ and $\textit{offset} = \tau_\mathcal{C}^\textit{seg}$. Its value at time $\Delta t$ is $sc_\mathcal{C}^\textit{seg}(\Delta t) = \max(0, \textit{slope} \cdot (\Delta t - \textit{offset}))$. More importantly, its inverse is give by:
\begin{equation}
  \mathtt{findt}_\mathcal{C}^\textit{seg}(y) = y / \textit{slope} + \textit{offset}.
\end{equation}

The full arrival curve, as defined in Equation~\ref{eq:sp-arrival-curve-full}, is represented by the list $AC_\mathcal{C}^\textit{seg} = \lbrack ac_{\hat{\mathcal{S}},1}, ac_{\hat{\mathcal{S}},2}, \dots, ac_{\hat{\mathcal{S}},n} \rbrack$, where the $k$-th tuple represents streams from the $k$-th ingress port: $\mathcal{S}_\mathcal{C}^{\textit{prv},k}$. The downstream switch must know this stream-to-port mapping and the corresponding link speeds $R^\textit{prv} = ac_{\hat{\mathcal{S}},k}.\textit{slope1}$, which must be provided during stream reservation, cf. Table~\ref{tab:35-info-flow}.

Given the list of tuples $AC_\mathcal{C}^\textit{seg}$, the function $\mathtt{slope}_\mathcal{C}^\textit{seg}(\Delta t) = \sum_{1 \leq k \leq n} \mathtt{slope}_{\hat{\mathcal{S}}_k}(\Delta t)$ returns the aggregate slope at point $\Delta t$. Its inverse, $\mathtt{find\_slope}_\mathcal{C}^\textit{seg}(s)$, returns the earliest time $\Delta t$ where the aggregate slope is at most~$s$. Algorithm~\ref{alg:findslope} implements this by looping through a sorted list of bending positions until the aggregate slope falls below~$s$.

\begin{algorithm}
  \caption{$\mathtt{find\_slope}$}
  \label{alg:findslope}
  \SetKwInOut{Input}{Input}

    \Input{Class $\mathcal{C}$,
           Delay Segment $\textit{seg}$,
           Slope $s$,
           List of arrival curve tuples $AC$}

    \texttt{// slope tuples: (slope, bendpos)} \;
    $SL \leftarrow \lbrack ~ \rbrack$ \tcp*[f]{empty list of tuples} \;
    \texttt{// SL = [(s1,b1), (s2,b2), ...]} \;
    \texttt{// initialize first tuple at t=0} \;
    $SL.\mathtt{append}\left(\left( \sum_{ac \,\in\, AC} ac.\textit{slope1}, 0 \right)\right)$ \;
    \For{$ac \in AC$}{
      $bendpos \leftarrow ac.\textit{bendpos}$ \;
      $slope \leftarrow \mathtt{slope}_\mathcal{C}^\textit{seg}(bp)$ \tcp*[f]{$\sum$slopes(bp)} \;
      $SL.\mathtt{append}((slope, bendpos))$ \;
    }
    $\mathtt{sort}(SL, \text{key} = bendpos)$ \;

    $i \leftarrow 0$ \;
    \While(\tcp*[f]{search for $s$}){$SL[i].\textit{slope} > s$}{
      $i \leftarrow i + 1$ \;
      \lIf{$i \geq |SL|$}{
        \Return $-1$ \tcp*[f]{error}
      }
    }
    \Return $SL[i].\textit{bendpos}$
\end{algorithm} 

Ultimately, using arrival curves $AC_\mathcal{C}^\textit{seg}$ and service curve $sc_\mathcal{C}^\textit{seg}$, the two points $t_h^\alpha$ and $t_h^\beta$ via $\mathtt{findt}$ must be found, cf. Figure~\ref{fig:40-sp-delay-bound}. Here, $t_h^\alpha$ is given when the arrival rate falls below the service rate and can be obtained via $\mathtt{find\_slope}$. Then, $t_h^\beta$ can be found via $\mathtt{findt}_\mathcal{C}^\textit{seg}()$. Finally, the delay bound is $d_\mathcal{P}^\textit{seg} \leq t_h^\beta - t_h^\alpha$, as implemented in Algorithm~\ref{alg:wc-delay-sp}.

\begin{algorithm}
  \caption{$\mathtt{wc\_delay\_sp}$}
  \label{alg:wc-delay-sp}
  \SetKwInOut{Input}{Input}

    \Input{Priority $\mathcal{P}$,
           Delay Segment $\textit{seg}$,
           List of streams per ingress port $LS_\mathcal{C}^\textit{seg} = [\hat{\mathcal{S}}_1, \dots, \hat{\mathcal{S}}_n]$}

    $\mathcal{C} \leftarrow \mathcal{C}_\mathcal{P}^\textit{up}$ \tcp*[f]{upstream traffic class} \;
    $AC \leftarrow [\,]$ \tcp*[f]{empty list} \;
    \For{$\hat{\mathcal{S}}_k \in LS$}{
      $ac \leftarrow (0,0,0,0)$ \tcp*[f]{prepare tuple} \;
      $ac.\textit{burst} \leftarrow \max_{i \in \hat{\mathcal{S}}_k} \ell_i^\textit{max}$ \;
      $ac.\textit{slope1} \leftarrow R^{\hat{\mathcal{S}}_k.\textit{prv}}$ \tcp*[f]{$R^\textit{prv}$} \;
      $ac.\textit{slope2} \leftarrow \sum_{i \in \hat{\mathcal{S}}_k} r_i$ \;
      $ac.\textit{bendpos} \leftarrow t_{\hat{b}}$ \tcp*[f]{Eq.\,\ref{eq:tbhat}} \;
      $AC.\mathtt{append}(ac)$ \;
    }
    $sc \leftarrow (0,0)$ \tcp*[f]{slope, offset} \;
    $sc.\textit{slope} \leftarrow R^\textit{seg} - \sum_{i \in \mathcal{H}_\mathcal{C}} r_i$ \;
    $sc.\textit{offset} \leftarrow \dfrac{ \ell^\textit{max} + \sum_{i \in \mathcal{H}_\mathcal{C}} b_i^\textit{eff} }{ R^\textit{seg} - \sum_{i \in \mathcal{H}_\mathcal{C}} r_i }$ \;

    $t_h^\alpha \leftarrow \mathtt{find\_slope}_\mathcal{C}^\textit{seg}(sc.\textit{slope}, AC)$ \;
    $d_\mathcal{P}^\textit{seg} \leftarrow \mathtt{findt}_\mathcal{C}^\textit{seg}(\alpha_\mathcal{C}^\textit{seg}(t_h^\alpha)) - t_h^\alpha$ \;
    \Return $d_\mathcal{P}^\textit{seg}$
\end{algorithm}

In summary, the delay bound in Algorithm~\ref{alg:wc-delay-sp} is tighter than the simple bound in Algorithm~\ref{alg:spSimple}, but it is more complex, slower to compute, and requires more information from the reservation process, cf. Table~\ref{tab:35-info-flow}. The comparison is picked up in Section~\ref{sec:evaluation} in more detail.

\begin{figure*}
  \centering
  \includegraphics[scale=0.50,page=1]{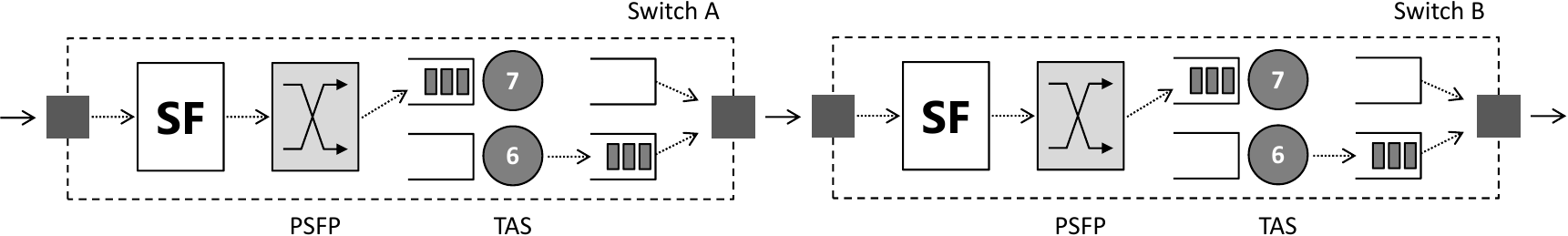}
  \caption{Snapshot of a Cyclic Queuing and Forwarding configuration in this work's delay model (first time slot from Figure~\ref{fig:40-cqf-2})}
  \label{fig:40-cqf}
\end{figure*}

\begin{figure}
  \centering
  \includegraphics[scale=0.50,page=2]{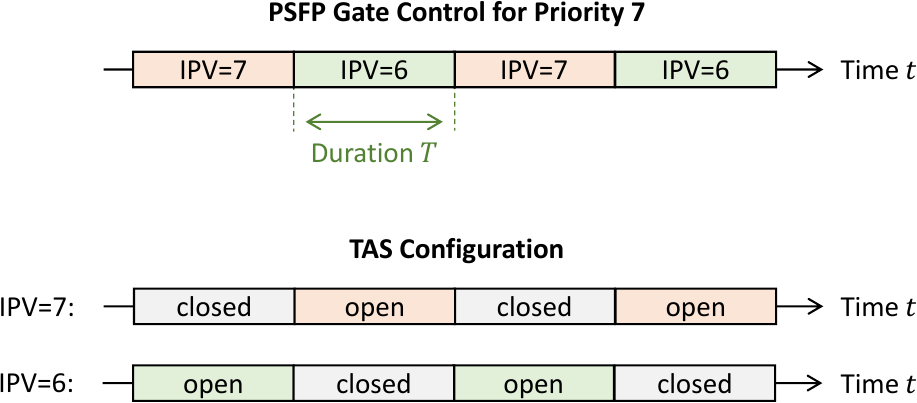}
  \caption{Example configuration of PSFP and TAS gates for one external priority in CQF with Internal Priority Values (IPVs) 6~and~7}
  \label{fig:40-cqf-2}
\end{figure}

\subsection{Cyclic Queuing and Forwarding}
\label{ssec:cqf}

Cyclic Queuing and Forwarding (CQF)~\cite[Annex~T]{8021Q} is the first shaper addressed in this work. This section outlines the modifications introduced by CQF, and then addresses how the latency model of SP can be adapted to that.

Standardization-wise, CQF is not a standalone shaper but a specific configuration of Time-Aware Shaping (TAS) and Per-Stream Filtering and Policing (PSFP), as illustrated in Figure~\ref{fig:40-cqf}. For detailed derivations of these TSN mechanisms, readers are referred to the standard~\cite{8021Q} and comprehensive surveys~\cite{Nasrallah_2019_UltralowLatency}. Briefly, \emph{cyclic queuing} periodically alternates traffic of the same priority between multiple queues. Incoming streams are directed to one queue by PSFP, while another queue is selected by TAS for transmission. Both gating mechanisms require globally synchronized time, and identical configurations are deployed on every network device. An exemplary configuration is shown in Figure~\ref{fig:40-cqf-2} and analyzed below.

The primary objective is to ensure that every hop-to-hop transmission takes exactly~$T$ time units, even under light traffic load. To achieve this, the gating mechanisms of PSFP and TAS are operated out-of-phase. When PSFP selects an Internal Priority Value (IPV) of~7, the corresponding transmission gate is closed by TAS, while the gate for~6 is opened. This state is illustrated in Figure~\ref{fig:40-cqf}. After a fixed slot duration of~$T$, PSFP assigns an IPV of~6 to incoming packets of the same stream, while TAS closes gate~6 and opens gate~7 (depicted in Figure~\ref{fig:40-cqf-2}). Consequently, as soon as TAS opens a gate on Switch~A, PSFP on Switch~B classifies these frames to its closed queue. Because this closed gate is opened exactly~$T$ time units after the upstream gate on Switch~A, each hop between shapers takes exactly~$T$ time units. Thus, the latency threshold~$\delta$ (cf. Sec.\,\ref{ssec:a-priori-info}) must be configured as $\delta \leq T - d_\textit{misc}$, where $d_\textit{misc}$ accounts for synchronization inaccuracies. If the admission control mechanism ensures that the worst-case delay~$d$ remains below~$\delta$, the queue is guaranteed to be empty at the end of window~$T$. Hence, a downstream shaper considers a frame eligible for transmission exactly~$T$ time units after the upstream shaper did. Leveraging this property is a key benefit of the shaper-to-shaper delay segments presented in Section~\ref{ssec:delay-segments}.

\paragraph*{Latency Analysis}
Adapting the simple SP latency model from Section~\ref{ssec:sp-simple} to CQF involves three key modifications.

(i) The number of usable traffic classes is halved, because each external class requires two internal queues. Without loss of generality, the evaluation in Section~\ref{sec:evaluation} assumes eight available traffic classes, where classes 4~to~7 are allocated to reserved traffic. Under CQF, classes are consolidated to the higher priorities, i.e., $\mathcal{P}_{CQF}(7) = 7, \mathcal{P}_{CQF}(6) = 7, \mathcal{P}_{CQF}(5) = 5, \mathcal{P}_{CQF}(4) = 5$. This consolidation increases traffic interference in high-priority classes, which raises their worst-case delay. Fore a more sophisticated evaluation, priority mappings and delay bounds could be optimized specifically for CQF.

(ii) In the shaper-to-shaper delay model, the minimum latency becomes equal to the maximum latency.
Equation~\ref{eq:acc-min-latency} for~$\mathcal{L}^-$ is thus updated to match~$\mathcal{L}^+$:
\begin{align}
  \mathcal{L}_{i,CQF}^+(\textit{Downstream}) = & ~
    \mathcal{L}_{i,CQF}^+(\textit{Upstream}) + T_{\mathcal{P}_i},
  \\
  \mathcal{L}_{i,CQF}^-(\textit{Downstream}) = & ~
    \mathcal{L}_{i,CQF}^-(\textit{Upstream}) + T_{\mathcal{P}_i}.
\end{align}
Consequently, in Equation~\ref{eq:effective-burstiness}, the effective burst~$b_i^\textit{eff}$ is nearly identical to the original burst~$b_i$, depending solely on the difference between $\mathcal{L}^-$ and $\mathcal{L}^+$ at the first hop.

(iii) The SP model assumes that backlog is cleared faster than new traffic accumulates. This assumption does not hold for CQF, where the gating mechanism effectively stops transmissions until queues alternate. Consequently, extra traffic can accumulate in the queue of priority~$\mathcal{P}$ for up to~$\delta_{\mathcal{P}}^\textit{seg}$ time units. This additional volume is calculated similarly to the higher-priority case as $\delta_\mathcal{P}^\textit{seg} \cdot r_i$.

\begin{algorithm}
  \caption{$\mathtt{max\_interf\_cqf}$}
  \label{alg:maxInterferenceCQF}
  \SetKwInOut{Input}{Input}

    \Input{Stream $i$, Priority $\mathcal{P}_\textit{orig}$, Delay Segment $\textit{seg}$}

    $\mathcal{P} \leftarrow \mathcal{P}_{CQF}(\mathcal{P}_\textit{orig})$ \;

    \uIf(\tcp*[f]{higher class}){$\mathcal{C}_i^\textit{up} \geq \mathcal{C}_\mathcal{P}^\textit{up}$}{
      \Return $w_i(\mathcal{L}_i^+ - \mathcal{L}_i^- + \delta_\mathcal{P}^\textit{seg})$ \;
    }
    \Else(\tcp*[f]{lower class}){
      \Return $0$ \;
    }
\end{algorithm}

Accounting for these adjustments, Algorithm~\ref{alg:maxInterferenceCQF} details the adaptation of the original $\mathtt{max\_interf}$ (Algorithm~\ref{alg:maxInterference}) to the CQF priority mapping. The remainder of the SP latency analysis from Section~\ref{ssec:sp-simple} remains unchanged. Detailed formulas and pseudo-codes are omitted. Note that the simple SP formula must be used for CQF, as gating negates the serialization gains modeled by the improved formula. The enhancements from Section~\ref{ssec:sp-nc} are inapplicable here, as the final TAS stage in the segment would require a very different service curve. Furthermore, periodic streams can be penalized by the token bucket traffic model. For instance, a periodic stream transmitting \SI{120}{bytes} packets every \SI{2}{ms} would be mapped to a burst of \SI{120}{bytes} and a rate of \SI{480}{kbit/s}. If the CQF window is configured to be \SI{2}{ms} as well, in theory, there should only be one packet per CQF window. However, Algorithm~\ref{alg:maxInterferenceCQF} yields at least \SI{240}{bytes} of worst-case interference. This is a valid assumption under the token bucket model, which permits a second packet of, e.g., \SI{110}{bytes} after only \SI{1.9}{ms}, before the queues alternate. For such scenarios, incorporating specialized traffic models for periodic streams would be highly beneficial.


\subsection{Asynchronous Traffic Shaping}
\label{ssec:ats}

The shift of delay segments described in Section~\ref{ssec:delay-segments} was first proposed in the latency analysis of the Asynchronous Traffic Shaper (ATS)~\cite{8021Qcr} in~\cite{specht2016urgency}. This section briefly outlines the changes introduced by ATS, and presents its latency analysis results. Further details are available in~\cite{specht2016urgency}.

ATS enforces the token-bucket constraint from Equation~\ref{eq:token-bucket} using per-stream shaping. Within each switch, each stream~$i$ is assigned a virtual token bucket with capacity~$b_i$ and token generation rate~$r_i$. A frame~$f_i$ of size $\ell_{f_i}$ can only be transmitted if $\ell_{f_i}$ tokens
can be subtracted from the bucket without going negative. Every shaper in this work is assumed to be configured to re-enforce the end device's original burst and rate constraints. Individual shaping combined with per-class queuing can cause head-of-line blocking delays, therefore ATS mitigates this by applying three Queue Allocation Rules to separate traffic into virtual queues before the shaper. Specifically, frames are isolated into separate virtual queues if they have different ingress port, egress port, or traffic class on either the upstream or downstream switch.

\paragraph*{Latency Analysis}
Under these constraints, the worst-case latency with ATS is bounded by:
\begin{equation}
  \label{eq:ats-delay}
  d^\textit{seg}_{\mathcal{P}} \leq \frac{\sum_{i \in \mathcal{H}_\mathcal{C}} b_i + \sum_{i \in \mathcal{C}} b_i - \ell_{\mathcal{P}}^\textit{min} + \ell^\textit{max}}{R^\textit{seg} - \sum_{i \in \mathcal{H}_\mathcal{C}} r_i} + \frac{\ell_{\mathcal{P}}^\textit{min}}{R^\textit{seg}}.
\end{equation}
This formulation simplifies the original equation in~\cite[Eq.\,21]{specht2016urgency}. Unlike the simple SP model in Equation~\ref{eq:sp-simple-delay}, the ATS delay bound uses the original burstiness from the end devices' Traffic Specification. Furthermore, ATS avoids the need for a pre-configured latency threshold~$\delta_\mathcal{P}^\textit{seg}$, though its queuing rules require more complex hardware implementations. Finally, the serialization-based tighter bounds from Section~\ref{ssec:sp-nc} are not applicable to ATS, as a limited amount of head-of-line blocking can still occur within the virtual queues.

\paragraph*{Pseudo Code}
Because this latency model is independent of accumulated latencies ($\mathcal{L}^-, \mathcal{L}^+$) and latency thresholds ($\delta_\mathcal{P}^\textit{seg}$), its implementation is simpler than Algorithms~\ref{alg:maxInterference} and~\ref{alg:spSimple}. The corresponding logic is detailed in Algorithm~\ref{alg:ats}.
\begin{algorithm}
  \caption{$\mathtt{wc\_delay\_ats}$}
  \label{alg:ats}
  \SetKwInOut{Input}{Input}

  	\Input{Priority $\mathcal{P}$, Streams $\mathcal{S}^\textit{seg}$, Additional Stream $s$}

  	$interf \leftarrow \ell^\textit{max}$ \;
    $r_{\mathcal{H}_\mathcal{C}} \leftarrow 0$ \;
    $\ell^\textit{min}_\mathcal{P} \leftarrow \ell^\textit{max}$ \;
    \For{Stream $i \in \mathcal{S}^\textit{seg} \cup \{s\}$}{
    	\If{$\mathcal{C}_i^\textit{up} == \mathcal{C}_\mathcal{P}^\textit{up}$}{
        $interf \leftarrow interf + b_i$ \;
        $\ell^\textit{min}_\mathcal{P} \leftarrow \min(\ell^\textit{min}_\mathcal{P}, \ell^\textit{min}_i)$ \;
      }
      \ElseIf{$\mathcal{C}_i^\textit{up} > \mathcal{C}_\mathcal{P}^\textit{up}$}{
        $interf \leftarrow interf + b_i$ \;
        $r_{\mathcal{H}_\mathcal{C}} \leftarrow r_{\mathcal{H}_\mathcal{C}} + r_i$ \;
      }
    }
    \Return $(interf - \ell^\textit{min}_\mathcal{P}) / (R^\textit{seg} - r_{\mathcal{H}_\mathcal{C}}) + \ell^\textit{min}_\mathcal{P} / R^\textit{seg}$ \;
\end{algorithm}
The implementation uses a single loop over all reserved streams, including the new reservation candidate, to update the total interfering burst ($interf$), the sum of higher-priority rates ($r_{\mathcal{H}_\mathcal{C}}$), and the minimum frame lengths of the target priority ($\ell_\mathcal{P}^\textit{min}$). These computed metrics are then used to evaluate and return the latency bound from Equation~\ref{eq:ats-delay}.


\subsection{Credit-Based Shaping}
\label{ssec:cbs}

Predating the Asynchronous Traffic Shaper, the Credit-Based Shaper (CBS) \cite[Sec.\,8.6.8.2]{8021Q} was originally designed to limit egress traffic to provide solid mathematical foundations for worst-case latency analysis. However, the latency model proposed in the 802.1BA profile~\cite[Sec.\,6.5]{8021BA} was proven inaccurate~\cite{boiger2010class,Maile_2023_OnThe}. Because CBS is a class-based shaper, multiple streams are shaped as an aggregate by a single instance. This aggregation introduces latency jitter despite shaping, which leads to an increased effective burstiness as described in Section~\ref{ssec:sp-simple}.

For each port and traffic class, CBS manages a credit value and a configured rate called the \emph{idle\_slope}. Frame transmission is only permitted when the credit value of that class is non-negative. This credit value increases over time with the \textit{idle\_slope}. During active transmission of a class~$\mathcal{C}$ frame, the credit decreases at rate~$R^\textit{seg}$, yielding an effective decrease rate of $R^\textit{seg} - \textit{idle\_slope}$. Although credits can become positive, the value resets to zero if a new frame arrives at an empty queue, preventing long-term credit accumulation. Consequently, positive credits only build up while the queue is non-empty. This behavior resembles the token-bucket model in ATS, with two exceptions: (i)~shaping applies to the entire traffic class rather than per-stream, and (ii)~the credit value is not upper bounded directly by the shaper parameters, but indirectly by the time that the queue can be non-empty and idle at the same time, due to higher-priority traffic.

\begin{figure}
  \centering
  \includegraphics[scale=0.50,page=3]{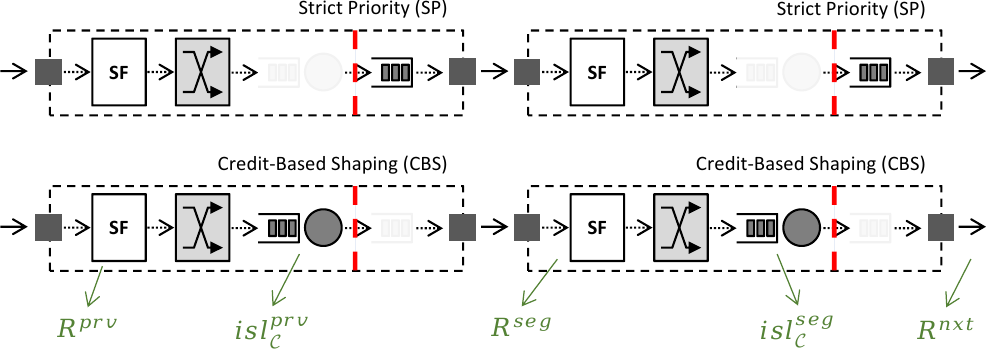}
  \caption{Comparison of effective delay segments in SP and CBS}
  \label{fig:40-delay-model-cbs}
\end{figure}

\paragraph*{Conformance with Delay Segment Definition}
The CBS is a unique case in this work. The delay segment definition in Section~\ref{ssec:delay-segments} requires strict separation between shaping and priority transmission selection. This separation is straightforward for ATS and CQF, and it benefits their analysis. However, it is challenging for CBS because its credit reset behavior depends on whether other frames of the same class remain in the queue. Thus, the shaper state is directly influenced by priority transmission selection.

As the shaper state changes continuously during transmission, the \emph{shaping} step from Figure~\ref{fig:30-delay-types} is only completed after a frame is fully transmitted. The \emph{priority queuing} step is empty, similar to the empty \emph{shaping} step in the SP mechanism. This relationship is shown in Figure~\ref{fig:40-delay-model-cbs}, along with the link speeds~$R^{\{\textit{prv},\textit{seg},\textit{nxt}\}}$ and \emph{idle\_slopes}~$isl_\mathcal{C}^{\{\textit{prv},\textit{seg},\textit{nxt}\}}$. Under CBS, the downstream switch calculates its own priority queuing delay within the latency model, similar to the behavior of most delay models in literature. Co-existence of CBS with other shapers therefore requires careful handling of transitions between these distinct delay models. This integration challenge is discussed in Section~\ref{sec:discussion}. Here, it is assumed that the upstream switch also employs CBS and has already considered its own queuing delay accordingly.

\paragraph*{Latency Analysis}
The latency analysis of CBS adapts the improved Network Calculus formula for SP from Section~\ref{ssec:sp-nc} with three main modifications. (i)~The delay segment incorporates the queuing delay of the downstream switch rather than the upstream switch. As a consequence, the transmission queue of a single egress port
is shared by several delay segments~\textit{seg} from different ingress ports. (ii)~The piecewise arrival curve of SP (Fig.\,\ref{fig:40-arrival-curve-sp}) is adapted to include a segment corresponding to the CBS \emph{idle\_slope}~$isl_\mathcal{C}^\textit{prv}$, which can reduce calculated delays due to restricted traffic arrival. (iii)~The rate-latency service curve of SP (Fig.\,\ref{fig:40-service-curve-sp}) is adapted to the reduced rate of the \emph{idle\_slope}~$isl_\mathcal{C}^\textit{seg}$, which typically increases calculated delays as the service rate decreases.

\begin{figure}
  \centering
  \subfloat[CBS arrival curve for $\hat{\mathcal{S}} = \mathcal{S}_\mathcal{C}^\textit{seg}$\label{fig:40-cbs-arrival-curve}]{%
    \includegraphics[scale=0.50,page=6,valign=t]{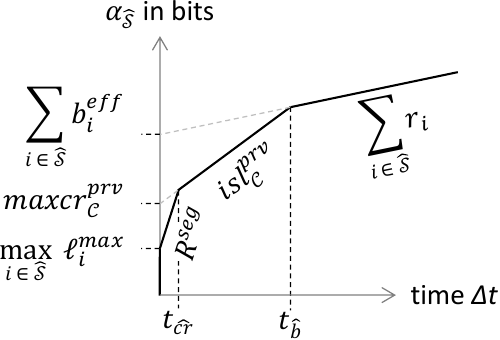}
  }
  \hspace{1em}
  \subfloat[CBS service curve for class $\mathcal{C}$\label{fig:40-cbs-service-curve}]{%
    \vphantom{\includegraphics[scale=0.50,page=6,valign=t]{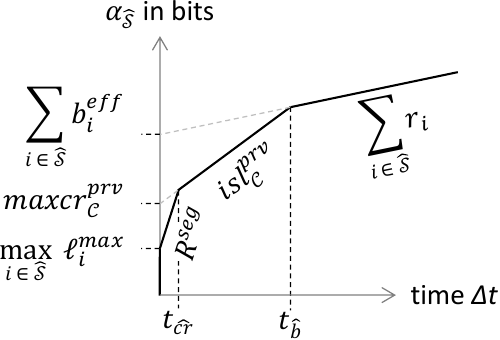}}
    ~~\includegraphics[scale=0.50,page=7,valign=t]{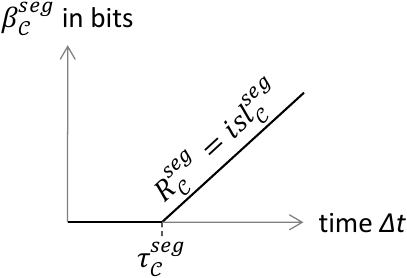}
  }
  \caption{Adjusted arrival and service curves for CBS}
  \label{fig:40-nc-curves-cbs}
\end{figure}

The resulting CBS curves are summarized in Figure~\ref{fig:40-nc-curves-cbs} and detailed below.

For the \textbf{arrival curve}, the model from Figure~\ref{fig:40-arrival-curve-single-port} is valid but can be refined using CBS bandwidth limits, as shown in Figure~\ref{fig:40-cbs-arrival-curve}. Class~$\mathcal{C}$ transmits at line rate~$R^\textit{seg}$ only while its credit is positive. Thereafter, even if the accumulated burst $\sum_{i \in \hat{S}} b_i^\textit{eff}$ is not depleted, the transmission rate is constrained by the upstream device's idle slope~$\textit{isl}_\mathcal{C}^\textit{seg}$. If this idle slope exceeds the aggregate stream rate $\sum_{i \in \hat{S}} r_i$, the rate drops to that aggregate rate at the moment~$t_{\hat{b}}$, just like in the SP model.

The arrival curve is modeled as the infimum of three upper bounds:
\begin{align}
  \alpha_{\hat{\mathcal{S}},1}(\Delta t) & = \max_{i \in \hat{\mathcal{S}}} \ell^\textit{max}_i + R^\textit{seg} \cdot \Delta t, \\
  \label{eq:spnc-alphaS2}
  \alpha_{\hat{\mathcal{S}},2}(\Delta t) & = maxcr_\mathcal{C}^\textit{prv} + isl_\mathcal{C}^\textit{prv} \cdot \Delta t, \\
  \alpha_{\hat{\mathcal{S}},3}(\Delta t) & = \sum_{i \in \hat{\mathcal{S}}} b_i^\textit{eff} + \sum_{i \in \hat{\mathcal{S}}} r_i \cdot \Delta t.
\end{align}
These curves intersect at the points~$t_{\hat{cr}}$ and~$t_{\hat{b}}$, as indicated in Figure~\ref{fig:40-cbs-arrival-curve}.
The intersection points are given by:
\begin{align}
  \label{eq:cbs-tcr}
  t_{\hat{cr}} & = \max \left( 0, \frac{maxcr_\mathcal{C}^\textit{prv} - \max_{i \in \hat{\mathcal{S}}} \ell^\textit{max}_i}{R^\textit{seg} - isl_\mathcal{C}^\textit{prv}} \right), \\
  \label{eq:cbs-tb}
  t_{\hat{b}} & = \max \left( t_{\hat{cr}}, \frac{\sum_{i \in \hat{\mathcal{S}}} b^\textit{eff}_i - maxcr_\mathcal{C}^\textit{prv}}{isl_\mathcal{C}^\textit{prv} - \sum_{i \in \hat{\mathcal{S}}} r_i} \right).
\end{align}
Here, the term $maxcr_\mathcal{C}^\textit{seg}$ represents the maximum positive credit for class~$\mathcal{C}$ on segment~\textit{seg}. Its computation is revisited below, especially in Equation~\ref{eq:maxcr}. Note that, if $maxcr_\mathcal{C}^\textit{seg} < \max_{i \in \hat{\mathcal{S}}} \ell^\textit{max}_i$ holds, then $t_{\hat{cr}} = 0$ is assumed, but $\alpha_{\hat{\mathcal{S}}}(\Delta t)$ is always at least equal to $\max_{i \in \hat{\mathcal{S}}} \ell^\textit{max}_i$ for every non-negative~$t$. This results in the following curve~$\alpha_{\hat{\mathcal{S}}}$ for the streams of a single ingress port $\hat{\mathcal{S}} = \mathcal{S}_\mathcal{C}^\textit{prv}$:
\begin{equation}
  \alpha_{\hat{\mathcal{S}}}(\Delta t) =
  \begin{cases}
    \max_{i \in \hat{\mathcal{S}}} \ell^\textit{max}_i + R^\textit{seg} \cdot \Delta t & \text{ if } 0 \leq \Delta t \leq t_{\hat{cr}}, \\
    maxcr_\mathcal{C}^\textit{prv} + isl_\mathcal{C}^\textit{prv} \cdot \Delta t & \text{ if } t_{\hat{cr}} < \Delta t \leq t_{\hat{b}}, \\
    \sum_{i \in \hat{\mathcal{S}}} b_i^\textit{eff} + \sum_{i \in \hat{\mathcal{S}}} r_i \cdot \Delta t & \text{ if } t_{\hat{b}} < \Delta t.
  \end{cases}
\end{equation}
As before, the full arrival curve~$\alpha_{\mathcal{C}}^\textit{seg}$ is the sum of individual ingress arrival curves~$\alpha_{\hat{\mathcal{S}}}$, cf. Equation~\ref{eq:sp-arrival-curve-full}.

The \textbf{service curve} follows the same rate-latency model. As shown in Figure~\ref{fig:40-cbs-service-curve}, the service rate~$R_\mathcal{C}^\textit{seg}$ matches the \emph{idle\_slope}~$isl_\mathcal{C}^\textit{seg}$ of the CBS. The shaper only allows to exceed this rate temporarily, when positive credit has accumulated. In this model, these positive credit phases are omitted, as these high-rate phases can be interrupted by various events. As before, the offset of the service curve can be obtained in two steps.

First, the offset~$\tau_{\hat{\mathcal{C}}}^\textit{seg}$ for the highest class~$\hat{\mathcal{C}}$ is bounded by one MTU-sized lower-priority packet, identical to the SP analysis: $\tau_{\hat{\mathcal{C}}}^\textit{seg} = \ell^\textit{max} / R^\textit{nxt}$. For lower classes~$\mathcal{C}$, the offset depends on the maximum duration during which higher classes~$\mathcal{H}_\mathcal{C}$ can continuously transmit. The adjacent higher class~$\mathcal{C}+1$ accumulates credit for at most~$\tau_{\mathcal{C}+1}^\textit{seg}$ time, reaching up to $\tau_{\mathcal{C}+1}^\textit{seg} \cdot isl_{\mathcal{C}+1}^\textit{seg}$ credits. These credits are consumed during transmissions at rate $R^\textit{nxt} - isl_{\mathcal{C}+1}^\textit{seg}$, but they can be interrupted by additional higher class packets at rate $r_{\mathcal{H}_{\mathcal{C}+1}} = \sum_{i \in \mathcal{H}_{\mathcal{C}+1}} r_i$. At zero credit, one additional frame up to size~$\ell_{\mathcal{C}+1}^\textit{max}$ may complete transmission. Together, this results in the following offset~$\tau_\mathcal{C}^\textit{seg}$:
\begin{align}
  \label{eq:maxcr}
  maxcr_{\mathcal{C}}^\textit{seg} & = \tau_{\mathcal{C}}^\textit{seg} \cdot isl_{\mathcal{C}}^\textit{seg} \\
  \tau_\mathcal{C}^\textit{seg} & = \tau_{\mathcal{C}+1}^\textit{seg} + \frac{maxcr_{\mathcal{C}+1}^\textit{seg}}{R^\textit{nxt} - isl_{\mathcal{C}+1}^\textit{seg} - r_{\mathcal{H}_{\mathcal{C}+1}}} + \frac{\ell_{\mathcal{C}+1}^\textit{max}}{R^\textit{nxt}}.
\end{align}

\paragraph*{Pseudo Code}
The CBS model is implemented based on the improved SP pseudo-code. Arrival curves~$\alpha_{\hat{\mathcal{S}}}$ for different ingress ports are represented by the tuple: $ac_{\hat{\mathcal{S}}} = (\textit{burst}, \textit{slope1}, \textit{slope2}, \textit{slope3}, \textit{bendpos1}, \textit{bendpos2})$. The service curve~$\beta_\mathcal{C}^\textit{seg}$ uses the same representation as before: $sc_\mathcal{C}^\textit{seg} = (\textit{slope}, \textit{offset})$. Algorithm~\ref{alg:wc-delay-cbs} maps this logic to pseudo-code. As input, it takes a list of streams which belong to class~$\mathcal{C} = \mathcal{C}_\mathcal{P}^\textit{down}$, grouped by ingress port at the downstream switch. If two streams were mapped to different traffic classes by upstream, they must be in separate groups $\hat{\mathcal{S}}_x, \hat{\mathcal{S}}_y$, too. Although they cannot arrive simultaneously from the same port, they have different ingress idle slopes, which cannot easily be combined.

\begin{algorithm}
  \caption{$\mathtt{wc\_delay\_cbs}$}
  \label{alg:wc-delay-cbs}
  \SetKwInOut{Input}{Input}

    \Input{Priority $\mathcal{P}$,
           Delay Segment $\textit{seg}$,
           List of streams per ingress port $LS_\mathcal{C}^\textit{seg} = [\hat{\mathcal{S}}_1, \dots, \hat{\mathcal{S}}_n]$}

    $\mathcal{C} \leftarrow \mathcal{C}_\mathcal{P}^\textit{down}$ \tcp*[f]{downstream traffic class} \;
    $AC \leftarrow [\,]$ \tcp*[f]{empty list} \;
    \For{$\hat{\mathcal{S}}_k \in LS$}{
      $\mathcal{C}_k \leftarrow \mathcal{C}_\mathcal{P}^\textit{up}$ \;
      $ac \leftarrow (0,0,0,0,0,0)$ \tcp*[f]{prepare tuple} \;
      $ac.\textit{burst} \leftarrow \sum_{i \in \hat{\mathcal{S}}_k} b_i^\textit{eff}$ \;
      $ac.\textit{slope1} \leftarrow R^\textit{seg}$ \;
      \vspace{-2pt}
      $ac.\textit{slope2} \leftarrow isl_{\mathcal{C}_k}^{\hat{\mathcal{S}}_k.\textit{prv}}$ \tcp*[f]{idle slope} \;
      $ac.\textit{slope3} \leftarrow \sum_{i \in \hat{\mathcal{S}}_k} r_i$ \;
      $ac.\textit{bendpos1} \leftarrow t_{\hat{cr}}$ \tcp*[f]{Eq.\,\ref{eq:cbs-tcr}}  \;
      $ac.\textit{bendpos2} \leftarrow t_{\hat{b}}$ \tcp*[f]{Eq.\,\ref{eq:cbs-tb}} \;
      $AC.\mathtt{append}(ac)$ \;
    }
    \vspace{-2pt}
    $c \leftarrow \hat{\mathcal{C}}$ \tcp*[f]{highest class} \;
    $\tau_c \leftarrow \ell^\textit{max} / R^\textit{seg}$ \tcp*[f]{initial offset} \;
    \While(\tcp*[f]{recursive function}){$c > \mathcal{C}$}{
      $c \leftarrow c - 1$ \;
      $\tau_c \leftarrow \tau_{c+1}^\textit{seg} + \dfrac{\tau_{c+1}^\textit{seg} \cdot isl_{c+1}^\textit{seg}}{R^\textit{nxt} - isl_{c+1}^\textit{seg} - r_{\mathcal{H}_{c+1}}} + \dfrac{\ell_{c+1}^\textit{max}}{R^\textit{nxt}}$ \;
    }
    $sc \leftarrow (0,0)$ \tcp*[f]{slope, offset} \;
    $sc.\textit{slope} \leftarrow isl_\mathcal{C}^\textit{seg}$ \;
    $sc.\textit{offset} \leftarrow \tau_\mathcal{C}$ \;

    $t_h^\alpha \leftarrow \mathtt{find\_slope\_cbs}_\mathcal{C}^\textit{seg}(sc.\textit{slope}, AC)$ \;
    $d_\mathcal{P}^\textit{seg} \leftarrow \mathtt{findt}_\mathcal{C}^\textit{seg}(\alpha_{\mathcal{C}}^\textit{seg}(t_h^\alpha)) - t_h^\alpha$ \;
    \Return $d_\mathcal{P}^\textit{seg}$
\end{algorithm}

Algorithm~\ref{alg:wc-delay-cbs} is similar to the previous Algorithm~\ref{alg:wc-delay-sp}, but it features a larger arrival curve setup. Additionally, the service curve calculation (lines 13--20) is now recursive because $\tau_c$ depends on $\tau_{c+1}$. The final step uses the identical inverse function $\mathtt{findt}_\mathcal{C}^\textit{seg}(y) = y / slope + offset$, while $\mathtt{find\_slope\_cbs}_\mathcal{C}^\textit{seg}$ extends Algorithm~\ref{alg:findslope} by processing two bending positions per arrival curve, so the contents of the for-loop (lines 7--9) must be done for both of them.

\subsection{Lessons learned}
In simplified terms, worst-case delay computation for SP, CQF, ATS and CBS follows the same pattern: using Traffic Specifications and accumulated latencies to get an upper bound for the workload in the system, and using the link speed or shaper rate to calculate the service time for that workload. This section provided formal proofs for worst-case delays with all four shapers in decentralized architectures. For SP and CBS, additional Network Calculus models were derived that provide tighter bounds at the expense of more required information and more complex computation. They benefit from the serialization effect: additional workload cannot appear instantaneously, it is received over time, while the previous workload can already be serviced. To bridge the gap between formal theory and applicability, pseudo code implementations are suggested for each discussed model.

Using Network Calculus theorems like \emph{leftover service} or \emph{output characterization}~\cite[Sec.\,2.3]{jiang2008stochastic} directly with the service curves is avoided here, as that would require frequent re-calculations when streams change anywhere in the network. The accumulated maximum latency is used with the Traffic Specification to construct local arrival curves instead.


\section{Evaluation}
\label{sec:evaluation}

This section presents a comparison of the five different latency models. A summary of the evaluated Key Performance Indicators (KPIs) and scenario parameters is provided in Table~\ref{tab:50-summary}. The KPIs used to quantify model performance in practice are explained in Section~\ref{ssec:kpis}. The network topologies used for the evaluations are detailed in Section~\ref{ssec:topologies}. The testing methodology and results are then presented for three distinct scenarios in Sections \ref{ssec:eval-preconf}, \ref{ssec:eval-autoconf}, and \ref{ssec:eval-single}: (\ref{ssec:eval-preconf})~a pre-configured scenario with fixed delay thresholds where stream acceptance is analyzed, (\ref{ssec:eval-autoconf})~an auto-configured scenario accepting all streams to evaluate end-to-end guarantees, and (\ref{ssec:eval-single})~a single-switch parameter study observing individual parameter effects. These scenarios are also summarized in Table~\ref{tab:50-summary}, and their implementation is available on GitHub\footnote{\url{https://github.com/lsinfo3/tsn-decentralized-network-planning}}. Finally, a summary and shaper selection recommendations are provided in Section~\ref{ssec:eval-summary}.

\begin{table}
	\centering
	\caption{Overview of KPIs and evaluation scenarios}
	\label{tab:50-summary}
	\begin{tabular}{>{\hangindent=0.7em}p{2em}
									>{\hangindent=0.7em}p{4em}
	                >{\hangindent=0.7em}p{19em}} \toprule
		\textbf{KPI} & \textbf{Scope} & \textbf{Description} \\ \midrule
		$|\mathcal{S}^\mathcal{P}_a|$ & priority & Number of accepted streams \\
		$r_a^\mathcal{P}$ & priority & Sum of data rates of accepted streams \\
		$b_a^\mathcal{P}$ & priority & Sum of bursts of accepted streams \\
		$d_\mathcal{P}^\textit{seg}$ & prio., link & Latency bound for priority~$\mathcal{P}$ and segment~\textit{seg} \\
		$\mathcal{L}^+_i$ & stream & Accumulated maximum latency of stream~$i$ \\
		$\mathcal{L}^\Delta_i$ & stream & Jitter: difference between max and min latency \\
		$b_i^\textit{eff}$ & stream & Effective burstiness (at the receiver device) \\
		$t_\textit{comp}$ & total & Computation time for entire scenario \\
	\end{tabular}
	\\ \vspace{2pt}
	\begin{tabular}{>{\hangindent=0.7em}p{14em}
	                >{\hangindent=0.7em}p{12.5em}} \toprule
		\textbf{Scenario} & \textbf{KPIs} \\ \midrule
		preconf: fixed delay thresholds & $|\mathcal{S}^\mathcal{P}_a|$, $r_a^\mathcal{P}$, $b_a^\mathcal{P}$, $d_\mathcal{P}^\textit{seg}$ \\[0.2em]
		autoconf: accepting all streams & $\mathcal{L}^+_i$, $\mathcal{L}^\Delta_i$, $b_i^\textit{eff}$, $t_\textit{comp}$ \\[0.2em]
		single switch parameter study & $d_\mathcal{P}^\textit{seg}$ \\
		\bottomrule
	\end{tabular}
\end{table}

\subsection{KPIs}
\label{ssec:kpis}

Resource reservation for deterministic networking is fundamentally an admission control process, where streams requesting allocation in high-priority queues are either accepted or rejected along their path. Consequently, the most direct performance indicator is the \textbf{number of successfully admitted stream reservations}~$|\mathcal{S}^\mathcal{P}_a|$ in a given scenario. This metric is mapped to underlying traffic characteristics: the \textbf{sum of accepted data rates}~$r_a^\mathcal{P}$ and \textbf{bursts}~$b_a^\mathcal{P}$, since larger traffic volumes demand more resources. Furthermore, the \textbf{calculated latency bound}~$d_\mathcal{P}^\textit{seg}$ is a direct indicator of the these measures, as lower latency predictions permit more admissions.

However, this comparison is not straightforward. Most decentralized latency models presented here require pre-configured latency thresholds~$\delta_\mathcal{P}^\textit{seg}$ to function. A latency bound~$d_\mathcal{P}^\textit{seg}$ remains valid only if every network link stays below its configured threshold~$\delta_\mathcal{P}^\textit{seg}$. If excessive stream admissions cause $d_\mathcal{P}^\textit{seg}$ to exceed $\delta_\mathcal{P}^\textit{seg}$ at any point, the underlying assumptions of the latency models behind Section~\ref{sec:latency-bounds} fail, invalidating all calculated bounds across the network. Consequently, in scenarios where reservations are rejected, the resulting latency bounds $d_\mathcal{P}^\textit{seg}$ cannot properly reflect the performance of the model. They simply reflect the pre-defined configuration rather than actual model performance.

To resolve this issue, the `autoconf' scenarios do not use pre-configured thresholds. Instead, thresholds~$\delta_\mathcal{P}^\textit{seg}$ are increased iteratively until they closely match the calculated delays~$d_\mathcal{P}^\textit{seg}$ and all streams are accepted. Under this near-optimal configuration, the streams' performance is represented by the \textbf{accumulated maximum latency}~$\mathcal{L}^+_i = \sum_{\textit{seg}} \delta_\mathcal{P}^\textit{seg}$, at the receiver. Other bounds may be required depending on the use case: distributed control loops could prioritize an upper bound for \textbf{latency jitter}~$\mathcal{L}^\Delta_i = \mathcal{L}^+_i - \mathcal{L}^-_i$, while low-tier devices with small buffers may require low \textbf{effective burstiness}~$b_i^\textit{eff}$.

Finally, the total \textbf{computation time}~$t_\textit{comp}$ of all latency calculations within a scenario is measured. Note that this is not a sophisticated measurement, it can only provide a rough comparison. The execution time in Python is subject to random effects such as Garbage Collection, and there are no optimizations considered, such as memorizing intermediate results, which can have significant effects as demonstrated in~\cite[Fig.\,6c]{grigorjew2021distributed}. However, the timings still provide a rough overview of the relative complexity of the different algorithms.

\begin{figure}
  \centering
  \subfloat[Small\label{fig:50-industrial-example-small}]{%
    \includegraphics[scale=0.2,page=1,valign=t]{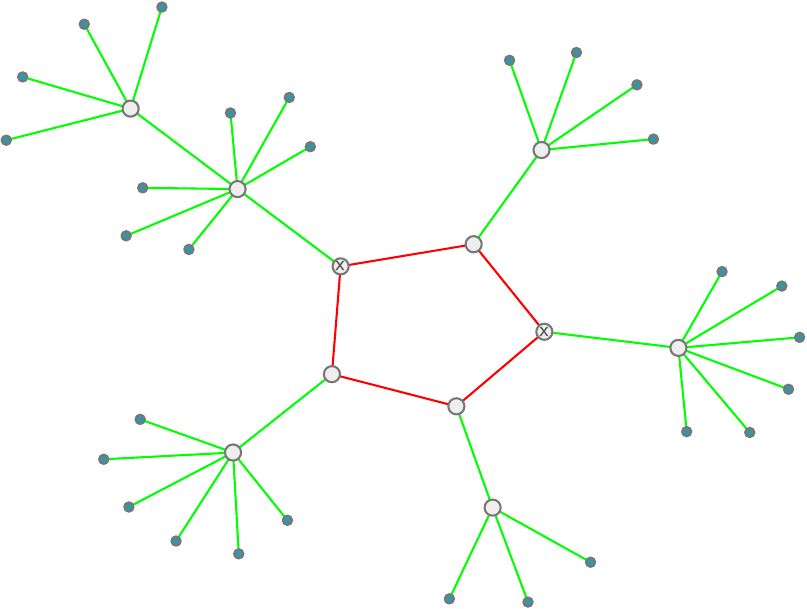}
  }
  \subfloat[Medium\label{fig:50-industrial-example-medium}]{%
    \includegraphics[scale=0.2,page=2,valign=t]{figs/50-industrial-examples.pdf}
  }
  \subfloat[Big\label{fig:50-industrial-example-big}]{%
    \includegraphics[scale=0.2,page=3,valign=t]{figs/50-industrial-examples.pdf}
  }
  \caption{Randomly sampled industrial scenarios; link colors indicate relative bandwidth (100M-1G in small, 100M-10G in medium and big topologies)}
  \label{fig:50-industrial-examples}
\end{figure}

\subsection{Network Topologies}
\label{ssec:topologies}

The performance evaluation is conducted on three network topology types: (i)~a linear topology representing daisy-chained industrial automation networks, (ii)~a set of synthetic industrial topologies with varying sizes, link speeds, stream paths, and Traffic Specifications, and (iii)~a single-switch topology with varying amounts of connections for the parameter study.

\paragraph*{Linear Topology}
Since realistic industrial networks are heavily based on linear chains of devices~\cite{profinet-design-guidelines,industrial-profile-use-cases} to facilitate factory deployment, a \emph{linear topology} is used to establish a baseline scenario without random variance. A chain of 25~switches is modeled, with each switch connected to four end devices. The first switch is connected to a Programmable Logic Controller (PLC) that receives traffic from all 100~devices, resulting in path lengths of 2 to 26~hops. Each device generates a single stream with a burst size of \SI{120}{bytes} and a data rate of \SI{150}{kbits/s}, representing periodic industrial control loop traffic, roughly one packet every \SI{6.4}{ms}.

\paragraph*{Industrial Topologies}
The industrial topologies extend the daisy-chain concept using synthetic networks modeled after common industrial automation patterns. These configurations are inspired by the IEC/IEEE 60802 industrial profile~\cite{industrial-profile-use-cases} and PROFINET design guidelines~\cite{profinet-design-guidelines}, which were combined into a scenario generation tool in prior work~\cite{grigorjew2023developing}.

The initial topology generator was based on traditional industrial infrastructures. Typically, the field level consists of narrow linear chains of low-port switches representing individual machines or sensor arrays, which connect to the control level via the built-in switch of their PLC. At the control level, switches are often arranged in a ring topology for redundancy, which connects to the plant level via redundant links supporting higher bandwidth. The plant level itself is often structured like the higher layers of a fat tree topology.

The tool from~\cite{grigorjew2023developing} was further extended to anticipate future industrial trends, such as the migration of physical PLC functions to virtual PLCs in plant-level mini-clouds. This shift increases the scale, heterogeneity, and stream count at the plant level. Technologies like TSN enable such real-time communication over Ethernet, beyond the traditional closed systems, which makes these heterogeneous scenarios highly suitable for evaluating TSN latency models.

Example topologies for the small, medium, and big profiles are depicted in Figure~\ref{fig:50-industrial-examples}. A small topology comprises a main chain of 2--5 switches (\SI{1}{Gbit/s}, red), connected to a ring in 50\% of the cases, with linear branches of length 1--3 (\SI{100}{Mbit/s}, green) and 3--6 end devices per branch switch. Note that the two main switches in Figure~\ref{fig:50-industrial-example-small} are marked with an `X': they serve as \emph{join points} that can be used to combine multiple topologies. Essentially, the medium scenario is just a combination of multiple small topologies, connected to a \SI{10}{Gbit/s} main ring; cf. Figure~\ref{fig:50-industrial-example-medium}. The big scenario scales this up by connecting a small fat-tree representing a local data center with \SI{1}{Gbit/s} links to the main ring.

Stream paths, bursts, data rates, and priorities are generated randomly. Short paths are generated more frequently than longer paths. Small, medium, and big topologies contain 110, 220, and 430 streams, respectively. Priorities are selected randomly from 4 to 7, burst sizes are uniformly distributed in the range $\lbrack \SI{64}{bytes}, \SI{1000}{bytes} \rbrack$, and data rates are exponentially distributed in the range $\lbrack \SI{10}{kbits/s}, \SI{50}{Mbits/s} \rbrack$ to favor lower data rates. All streams utilize shortest-path routing without assuming a spanning tree.

\begin{figure}
  \centering
  \includegraphics[scale=0.50,page=1]{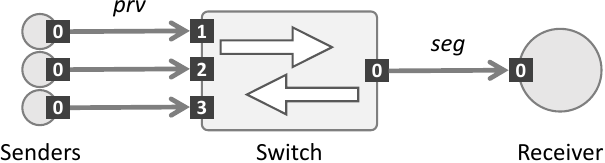}
  \caption{Single switch evaluation scenario}
  \label{fig:50-single-switch}
\end{figure}

\paragraph*{Parameter Study Topology}
The parameter study network is depicted in Figure~\ref{fig:50-single-switch}. It consists of a single switch with several senders transmitting to a single receiver. The number of senders varies dynamically, and the delay analysis focuses on the egress segment~\textit{seg} connecting the switch to the receiver.

\subsection{Pre-configured Scenario \& Accepted Streams}
\label{ssec:eval-preconf}

\begin{figure}
  \centering
  \includegraphics[scale=0.5,page=2]{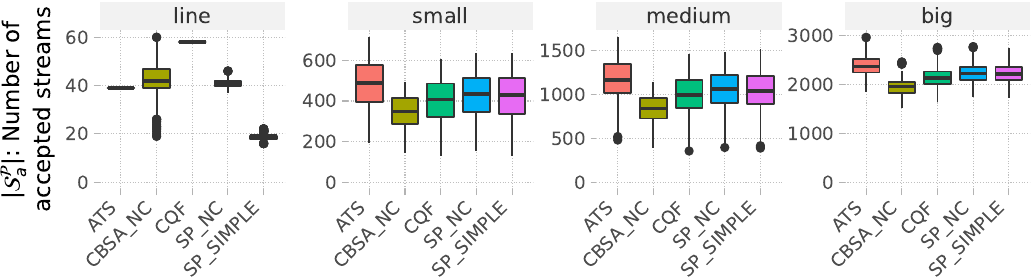}\\[0.5em]
  \includegraphics[scale=0.5,page=4]{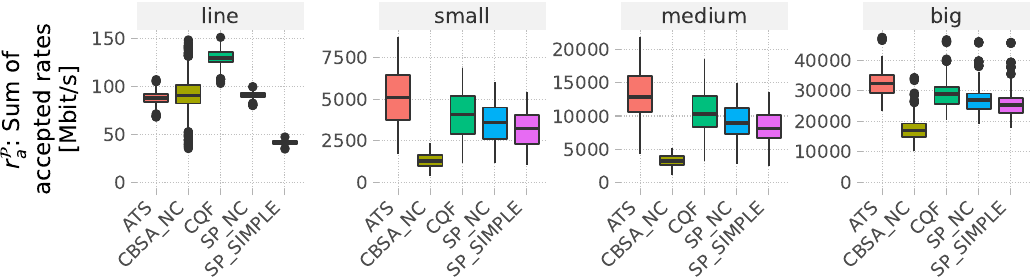}\\[0.5em]
  \includegraphics[scale=0.5,page=6]{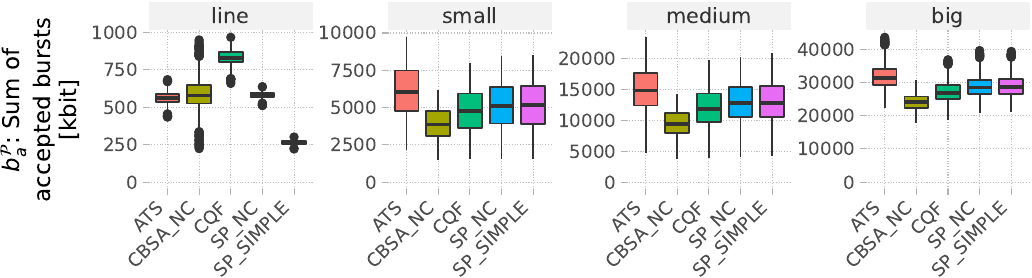}
  \caption{Comparison of accepted streams, throughputs, and bursts per scenario}
  \label{fig:50-eval-preconf-sizes}
\end{figure}

In the initial comparison, the four network sizes (linear, small, medium, and big) are pre-configured with static default latency and bandwidth utilization thresholds without optimizing for individual links. Stream reservations are simulated sequentially. This sequential order creates an implicit stream prioritization: earlier reservation requests are more likely to be accepted, as there are more resources available in the beginning.

This setup can create unexpected bottlenecks and uneven resource utilization, complicating model comparisons. For example, admitting a single high-burst stream with a long path consumes more resources than multiple short, low-burst streams. If one shaper successfully accommodates this larger stream, it may reject subsequent smaller reservations due to exhausted resources. Conversely, a \emph{slightly} less accurate shaper might reject the large stream immediately, allowing it to accept more small streams. Comparing only the total number of accepted reservations~$|\mathcal{S}_a^\mathcal{P}|$ would incorrectly suggest that the second shaper performed better despite predicting higher latencies.

To mitigate this bias, two actions are taken: (i)~the total accepted bursts~$b_a^\mathcal{P}$ and rates~$r_a^\mathcal{P}$ are evaluated alongside~$|\mathcal{S}_a^\mathcal{P}|$, and (ii)~the stream list is shuffled 100 times per scenario to perform reservation simulations under varying orders, mitigating the impact of unlucky orders.

Specifically, the evaluation procedure is as follows: For all scenario sizes, 100~generated topologies are analyzed across 100~random stream shuffles for each of the five latency models (SP\_SIMPLE, SP\_NC, CBSA\_NC, CQF, ATS), repeating the following steps:
\begin{enumerate}[leftmargin=1.3em]
	\item Generate the network topology. Generate 10~times the baseline stream count to minimize random statistical variation in acceptance metrics.

  \item Set per-hop latency thresholds for every delay segment in the network: $\delta_4 = \SI{5}{ms}$, $\delta_5 = \SI{2}{ms}$, $\delta_6 = \SI{1}{ms}$, and $\delta_7 = \SI{500}{\micro s}$. For CQF, priorities are consolidated, and thresholds are configured as: $\delta_4 = \delta_5 = \SI{3.5}{ms}$, and $\delta_6 = \delta_7 = \SI{750}{\micro s}$. These values are based on industrial \emph{network cycle} times~\cite{industrial-profile-use-cases} and standard profiles~\cite{8021BA}.

  \item For CBSA\_NC, set bandwidth thresholds (and idle slopes) to 20\% of the link bandwidth for classes 4--7. For other shapers, stream rates are not limited per class, they are only bounded by the total link bandwidth.

  \item For each stream, test whether any latency threshold would be exceeded by including that stream. If not, permanently admit the stream. Otherwise, mark that reservation as \emph{failed} and proceed without that stream.

  \item Record the list of all accepted streams for this iteration.
\end{enumerate}

\begin{figure}
  \centering
  \includegraphics[scale=0.5,page=1]{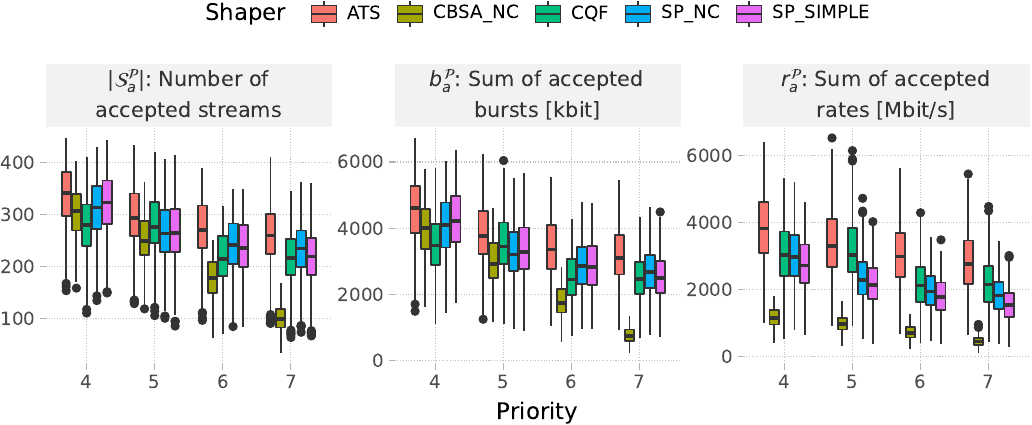}
  \caption{Comparison of priorities within the industrial medium scenario}
  \label{fig:50-eval-preconf-medium-priorities}
\end{figure}

\begin{figure}
  \centering
  \subfloat[accepted throughput utilization\label{fig:50-eval-preconf-small-rates}]{%
    \includegraphics[scale=0.5,page=1]{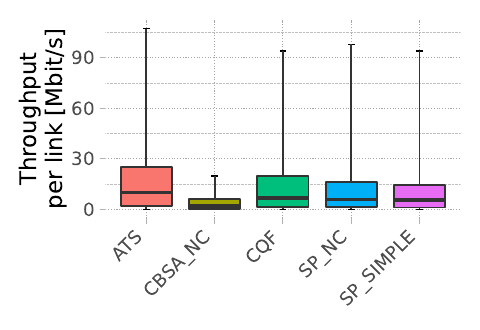}
  }
  \subfloat[accepted delay utilization\label{fig:50-eval-preconf-small-delay}]{%
    \includegraphics[scale=0.5,page=3]{figs/50-eval/preconf__links.pdf}
  }
  \caption{Accepted utilization per link in the industrial small topology;
    whiskers extend to the maximum due to an excessive amount of outliers}
  \label{fig:50-eval-preconf-small}
\end{figure}

\paragraph*{Evaluation Results}
Figure~\ref{fig:50-eval-preconf-sizes} summarizes the three KPIs: accepted streams~$|\mathcal{S}^\mathcal{P}_a|$, aggregate accepted rates~$r_a^\mathcal{P}$, and aggregate accepted bursts~$b_a^\mathcal{P}$. Each boxplot data point represents a single topology under the specified parameters. The plots are grouped by topology size, and colors indicate the latency models. In addition, Figures~\ref{fig:50-eval-preconf-medium-priorities} and~\ref{fig:50-eval-preconf-small} zoom into specific scenarios. For most plots except for Figure~\ref{fig:50-eval-preconf-small-delay}, higher values indicate better model performance. Key observations include:

(1) The linear topology behaves differently than the three industrial topologies (Figure~\ref{fig:50-eval-preconf-sizes}). Within the industrial subset (small, medium, big), identical trends occur across all three KPIs, indicating that network size has minimal impact on shaper comparison as long as structural properties are maintained. However, the \emph{network type} and shape, which affects path lengths, number of ingress ports, and fan-in/fan-out, can alter the preferred latency model. Specific parameters are analyzed in Section~\ref{ssec:eval-single}.

(2) In the linear topology, CQF appears to perform best, but this is most likely an artifact of the configuration. CQF uses a higher latency threshold to compensate for priority consolidation. While this configuration usually reduces stream capacity, the linear network evaluated here contains only priority~7 streams, neutralizing consolidation penalties. The higher threshold directly translates to more accepted streams, which does not occur in other topologies.

(3) In the linear topology, ATS, CBSA\_NC, and SP\_NC show similar performance across all KPIs. In contrast, SP\_SIMPLE accepts only half as many streams, demonstrating the capacity benefits of the other models' improvements.

(4) In industrial topologies, CBSA performs worst across all KPIs, particularly in the rate comparison, indicating high sensitivity to high-rate streams. Although CBSA\_NC is the only model requiring a bandwidth cap (20\% per priority), its lower performance is visible long before reaching this limit. As shown in Figure~\ref{fig:50-eval-preconf-small-rates}, while outliers are indeed capped at \SI{20}{Mbit/s}, the 75\% quantile of the CBSA box is significantly below that of other shapers.

(5) Generally, ATS accepts the most streams, throughput, and burst in industrial scenarios (Figure~\ref{fig:50-eval-preconf-sizes}). CQF, SP\_SIMPLE, and SP\_NC accept similar stream counts, with CQF favoring higher rates and SP favoring higher bursts. In this context, the difference between SP\_SIMPLE and SP\_NC is surprisingly small.

(6) To analyze this further, Figure~\ref{fig:50-eval-preconf-medium-priorities} separates the KPIs in the medium industrial scenario by priority. SP\_NC performs slightly better at high priorities, but the trend reverses at lower priorities. This occurs because admitting more high-priority streams increases lower-priority latencies, consuming their resources. Additionally, in pre-configured scenarios, all links are configured equally, but they are not utilized equally, and a single bottleneck switch can reject many streams. Figure~\ref{fig:50-eval-preconf-small-delay} shows the delay utilization distribution of the small industrial scenarios, i.e., the actual delay~$d_\mathcal{P}^\textit{seg}$ over the pre-configured threshold~$\delta_\mathcal{P}^\textit{seg}$. Although SP\_NC matched SP\_SIMPLE's overall acceptance in Figure~\ref{fig:50-eval-preconf-sizes}, it maintains a lower overall delay utilization. Optimizing thresholds for expected link loads would enable SP\_NC to perform better by using fewer resources. This is evaluated in the auto-configured scenarios.

\subsection{Auto-configured Scenario \& E2E Guarantees}
\label{ssec:eval-autoconf}

A drawback of pre-configured scenarios is that static configurations can introduce bias. A specific configuration that works well for one shaper could be disadvantageous for another. Furthermore, uniform link configuration creates highly utilized bottleneck switches while leaving the rest of the network under-utilized. In theory, having an \emph{'optimal'} configuration for each shaper could make their comparison more representative of their real performance, and reveal more differences.

To avoid these pitfalls, an auto-configured study is conducted where all streams are accepted, and per-hop thresholds~$\delta_\mathcal{P}^\textit{seg}$ are iteratively matched to calculated delays~$d_\mathcal{P}^\textit{seg}$ to eliminate overprovisioning. For all latency models in this paper, the computed delays~$d_\mathcal{P}^\textit{seg}$ are non-decreasing when the thresholds~$\delta_\mathcal{P}^\textit{seg}$ are increased. Therefore, optimal configurations can be achieved by increasing the thresholds~$\delta_\mathcal{P}^\textit{seg}$ in small increments until $d_\mathcal{P}^\textit{seg} \leq \delta_\mathcal{P}^\textit{seg}$.

Because all streams are accepted, performance is now measured using end-to-end latency guarantees~$\mathcal{L}_i^+$, latency jitter~$\mathcal{L}_i^\Delta$, and effective burstiness~$b_i^\textit{eff}$, alongside computation time~$t_\textit{comp}$. For each scenario type and each shaper, the following steps are performed over 100 topologies:
\begin{enumerate}[leftmargin=1.3em]
  \item Generate the network topology and streams as explained before.

  \item Ensure no link is overutilized, verifying that $\sum_{i \in \mathcal{S}^\textit{seg}} r_i \leq R^\textit{seg}$ on all segments. Remove streams if necessary to satisfy this condition.
  
  \item Configure all latency thresholds~$\delta_\mathcal{P}^\textit{seg}$ in the network:
    \begin{enumerate}
      \item Initialize all thresholds to a default value ($\delta_{4,5,6,7}=\SI{1}{ns}$).
      \item \label{enum:loop} Compute delay bounds~$d_\mathcal{P}^\textit{seg}$ based on current thresholds.
      \item Check if $d_\mathcal{P}^\textit{seg} \leq \delta_\mathcal{P}^\textit{seg}$ holds for all segments and priorities.
      \item If satisfied, terminate the loop.
      \item \label{enum:config} If not, update thresholds to $\delta_\mathcal{P}^\textit{seg} = d_\mathcal{P}^\textit{seg} \cdot 1.01$ and return to step~\ref{enum:loop}.
    \end{enumerate}


  \item Record the stream KPIs: $\mathcal{L}^+_i$, $\mathcal{L}^\Delta_i$, $b_i^\textit{eff}$, and $t_\textit{comp}$.
\end{enumerate}
Note that for CQF, per-link optimization is not possible since the shaper assumes equal hop delays. Step~\ref{enum:config} is therefore updated to use the maximum delay: $\delta_\mathcal{P} = \max_{seg} d_\mathcal{P}^\textit{seg} \cdot 1.01$. For CBSA, the idle slope is configured as $\textit{isl}_\mathcal{C}^\textit{seg} = \sum_{i \in \mathcal{S}_\mathcal{C}^\textit{seg}} r_i \cdot 1.01$. Alternative options are explored in Section~\ref{ssec:eval-single}.

\begin{figure}
  \centering
  \includegraphics[scale=0.5,page=4]{figs/50-eval/accept_all_paperplots2_ecdf_small-crop.pdf}\\[0.5em]
  \includegraphics[scale=0.5,page=1]{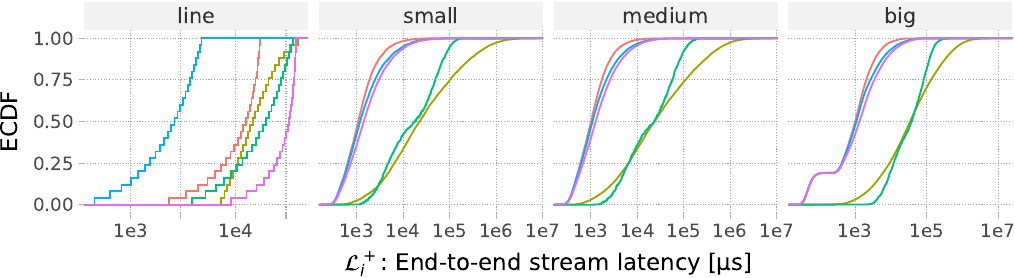}\\[0.5em]
  \includegraphics[scale=0.5,page=2]{figs/50-eval/accept_all_paperplots2_ecdf_small-crop.pdf}\\[0.5em]
  \includegraphics[scale=0.5,page=3]{figs/50-eval/accept_all_paperplots2_ecdf_small-crop.pdf}
  \caption{Comparison of end-to-end latency guarantee, jitter, and effective burst for every topology type}
  \label{fig:50-eval-acceptall-sizes}
\end{figure}

\paragraph*{Evaluation Results}
Figure~\ref{fig:50-eval-acceptall-sizes} presents the Empirical Cumulative Distribution Functions (ECDFs) for each topology size and KPI. Different colors indicate the different latency models. The x-axis is spaced logarithmically, as the results can be several orders of magnitude apart. Generally, the further the ECDF is on the left, the better the latency model performs. Key findings include:

(1) In industrial topologies, CQF and CBSA show significantly higher latencies ($\mathcal{L}_i^+$ and $\mathcal{L}_i^\Delta$) than the other shapers, occasionally by a factor of 100. This difference is much higher compared to the previous pre-configured evaluation. This is expected for CQF, as its global threshold configuration forces the bottleneck link delays to every other link, while all other shapers benefit from the per-hop auto-configuration. For CBSA\_NC, this poor performance stems from a reduced service rate compared to SP\_NC. CBSA can perform much better in general -- the conditions for that are explored later in Section~\ref{ssec:eval-single}. Here in the industrial scenarios, these conditions are rarely met.

(2) The maximum jitter~$\mathcal{L}_i^\Delta$ is dominated by the maximum latency~$\mathcal{L}_i^+$ since the minimum latency is negligible for most shapers. CQF is the only exception; it ensures that each hop takes exactly $\delta_\mathcal{P}^\textit{seg}$ time, leaving only the last-hop jitter. However, for the industrial scenarios, CQF jitter remains higher than that of locally optimized shapers because its last-hop threshold~$\delta_\mathcal{P}^\textit{seg}$ is still quite large in comparison.

(3) ATS performs best across all sizes and KPIs in industrial scenarios, particularly for receiver burstiness~$b_i^\textit{eff}$. This shows that mitigating the burstiness cascade is an effective strategy -- if the switches are actually capable of performing the expensive shaping mechanism. SP\_NC is the second-best option across all KPIs.

(4) Big topologies show a distinct step in latency curves at low values. For ATS and both SP models, roughly 18\% of end-to-end latencies fall between \SI{20}{\micro s} and \SI{100}{\micro s}. This is an artifact of the topology generation. The big topologies include a small fat-tree section where each end-device is connected with \SI{1}{Gbit/s}. It is modeled after anticipation of local mini data centers. All the other end devices are connected with \SI{100}{Mbit/s}. Streams within these high-bandwidth subnets experience shorter paths and lower latency. For CQF and CBSA, this step is not visible. Per-hop optimization is not applicable to CQF, so the delay thresholds are equal everywhere, and CBSA service curves rely on the idle slope rather than the link speed.

(5) Linear topology latency curves ($\mathcal{L}_i^+$) show exactly 25 steps, representing each switch. CBSA and CQF perform better here due to the lack of fan-out. All streams share the same egress port and all delay segment calculations have five ingress ports. SP\_NC outperforms ATS here, likely due to serialization gains. Most streams arrive from the same ingress port, and SP\_NC accounts for sequential arrival over ingress links. With ATS, the packets could theoretically leave the previous shaper simultaneously.

(6) In linear topology jitter metrics~$\mathcal{L}_i^\Delta$, CQF maintains a constant jitter bound over all streams. CQF's remaining jitter is localized to the last hop, and since all streams share the same last hop, their jitter is fixed by the last hop's threshold~$\delta_7^\textit{seg}$.

(7) Regarding linear topology burstiness~$b_i^\textit{eff}$, ATS and CQF provide constant receiver bursts. ATS reshapes traffic directly with the token bucket model, making receiver burstiness dependent on the source Traffic Specifications, independently of their path. CQF reshapes traffic to maintain a constant delay, which directly prevents the burstiness cascading from occurring, as the additional burstiness is based on jitter, cf. Equation~\ref{eq:effective-burstiness}.

\begin{figure}
  \centering
  \includegraphics[scale=0.5,page=2]{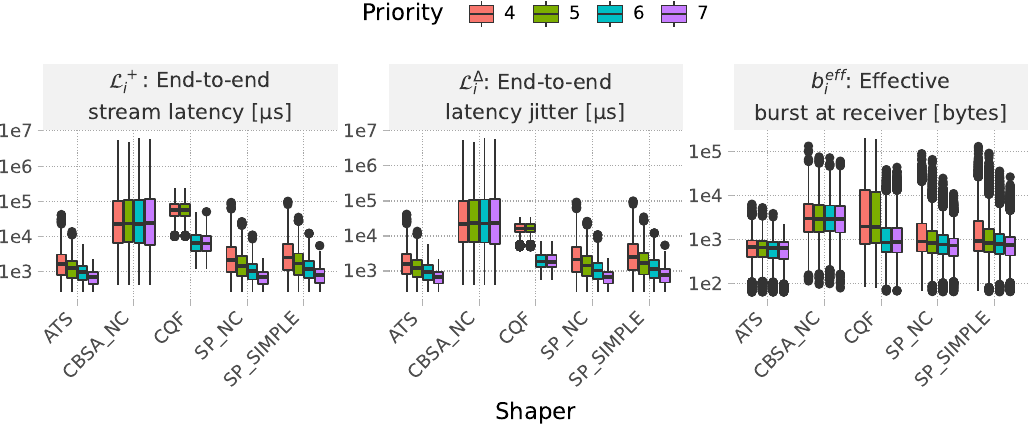}\\[0.5em]
  \caption{Influence of the shaper on the relative differences between priorities, zoomed into the medium topology size}
  \label{fig:50-eval-acceptall-priorities}
\end{figure}

(8) Between the industrial scenarios, the topology size appears to have little effect on the relative differences between shapers. Therefore, only the medium topology is analyzed for the remaining observations. Figure~\ref{fig:50-eval-acceptall-priorities} breaks down performance by priority. ATS, SP\_NC, and SP\_SIMPLE exhibit delays and jitter that align with priority levels: higher priorities show lower values. This trend disappears under CBSA, likely because the service rate is reduced to the stream data rates, making service rate limits more influential than the presence of higher-priority traffic. Under CQF, the consolidated priority pairs show similar performance. For effective bursts, most observations are similarly applicable, but ATS remains uniform across priorities. Due to per-stream shaping of ATS, the graph reflects the source Traffic Specifications across all priorities.

\begin{figure*}
  \centering
  \includegraphics[scale=0.5,page=9]{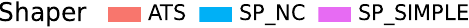}\\[0.5em]
  \subfloat[Relative end-to-end stream latency, normalized to SP\_NC\label{fig:50-eval-acceptall-relative-spnc-latency}]{%
    \includegraphics[scale=0.5,page=7]{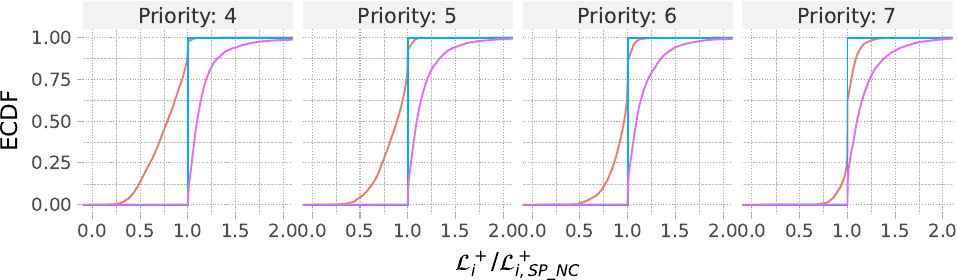}
  }\hspace{2em}
  \subfloat[Relative effective burst at receiver, normalized to SP\_NC\label{fig:50-eval-acceptall-relative-spnc-burst}]{%
    \includegraphics[scale=0.5,page=8]{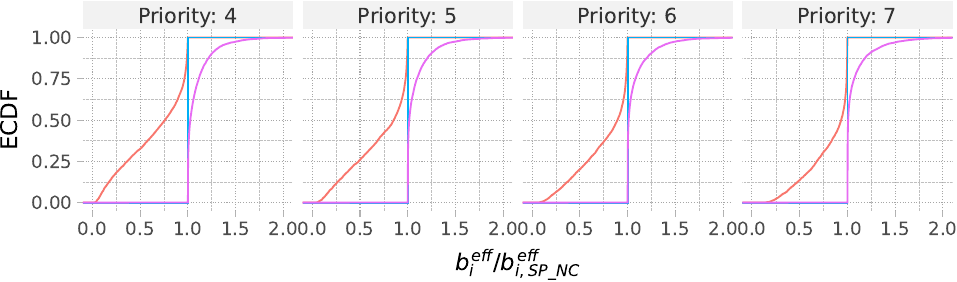}
  }
  \caption{Relative performance of ATS, SP\_SIMPLE and SP\_NC in the medium topologies}
  \label{fig:50-eval-acceptall-relative-spnc}
\end{figure*}

(9) In the previous figures, the scale of differences is difficult to recognize due to the logarithmic axes. Therefore, Figure~\ref{fig:50-eval-acceptall-relative-spnc} focuses on the three close shapers: ATS, SP\_NC and SP\_SIMPLE. The values for ATS and SP\_SIMPLE are normalized to SP\_NC to highlight relative differences. The jitter plot for $\mathcal{L}_i^\Delta$ was omitted, as it is very similar to the latency $\mathcal{L}_i^+$. For burstiness~$b_i^\textit{eff}$, ATS consistently matches or outperforms SP\_NC, while SP\_SIMPLE consistently performs worse than SP\_NC. This gap increases at lower priorities, as SP jitter grows. In latency metrics, SP\_NC strictly outperforms SP\_SIMPLE, but ATS and SP\_NC are highly competitive, especially at high priorities. For priority~7, 25\% of ATS streams outperform SP\_NC, while 30\% of SP\_NC streams outperform ATS. Thus, SP\_NC represents a direct upgrade to SP\_SIMPLE. The choice between ATS and SP\_NC depends on whether per-hop shaping or physical serialization is more dominant in the scenario.

\begin{figure}
  \centering
  \includegraphics[scale=0.5,page=3]{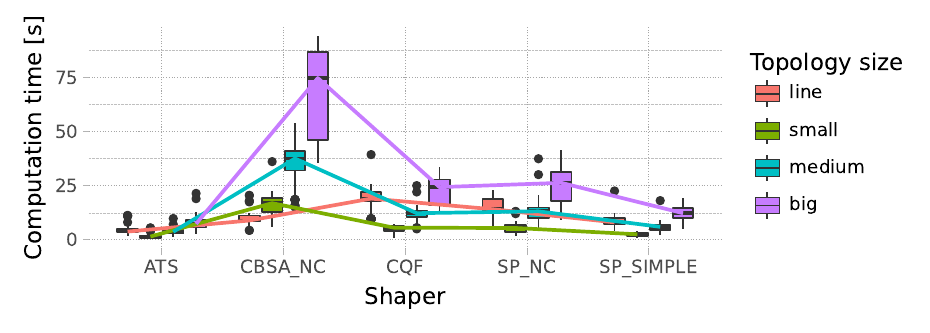}
  \caption{Required computation time for all latency guarantees}
  \label{fig:50-eval-acceptall-timings}
\end{figure}

(10) Finally, Figure~\ref{fig:50-eval-acceptall-timings} shows computation times. Each data point of the boxplot represents one scenario, while the lines connect the medians of each boxplot for improved readability. CBSA\_NC is the most demanding model for large topologies, whereas SP\_SIMPLE and ATS are highly efficient. Note that CQF is, in theory, also a simple model, but likely suffers here due to the straightforward implementation of the priority consolidation. In general, sophisticated optimization of the algorithms might shift the timing relations between the shapers, or otherwise reinforce them, but Figure~\ref{fig:50-eval-acceptall-timings} still provides a naive baseline overview of the algorithm complexity.

\begin{figure}
  \centering
  \includegraphics[scale=0.5,page=2]{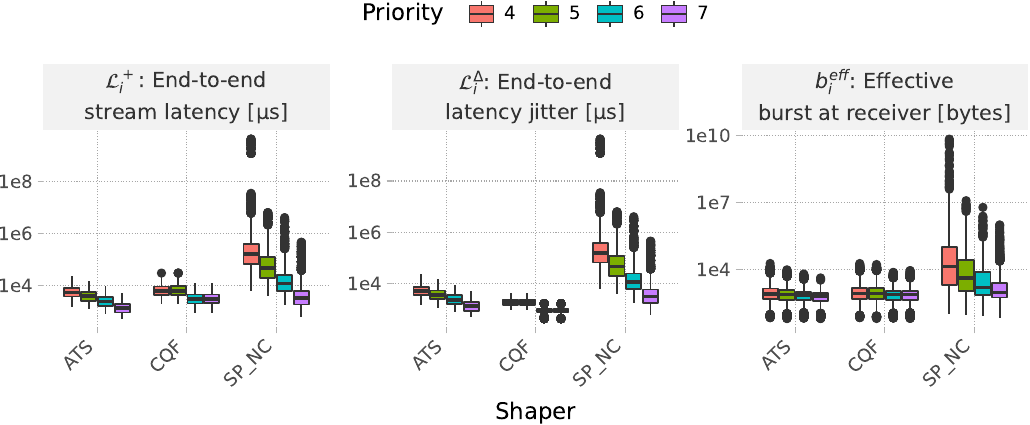}\\[0.5em]
  \caption{Global configuration performance of ATS, CQF and SP\_NC in the medium topology}
  \label{fig:50-eval-acceptall-priorities-gc}
\end{figure}

\paragraph*{Global Configuration Results}
Under global configuration, every link receives the same threshold, which is useful when per-hop optimization is undesired or impossible, e.g., for CQF. This restriction alters several observations from the previous evaluation.

Based on the evaluation procedure from Section~\ref{ssec:eval-autoconf}, step~\ref{enum:config} is modified as follows:
\begin{equation}
  \delta_\mathcal{P}^\textit{seg} = \max_\textit{oseg} \{ d_\mathcal{P}^\textit{oseg} \} \cdot 1.01.
\end{equation}
Each delay segment~\textit{seg} is now configured to the maximum delay $d_\mathcal{P}^\textit{oseg}$ out of all delay segments~\textit{oseg} in the network. Note that this can create circular dependencies: increasing any threshold~$\delta_\mathcal{P}^\textit{seg1}$ will also increase the accumulated latency downstream at other segments \textit{segN}, which will increase their local latency computation~$d_\mathcal{P}^\textit{segN}$. If that delay~$d_\mathcal{P}^\textit{segN}$ now exceeds the previous threshold~$\delta_\mathcal{P}^\textit{seg1}$, the next auto-config iteration will increase the configuration again, potentially rising indefinitely. For that reason, it is often impossible to obtain an optimal global configuration without rejecting any streams. CBSA\_NC and SP\_SIMPLE suffered from this effect and are omitted here.

Figure~\ref{fig:50-eval-acceptall-priorities-gc} shows the global configuration results for medium topologies. ATS outperforms SP\_NC here because SP\_NC burstiness is sensitive to the inflated delay thresholds of global configurations. While ATS end-to-end latency also increases, its burstiness remains bounded by per-hop reshaping. Since all shapers are now restricted to global configuration, CQF achieves the lowest jitter~$\mathcal{L}_i^\Delta$ here. This shows that the latency reduction of CQF is actually effective, although it is usually overshadowed by other effects if optimized per-hop configuration is possible.

\subsection{Single Switch Parameter Study}
\label{ssec:eval-single}

\begin{table}
    \centering
    \caption{Overview of parameter values for the single switch study}
    \label{tab:50-parameter-study}
    \begin{tabular}{>{\hangindent=0.7em}p{10em}
                    >{\hangindent=0.7em}p{4em}
                    >{\hangindent=0.7em}p{12em}} \toprule
        \textbf{Parameter} & \textbf{Default} & \textbf{Experiment Values} \\ \midrule
    
        bandwidth               & \SI{1}{Gbit/s}     & \SI{100}{Mbit/s}, ..., \SI{10}{Gbit/s} \\
        number of ingress ports & 5                  & 1, 2, ..., 30 \\
        total number of streams & 100                & 1, 2, ..., 1000 \\
        total burst             & \SI{32}{kB}        & \SI{6.4}{kB}, \SI{12.8}{kB}, ..., \SI{512}{kB} \\
        total rate percent      & 10\%               & 1\%, 10\%, ..., 75\% \\
        acc max delay           & \SI{500}{\micro s} & 0, \SI{50}{\micro s}, ..., \SI{10}{ms} \\
        acc min delay percent   & 50\%               & 0, 50\%, 99\% of max \\
        num high prio streams   & 50                 & 0, 50 \\
        idle slope ratio         & 1.0                & 0.1, 0.5, ..., 2.0 \\

        \bottomrule
    \end{tabular}
\end{table}

\begin{figure}
  \centering
  \includegraphics[scale=0.5,page=10]{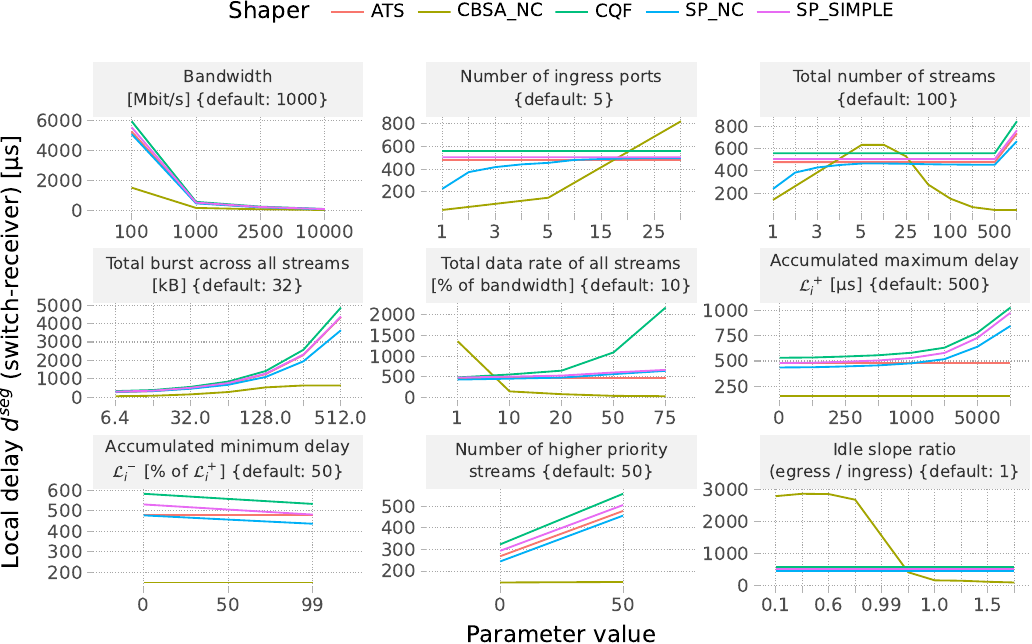}
  \caption{Single switch parameter study results; x-axis values are categorical and not to scale}
  \label{fig:50-eval-single-sw}
\end{figure}

The previous evaluations have shown that the performance of the five latency models differ significantly between the different topology types (linear and industrial small/medium/big). In order to isolate the effects and understand them better, this section offers a parameter study with the single switch from Figure~\ref{fig:50-single-switch}. A main effects plot is generated for the nine parameters listed in Table~\ref{tab:50-parameter-study}. In each experiment, one parameter is varied while the remaining factors are held at their default values.

The parameters are selected carefully to ensure self-contained scaling. For example, if there was a fixed `burst per stream' parameter, then modifying the number of streams would indirectly modify the total burstiness of each latency equation. The observed behavior would likely be dominated by that increased burst, not by the increased number of streams directly. Thus, the parameters from Table~\ref{tab:50-parameter-study} use \emph{total} amounts, or \emph{relative} definitions based on another parameter.

First, the \emph{bandwidth} defines both the bandwidth of ingress links~\textit{prv} and the egress link~\textit{seg}. The \emph{number of ingress ports} represents the count of upstream links~\textit{prv} feeding the switch. More ports increase the number of potentially simultaneous arrivals. The \emph{total number of streams} and \emph{total burst} are distributed evenly across these ingress ports. For example, 10~streams over 3~ports would be distributed as $\{4, 3, 3\}$. Individual bursts must still be kept above \SI{64}{bytes} to comply with minimum Ethernet frame lengths. Therefore, if the number of streams is too high, the effective total burst could be higher than configured. The \emph{total rate percent} scales relative to link bandwidth. The \emph{accumulated maximum delay} is the worst-case delay that every stream carries up to this point. In the previous full-network evaluations, this value was indirectly given by all per-hop thresholds $\delta_\mathcal{P}^\textit{seg}$ on the paths of the streams. Here, the accumulated maximum delay is controlled directly for a single hop to observe the direct effects. In contrast, the \emph{accumulated minimum delay} is given in percent based on the maximum delay, so it automatically scales. The \emph{number of higher priority streams} controls higher-class interference arriving via a dedicated link. The \emph{idle slope ratio} (\textit{isl}) dictates CBSA\_NC idle slope parameters. The default slope is equal to the sum of all streams' rates $\sum_{i \in \mathcal{S}^\textit{seg}} r_i$. For values $\textit{isl} \geq 1.0$, the egress idle slope is multiplied with that value. For values $\textit{isl} < 1.0$, the egress slope cannot be reduced further, as the system would become unstable. Here, the ingress idle slope is divided by \textit{isl} instead, so the parameter always represents the egress/ingress ratio.

\paragraph*{Evaluation Results}
The results of the main effects study are shown in Figure~\ref{fig:50-eval-single-sw}. Overall, CBSA\_NC behaves as an outlier, and it performs much better than in the previous evaluations. This came as a surprise and will be covered by the last parameter, the idle slope ratio. Detailed observations are presented in the following.

(1) Higher \emph{bandwidth} reduces latency across all shapers, as the queues are emptied faster.

(2) The \emph{number of ingress ports} has no effect on CQF, ATS, and SP\_SIMPLE, as it is not part of their model. For SP\_NC and CBSA\_NC, more ports increase the immediate burstiness as well as the slope of the arrival curves. SP\_NC converges toward the SP\_SIMPLE bound as the number of ports exceeds the stream count, which matches previous simulation results~\cite[Fig.\,8]{Grigorjew2020wpsp}. CBSA\_NC has lower arrival slopes because of shaped ingress links, but its service rate is also lowered by egress shaping. After roughly 17~ports, the concurrency penalty outweighs the ingress shaping benefits, and CBSA\_NC delays exceed SP\_NC.

(3) The \emph{number of streams} shows similar effects to the number of ports. The three non-NC models show no effect, except for the last value. For 1000~streams, the individual burst per stream would fall below \SI{64}{bytes}, which is corrected by the model and shows a higher total delay due to more total burstiness. For SP\_NC, delay increases up to 5 streams, which matches the 5 default ingress ports, as streams can occupy individual ports each. The delay decreases slightly beyond 5 streams because the total burstiness is split between all streams, which takes down the maximum frame lengths, too. CBSA\_NC exhibits the same trend on a larger scale due to rate limitations. Interestingly, the additional burstiness at 1000~streams does not show a significant increase in CBSA's latency -- which is consistent with the next observations.

(4) Higher \emph{total burst} increases latency for all models except CBSA\_NC, where delay plateaus after roughly \SI{128}{kB}. Presumably, at this point, the ratio of burst size~$b_i$ to maximum credits~$\textit{maxcr}_\mathcal{C}^\textit{prv}$ and frame sizes~$\ell_i^\textit{max}$ changes such that the linear piece
with rate~$R^\textit{seg}$ from Figure~\ref{fig:40-cbs-arrival-curve} becomes insignificant.

(5) The \emph{total data rate} shows three distinct behaviors. For ATS, the data rate of class~$\mathcal{C}$ has no influence on the delays of the same class, cf. Equation~\ref{eq:ats-delay}. Note that the delay of lower priorities would still be influenced. SP\_SIMPLE, SP\_NC, and CQF show increasing delay with higher rates. CQF exhibits the largest increase, partly due to its consolidated priority interference (Algorithm~\ref{alg:maxInterferenceCQF}). Partly, it is an artifact of the main effects analysis. In real scenarios, CQF's accumulated minimum latency~$\mathcal{L}_i^-$ would be much higher than the default 50\% of~$\mathcal{L}_i^+$ in this evaluation. Since the total data rate is multiplied by jitter~$\mathcal{L}_i^\Delta$ in Algorithm~\ref{alg:maxInterferenceCQF}, the effect here appears stronger than it would be in practice. Finally, CBSA\_NC latency decreases with higher rates, because the increased egress idle slope dominates the effect of ingress slopes in fan-in scenarios.

(6) and (7) -- The \emph{accumulated maximum delay} and \emph{accumulated minimum delay} affect the latency in the same way. Their difference $\mathcal{L}_i^+ - \mathcal{L}_i^-$ is part of most latency models, except for ATS. Larger maximum jitter increases delays for SP\_NC, SP\_SIMPLE, and CQF. While CBSA\_NC incorporates this jitter, its latency appears to be dominated by other arrival curve segments -- at least when the idle slope is given by $\textit{isl}_\mathcal{C}^\textit{seg} = \sum_{i \in \mathcal{S}_\mathcal{C}^\textit{seg}} r_i$.

(8) \emph{Higher priority streams} increase the delay of the observed class similarly across all shapers except CBSA\_NC. Since CBSA service rate is limited to its idle slope, it remains unaffected. The accumulated positive credit~$\textit{maxcr}$ and the initial latency of the service curve~$\tau$ could be affected, but the effect appears to be insignificant in this case.

\begin{figure*}
  \centering
  \rotatebox[origin=l]{90}{\usefont{OT1}{cmss}{m}{n}\scriptsize \hspace{1.5em} curve($\Delta$t) [kB]}%
	\includegraphics[scale=0.45]{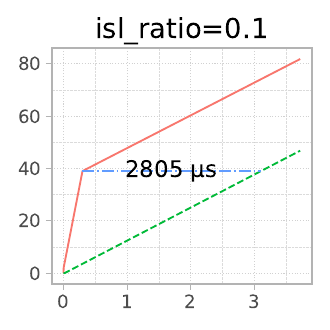}%
	\includegraphics[scale=0.45]{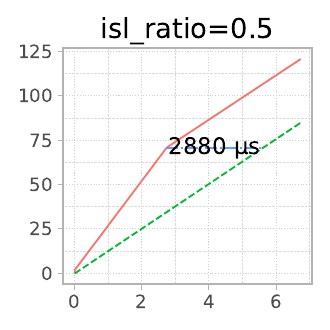}%
	\includegraphics[scale=0.45]{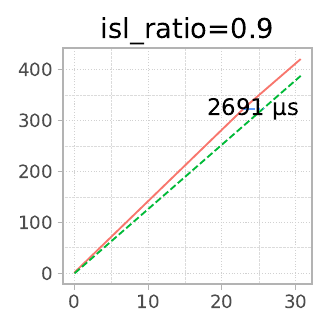}%
	\includegraphics[scale=0.45]{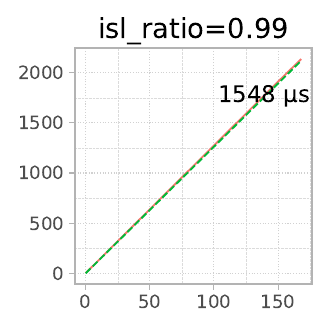}%
	\includegraphics[scale=0.45]{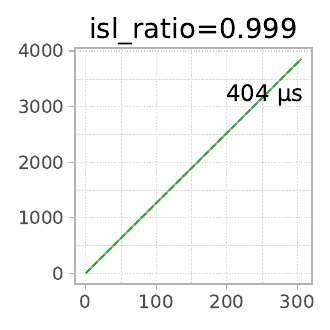}%
	\includegraphics[scale=0.45]{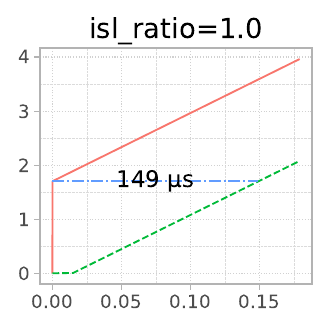}%
	\includegraphics[scale=0.45]{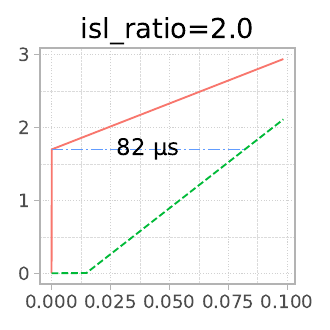}%
	\\ \usefont{OT1}{cmss}{m}{n}\scriptsize time $\Delta$t [ms]
  \caption{Overview of network calculus arrival (red) and service curves (green) for different CBSA idle slope ratios}
  \label{fig:50-eval-cbsa-curves}
\end{figure*}

(9) Finally, the \emph{idle slope ratio} reveals an interesting change of order between the shapers. Non-CBS models are unaffected. For $\textit{isl\_ratio} \geq 1.0$, CBSA\_NC delay is smaller than other shapers. However, for $\textit{isl\_ratio} < 1.0$, when ingress idle slopes exceed egress, CBSA delay rises to nearly 6~times the value of other shapers. This ratio appears to be the primary performance factor for CBSA, explaining why it excels in linear configurations but struggles in fan-out scenarios. In fan-out networks, streams diverge across multiple egress ports, reducing egress idle slopes relative to ingress, which degrades performance in the industrial topologies. This might also explain why CBSA appears to perform much better in shaper comparisons of other papers \cite{Zhao_2022_QuantitativePerformance}. This turning point around $\textit{isl\_ratio} = 1.0$ can be visualized in the network calculus arrival and service curve plots for CBSA\_NC.

Figure~\ref{fig:50-eval-cbsa-curves} shows snapshots of the arrival curves (red) and services curves (green) of the CBSA Network Calculus model for different idle slope rates. It also shows the maximum horizontal difference between both curves, representing the calculated worst-case delay. For the first five plots of Figure~\ref{fig:50-eval-cbsa-curves} where $\textit{isl\_ratio} < 1.0$, the arrival rate of the second segment~$\alpha_{\hat{\mathcal{S}},2}(\Delta t)$ is $isl_\mathcal{C}^\textit{prv}$, which is bigger than the service rate $isl_\mathcal{C}^\textit{seg}$, so the later bending position~$t_{\hat{b}}$ must be selected -- cf. Figure~\ref{fig:40-cbs-arrival-curve}. The bigger $\textit{isl\_ratio}$ becomes, the smaller $isl_\mathcal{C}^\textit{prv}$ becomes in comparison, until it is almost identical to $isl_\mathcal{C}^\textit{seg}$ for $\textit{isl\_ratio} = 0.999$. For the last two plots, at $\textit{isl\_ratio} \geq 1.0$, it follows $isl_\mathcal{C}^\textit{seg} \geq isl_\mathcal{C}^\textit{prv}$ (the green slope is bigger than the red slope), so the first bending position~$t_{\hat{cr}}$ from Figure~\ref{fig:40-cbs-arrival-curve} is selected -- much earlier than before. The delay is now given by the time it takes to service the immediate bursts~$\textit{maxcr}_\mathcal{C}^\textit{prv}$ from the predecessors, and no longer based on the original burstiness of the Traffic Specifications.

These results show that CBS can be a good option if the egress-to-ingress idle slope ratio is maintained above 1.0. This can be achieved via flow steering to avoid fan-out, or by overriding the configuration of idle slopes such that they never become smaller along the path of streams. It strongly depends on the network topology whether this is feasible. If these criteria cannot be met, it should be possible to switch to SP\_NC at any switch on the path without additional delay penalties. Future work could focus on automating these transitions.


\subsection{Lessons learned}
\label{ssec:eval-summary}

It is difficult to provide generalized recommendations for shapers, their relative performance is highly situational. While using default global configurations showed very little differences between shapers (cf.\,Fig.\,\ref{fig:50-eval-preconf-sizes}), optimal per-hop configuration reveals orders of magnitude between them (cf.\,Fig.\,\ref{fig:50-eval-acceptall-sizes}), and a factor of 2.0 (cf.\,Fig.\,\ref{fig:50-eval-acceptall-relative-spnc}) between points that previously looked similar. This section demonstrated how different shapers can still be compared fairly, without introducing bias through non-optimal configurations. The following paragraphs summarize the results for the industrial and linear topologies.

First, the improved SP\_NC outperforms SP\_SIMPLE across all scenarios and KPIs, except for computation time. If signaling and algorithmic complexity are acceptable, SP\_NC should always be preferred (cf.\,Fig.\,\ref{fig:50-eval-acceptall-relative-spnc}). Especially in global configuration scenarios, SP\_SIMPLE often failed to provide \emph{any} solution that would accept all streams due to cyclic dependencies.

Second, the optimal choice of shaper depends on the target KPI and topology. CBS can outperform other shapers if the egress idle slope exceeds the aggregate ingress slopes (cf.\,Fig.\,\ref{fig:50-eval-single-sw}). In scenarios with high fan-out, SP\_NC was always a better choice. Similarly, CQF excels at minimizing jitter under global configurations (cf.\,Fig.\,\ref{fig:50-eval-acceptall-priorities-gc}), but with local threshold optimization, SP\_NC performed better with respect to latency \emph{and} jitter. Finally, ATS performed well throughout all evaluations as it negates burstiness cascading. It showed the best results with respect to burstiness at the receiver, and it performed similarly to SP\_NC regarding worst-case latency. The serialization effect of SP\_NC can still outperform ATS (cf.\,Fig.\,\ref{fig:50-eval-acceptall-relative-spnc}, priority~7), especially when there are switches with very few ports. At lower priorities, ATS consistently outperforms SP\_NC (cf.\,Fig.\,\ref{fig:50-eval-acceptall-relative-spnc}, priority~4). The primary disadvantage of ATS is hardware cost and availability of the shaper.

Overall, for decentralized architectures, SP\_NC is a robust choice across all scenarios. The use of specialized shapers (CBS, CQF, ATS) remains situational, but can be beneficial, and should be verified per use case.

\section{Discussion and Future Work}
\label{sec:discussion}

\newcounter{discussct}
\newcommand\discussctpp{\refstepcounter{discussct}\thediscussct}

While this paper attempts to be self-contained and comprehensive, some caveats must be understood to put the results in full context. This tutorial is centered on the decentralized applicability of the presented concepts. Consequently, the shaper-to-shaper delay segment model (Fig.\,\ref{fig:30-delay-segments}) from the ATS delay analysis was adopted for all mechanisms. While this adaptation enables decentralized CQF applicability, it introduces some distinct tradeoffs.

(\discussctpp) One major tradeoff is already covered in Table~\ref{tab:35-info-flow} (Section~\ref{sec:information-flow}). To compute the worst-case queuing latency of the upstream switch, the downstream switch requires additional information that must be signaled during reservation. This requirement complicates the design and standardization of reservation protocols, but it has negligible consequences afterwards.

(\discussctpp) The delay of a single segment can no longer be measured directly. Although the delay of an entire device can be captured by tapping links before and after that device, it is not so simple to derive the exact time when the shaping step is finished \emph{inside} the device. Verification testbeds would require access to internal component logs, and even then, time stamps across different devices remain difficult to compare without tight time synchronization.

(\discussctpp) Hybrid deployments and horizontal interoperability are simplified by a shared delay segment definition. For example, SP\_NC can be deployed globally, while a few critical switches utilize ATS to reshape traffic and counter burstiness cascading. Such hybrid topologies offer targeted performance benefits at a lower deployment cost, however, they require theoretical adjustments. First, current latency models still assume homogeneous upstream shaping. The ATS model must be adapted to include additional shaping delay if the predecessor is non-shaping. Second, current SP and SP\_NC models recompute the effective burstiness $b_i^\textit{eff}$ at each hop, independently of other shapers along the path. To resolve this, reservation protocols should propagate both the current \emph{effective} Traffic Specification along with the original \emph{talker} Traffic Specification. This allows shapers to signal their changes to the effective burstiness further downstream. Recent RAP drafts already introduce separate Talker and Network TSpecs for this purpose (cf. \cite[Sec.\,6.3.10]{8021Qdd}).

(\discussctpp) Beyond horizontal interoperability, i.e., different shapers used in different delay segments, vertical interoperability should also be considered, i.e., different shapers used by different traffic classes of the same device. While the logic from Section~\ref{sec:latency-bounds} can be extended to derive combined class models, as demonstrated in~\cite{Zhao_2022_QuantitativePerformance}, this was omitted here due to the vast combination space. When combined shaper setups are desired, the latency models should be adapted accordingly.

(\discussctpp) The shaper-to-shaper model is not fully compatible with all common networking mechanisms, including some analyzed in this work. As noted in Section~\ref{ssec:cbs}, CBS is non-compliant because its shaping state transitions occur during frame transmission, which is subject to queuing delays. Consequently, for CBS, the formal \emph{shaping} step includes the queuing and transmission phases, leaving the literal \emph{queuing} step empty (cf. Fig.\,\ref{fig:40-delay-model-cbs}). Unlike the other four models, CBS requires each device to compute \emph{its own} queuing latency. While this fits homogeneous CBS networks, hybrid transitions (e.g., ATS to CBS) require downstream bridges to compute both upstream and downstream queuing delays. Exposing these transition boundaries within the reservation protocol could provide additional flexibility. While CBSA must always compute its own queuing delay downstream, and ATS and CQF always compute the upstream's queuing delay, the SP models are free to be designed either way for full compatibility. In \cite[p.\,14]{grigorjew2023latencymodel}, we have already suggested to include an indicator in the reservation protocol to specify whether the upstream device has already dealt with its own queuing delay, or left it to the downstream device to consider. This means that the queuing process no longer needs to be priority based. For example, end devices that are not capable of priority transmission selection could employ weighted fair queuing instead. As long as a proven queuing delay bound exists, the upstream device can use it, indicate that it has dealt with the queuing delay, and the downstream switch no longer has to bother.

(\discussctpp) Special care is required when using reservation protocols with redundancy. In Frame Replication and Elimination for Redundancy (FRER), a sequence number is encoded into the frame by adding a tag, which increases the frame's length and the stream's burst size. Furthermore, even after duplicate elimination, the burst size of the merged stream must often be doubled to consider worst-case events. If maximum latency jitter is high enough on both redundant paths, the frames may experience different delays on both paths. This is usually fine, as the second occurrence of every frame is eliminated anyways, but it creates a problem if the quicker path is temporarily disrupted. After the link is re-established, the merging point receives older frames from the slower link, and more recent frames from the quicker link. Both of these are valid and will not be eliminated, which temporarily causes a higher burst after merging, until all of the older frames from the slower link are exhausted and the merging point begins to eliminate duplicates again. This is explained in \cite{grigorjew2023resilience} and \cite[Sec.\,C.9]{8021CB} in more detail. Under our framework, ATS could be deployed at the merging point to immediately reshape the traffic, though the combination of FRER and ATS remains subject to ongoing standardization.

(\discussctpp) This work provides the theoretical foundation for resource reservation, but the ongoing protocol design, configuration, and implementation of resource reservation frameworks are subject of other works. We have conducted complementary studies in some of these aspects. Contributions were submitted to the IEEE 802.1DD RAP standard \cite{grigorjew2020bridgelocal,grigorjew2021trafficspecification,grigorjew2022measurementpoints,grigorjew2023latencymodel,grigorjew2023somenecessary}, and decentralized reservation performance was evaluated via discrete event simulation \cite{Raunecker_2024_CentralizedVs}. Recently, a proof-of-concept implementation was presented in a demo, including a management API for visualization and root cause analysis \cite{Grigorjew_2025_TsnReservation}. Additionally, this study assumes that stream paths and priorities are known a priori. In practice, assigning streams to priorities is an optimization problem itself. Configuration heuristics exist \cite{grigorjew2021mlassisted_MlassistedLatency,maile2023combining}, but they typically assume that all streams are perfectly known a priori. Ideally, a network configuration tool could also work with rough estimates or categories of utilization. A fully decentralized and automated configuration of priority assignments and delay thresholds could significantly simplify the installation process of real time networks.

(\discussctpp) The latency models in this work do not cover all sources of delay. As categorized in Section~\ref{ssec:types-of-delay}, only transmission and queuing delays are covered. Propagation and internal processing delays must be bounded separately. Decentralized reservation benefits here, as manufacturers can integrate device-specific upper bounds directly into their local protocol stacks. For centralized architectures using a CNC, these bounds must instead be exposed externally, such as through the \emph{digital data sheets} proposed in the IEEE/IEC 60802 industrial profile \cite[Sec.\,6.4.8.2]{iecieee60802}, which may not fit all custom implementations and can require extensive measurements.

(\discussctpp) While Time-Aware Shaping (TAS) is widely studied, it has not been covered in this work (except for CQF) due to its reliance on central orchestration. TAS can achieve full frame isolation and zero interference, but it is highly resource-intensive as entire links are reserved for specific streams for a certain time. Consequently, low latency under TAS does not inherently maximize stream density. Future work could evaluate the tradeoffs between asynchronous decentralized concepts and TAS scheduling. Further, both concepts could be combined, e.g., decentralized reservation could bridge the gap between two individual, differently configured TAS domains. This approach could also provide end-to-end guarantees beyond layer~2 networking, across heterogeneous network domains. For example, this has previously been suggested for joint 5G/TSN deployments as an alternative to the logical TSN bridge abstraction \cite{grigorjew2024time}.

\subsection*{Lessons learned}
The latency models presented in this work lay the groundwork for flexible and decentralized resource reservation. This section pointed out opportunities and requirements for improved heterogeneity. They include additional edge cases in the latency computation, or they add extra information to the reservation protocol. Concrete suggestions include the addition of an effective \emph{Network TSpec} to the original \emph{Talker TSpec}, and an indicator flag identifying whether the upstream device has already accounted for its local queuing delay. Together, these extensions support seamless shaper transitions without sacrificing their individual benefits.

\section{Conclusion}
\label{sec:conclusion}

In fields like industrial automation and automotive networks, legacy field bus technologies are slowly replaced by Ethernet to leverage higher bandwidth and standardized support. This transition enables converged networking, where real-time and Best-Effort traffic coexist on a single infrastructure. However, current TSN implementations remain monolithic, similar to the old field bus systems. True interoperability can be achieved with decentralized protocols supported by robust latency models.

This work addresses the challenges of decentralized, heterogeneous deployments with hard real-time guarantees. By formalizing delay segments as the borders of responsibility, shapers like ATS and CQF can be integrated into decentralized architectures. The necessary information flows are identified, including metadata from the upstream devices. Using these foundations, five latency models were derived, ranging from simple formulations (SP\_SIMPLE, ATS, CQF) to Network Calculus implementations (SP\_NC, CBSA\_NC). The models were evaluated using a comprehensive suite of KPIs across linear and industrial topologies.

The evaluations show that priority transmission selection without active shaping remains highly effective. The SP\_NC model consistently ranked among the top-performing approaches. If its computational complexity is acceptable, SP\_NC should be preferred over SP\_SIMPLE. Furthermore, ATS is a highly robust choice for transmission selection, offering simple delay calculations and excellent receiver burstiness bounds. In contrast, CQF and CBSA\_NC are highly situational. When per-hop delay threshold optimization is possible, SP\_NC outperforms CQF, as the benefits of threshold optimization exceed the jitter reduction of global CQF configurations. Finally, CBSA\_NC is effective in topologies with minimal fan-out, where the egress idle slope consistently matches or exceeds the aggregate ingress idle slopes. If the idle slope ratio falls below 1.0 on any link, CBSA\_NC performance degrades rapidly. Because this occurred in almost all tested industrial topologies, SP\_NC is generally recommended instead of CBSA\_NC here. Note that these results highly depend on the topology and stream configuration, but the presented framework can be used to evaluate any given topology.

The contributions in this work are specifically targeted for deterministic delay guarantees in heterogeneous networks. By formalizing the boundaries between different shapers and clarifying the necessary signaling, this tutorial establishes the mathematical basis for decentralized resource reservation protocols and technological transitions. Future work could branch into several directions, from improved latency models with mixed shapers, to fully self-configuring latency thresholds in decentralized networks. Finally, this work ties directly into the standardization efforts of the IEEE 802.1DD Resource Allocation Protocol and can be used as a direct reference there.

\printbibliography

\end{document}